\documentclass{article}
\usepackage[utf8]{inputenc}
\usepackage{amsmath}
\usepackage{mathtools}
\usepackage{breqn}
\usepackage{authblk}
\usepackage{amsfonts}
\usepackage{enumerate}
\usepackage{enumitem}
\usepackage{amssymb}
\usepackage[pdf]{pstricks}
\usepackage{comment}
\usepackage{graphics}
\usepackage{graphicx,url}
\usepackage{bm}
\usepackage{empheq}
\usepackage[top=2.5cm,bottom=2.5cm,right=2cm,left=2cm]{geometry}
\usepackage{float} 
\usepackage{amsthm}
\usepackage{bigints}
\usepackage{stackengine}
\usepackage{color} 
\usepackage{bbm}
\usepackage{appendix}
\usepackage{bbold}
\usepackage[nottoc]{tocbibind}
\usepackage{braket}
\usepackage{algorithm}
\usepackage[noend]{algpseudocode}
\usepackage{textcomp}
\usepackage{listings}
\usepackage{cite}

\usepackage{mathrsfs}

\usepackage{hyperref}
\usepackage{cleveref}
\crefname{appsec}{Appendix}{Appendices}

\stackMath

\usepackage{enumerate}

\usepackage{caption}

\usepackage{tikz}
\usetikzlibrary{arrows}

\makeatletter
\def\BState{\State\hskip-\ALG@thistlm}
\makeatother

\DeclareMathOperator{\tr}{Tr}

\newcommand{\mb}{\mathbf}

\newcommand{\Le}{\left}
\newcommand{\Ri}{\right}
\newcommand{\lla}{\left \langle}
\newcommand{\rra}{\right \rangle}
\newcommand{\p}{\partial}
\newcommand{\f}{\frac}

\newcommand{\mc}{\mathcal}

\newcommand{\mr}{\mathrm}
\newcommand{\bs}{\boldsymbol}
 
\newcommand{\nn}{\nonumber} 
\newcommand{\llv}{\left\lvert}
\newcommand{\rrv}{\right\rvert}

\newcommand{\tcb}{\textcolor{blue}}

\DeclarePairedDelimiter\floor{\lfloor}{\rfloor}
\def\multiset#1#2{\ensuremath{\left(\kern-.3em\left(\genfrac{}{}{0pt}{}{#1}{#2}\right)\kern-.3em\right)}}

\newcommand\numeq[1]%
{\stackrel{\scriptscriptstyle(\mkern-1.5mu#1\mkern-1.5mu)}{=}}

\renewcommand\bra[1]{{\langle{#1}|}}
\renewcommand\ket[1]{{|{#1}\rangle}}

	\title{Structural localization in the Classical and Quantum Fermi-Pasta-Ulam Model}
		\date{\today}
	\author[1]{Graziano Amati}
	\author[2]{Tanja Schilling}
	\affil[1,2]{Physikalisches Institut, Albert-Ludwigs-Universität, 79104 Freiburg, Germany }

\begin{document}
\maketitle
\section*{Abstract}
We study the statistics and short-times dynamics of the classical and the quantum Fermi-Pasta-Ulam chain in thermal equilibrium. We analyze the distributions of single-particle configurations by integrating out the rest of the system. At low temperatures we observe a systematic increase in the mobility of the chain when transitioning from classical to quantum mechanics due to zero-point energy effects. We  analyze the consequences of the quantum dispersion on the dynamics at short times of configurational correlation functions. 
\\ \\
{\bf{The Fermi-Pasta-Ulam system is a chain of particles with non-linear interactions between nearest neighbors. This model serves as a prototype for a plethora of complex systems, from DNA structures to polymer chains. The chain has been extensively studied within classical mechanics, and non-trivial localization between the fundamental components has been observed. This phenomenology is associated to anomalous behavior in the energy transfer between the vibrational modes of the chain. In many microscopic systems, however, quantum effects play a major role, especially at low temperature. Therefore, the behavior of the model in the quantum regime can critically differ from the classical counterpart. In the present work we study by numerical means the effects of quantum dispersion on thermal fluctuations, and the implications on the short-times dynamics of the system.}}

\section{Introduction}
Since its first formulation in 1955 \cite{fermi_pasta_ulam1955}, the Fermi-Pasta-Ulam (FPU) model has been the subject of fundamental discussions in the field of nonlinear dynamics. The system consists of a one-dimensional chain of classical particles, interacting between nearest neighbors (n.n.) with a weakly nonlinear interaction. The fame of the model is due to the fact that, despite its simplicity, it highlights the limits of Ergodic Theory. In particular, the expected thermalization of energy between the normal modes is hindered, provided only the lowest frequencies of the system are excited, and the total energy is sufficiently low. This apparent paradox has been discussed in terms of perturbation theory, applications of the KAM Theorem and the propagation of solitons (comprehensive reviews can be found in \cite{gavallotti} and \cite{berman2005}). A number of extensions of the original system have been conceived since its formulation, e.g.~by considering longer-ranged interactions \cite{bachelard2008} or poly-atomic particles \cite{gaison2014}. Analogues of the original phenomenology have been observed in these generalizations, elevating the so-called ``FPU paradox'' to a more general problem in Physics. The non-trivial behaviour of the model can also be observed under more typical initial conditions than the strongly out-of-equilibrium ones originally chosen; e.g.~it has been shown that the system in thermal equilibrium exhibits anomalous relaxation of the time-dependent specific heat of the modes\cite{carati2007}. Memory effects can identified in the dynamics of the chain at equilibrium by studying the intermediate scattering function \cite{amati2019}. \\
There are several reasons behind an increasing interest in recent years in the quantum mechanical extension of the classical FPU Model. Firstly, the takeover of chaos in the quantum system can be directly connected to the discreteness of the energy levels \cite{burin2019}, \cite{edwards1971}, \cite{leitner1997}. Additionally, the quantization of the chain into a bosonic system allows to identify and study discrete breathers in a quantum mechanical setting \cite{ivic2006}, \cite{kibey2015}. And a connection between the FPU and the Bose-Hubbard model has been drawn \cite{danshita2014}. Most of the works on the quantum FPU model rely on the reformulation of the Hamiltonian of the system in terms of creation and annihilation operators. As an alternative quantization procedure, in the present work we stick to the direct quantization of the configurations and momenta of the FPU chain. We construct and analyze correlation functions expressed in terms of path integrals \cite{feynman1948}, both in real and imaginary times. These path integrals are solved numerically via techniques based on their discretization, from the seminal work of \cite{tuckerman1993}. We can then draw a direct comparison with the corresponding classical results. In particular, we focus on the thermal distribution of the configurations and their auto-correlation functions in time. The results show that the quantum system exhibits increased thermal fluctuations with respect to the classical limit. This property allows to identify a temperature threshold in the nonlinear system below which the  zero-point-energy (ZPE) effects are statistically relevant. We then analyze the consequence of the increased mobility in a dynamical framework, by connecting thermal correlations to real-times auto-correlation functions. We show that the increased thermal fluctuations imply a speed up of the correlation loss for the system at short times. \\
The present work is organized as follows: In \cref{sect:model} we define the quantum FPU Model, and we show why this system is naturally suited to be identified with a classical isomorphism for an effective numerical implementation. In \cref{sect:statistics_theo} we discuss a scheme for the calculation of the microscopic statistics of the system. The method is applied to the analysis of the canonical displacements in \cref{subsec:distrib_q}. Finally, in \cref{dynamics} we analyze the short-times dynamics of configurational auto-correlation functions. The appendices contain the technical details behind the numerical implementation and provide further analytical insights.
\section{The Quantum FPU Model}\label{sect:model}
\begin{figure}[H]
	\centering
		\includegraphics[scale=0.8]{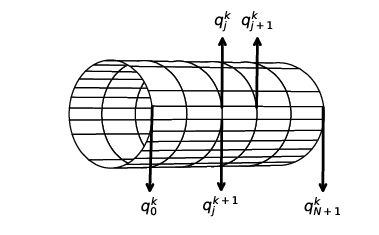}\caption{Classical isomorphism of the quantum FPU model; each horizontal line denotes a classical anharmonic chain, with fixed b.c. In the orthogonal direction there is a periodic harmonic interaction of infinitely many classical instances.}\label{fig:sketch}
              \end{figure}

\begin{figure}[H]
	\centering
	\includegraphics[scale=0.6]{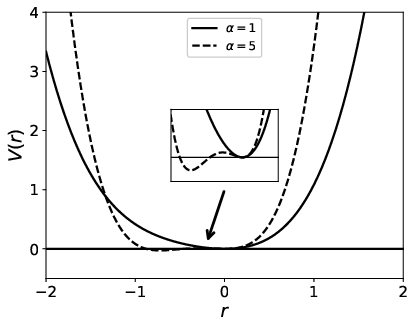}
	\caption{Two body potential in \cref{def_H}, for $\alpha=\beta$ }\label{fig:potential}
\end{figure}

The quantum FPU Hamiltonian for a system of $2N$ distinguishable d.o.f. ($N$ in configurations and $N$ in momenta) is defined by
 \begin{equation}\label{def_H}
\hat H(\mb{\hat q}, \mb{\hat p}) = \sum_{j=0}^N\left[\frac{\hat p_j^2}{2}+\f 12 (\hat q_{j+1}-\hat q_j)^2+\frac \alpha 3 (\hat q_{j+1}-\hat q_j)^3+\frac{\beta}{4}(\hat q_{j+1}-\hat q_j)^4 \right]\equiv \hat T(\mb{\hat p})+\hat V(\mb{\hat q})
\end{equation}
with Dirichlet boundary conditions (b.c.)
\begin{equation*}
\hat q_0 = \hat p_0 = \hat q_{N+1} =  \hat p_{N+1} =  \hat 0
\end{equation*}
where $\hat 0 = 0 \hat I$ is the null operator. Both the anharmonic parameters $\alpha$ and $\beta$ and the phase space coordinates and are dimensionless. The latter satisfy the canonical commutation relations (CCR)
\begin{equation}\label{def_CCR}
[\hat q_i, \hat p_j] = \hat \delta_{ij}=\hat I \delta_{ij} \hspace{5mm} [\hat q_i, \hat q_j]=[\hat p_i, \hat p_j] = \hat 0 \hspace{10mm} i,j=1\cdots N
\end{equation} 
In \cref{app:deriv_H} we discuss how \cref{def_H} can be constructed as a fourth order expansion from a physical potential with n.n. interaction. The partition function for the system in thermal equilibrium is defined by
\begin{align}\label{def_Z_T}
&Z_T \equiv \tr\left\{ e^{-\hat H/T}\right\} =\left. \int_{\mathbb R^N} \mathrm d \mathbf q\; \bra {\mathbf {q}} e^{-\hat H/T}  \ket {\mathbf q} \right|_{\hat q_0=\hat q_{N+1}=\hat 0} 
\end{align}
where $T$ is a dimensionless temperature and the trace is taken over the continuous basis of the configurations. (The ensemble stems from a canonical distribution, as defined in \cref{def_Xi_theta} in \cref{app:canonical}, where we discuss the notion of constant volume for the quantum mechanical case.) In order to allow for a numerical treatment of path integrals as in \cref{def_Z_T}, a number of discretization techniques have been developed under the name of Path Integral Molecular Dynamics (PIMD) \cite{tuckerman1993}. As discussed in detail in \cref{app:class_iso} and \cref{app:PIMD}, these methods are based on the concept that the quantum trace in \cref{def_Z_T} is isomorphic to the one of an infinite-dimensional Newtonian system; this consists of a bundle of classical chains placed in a periodically connected network, where correspondent particles of neighboring systems interact via harmonic springs. The numerical calculation of quantum thermal traces is then approximated by considering a sufficiently large number of $P\gg 1$ instances of the classical systems, and computing thermal averages with the two-dimensional Hamiltonian 
\begin{equation}\label{def_H_cl_scalar}
\mathcal H^{\mathrm{cl}}(q^1_1,\cdots, q_N^P, p_1^1, \cdots, p_N^P) \equiv  \sum_{j=0}^N\sum_{k=1}^P \Le[ \frac 1 2 (p_j^k)^2+ \frac{\omega_P^2}2 \left ( q_j^{k+1}- q_j^k \right )^2 +\frac 1 P V(q_{j+1}^k-q_j^k)\Ri] 
\end{equation}
with b.c.
\begin{equation*}
q_0^k=q_{N+1}^k=0 \hspace{5mm} k = 1\cdots P\hspace{10mm}q_j^{P+1}=q_j^1 \hspace{5mm} j = 1\cdots N
\end{equation*}
We will refer to this mapping as \textit{classical isomorphism} \cite{chandler1981}. \cref{def_H_cl_scalar} defines a classical Hamiltonian system in two dimensions, with nonlinear forces in the ``direction'' $j\in \{1,\cdots N\}$, plus harmonic springs along the dimension identified by the indices $k\in \{1,\cdots P\}$. Fixed boundary conditions are taken along $j$ and periodic ones along $k$, generating an effective cylindrical geometry. The relevance of the mapping is that the quantum statistics is exactly recovered in the $P\to+\infty$ limit. Conversely, in the limit of $P=1$, the isomorphism collapses to the classical model. A schematic representation of the system described by \cref{def_H_cl_scalar} is sketched \cref{fig:sketch}.
              In the following, for the two body interaction we will consider identical values for the harmonic parameters $\alpha=\beta>0$, as motivated in \cref{alpha_eq_beta}. A positive value for the quartic term ensures a global confinement for the system, thus preventing the breakdown of the chain \cite{carati2018}. The shapes of the n.n. potentials used in the work are shown in \cref{fig:potential}.
For $ \alpha \ge 4$ the potential obtains a second stationary point apart from the one at $r=0$. The non-monotonicity of $V'(r)$ in this regime increases the complexity of the global potential energy surface. This enhances the probability of trapping the classical system at low temperature in one of the many available disordered minima \cite{carati2015}. In the next section we present a numerical approach aimed at identifying localization of the configurations in the Canonical ensemble; this will allow us to compare the classical and quantum statistics on an equal footing, and to asses analogies and differences between them.

\section{Statistical analysis}\label{sect:statistics_theo}
Quantum states in the canonical ensemble can be efficiently sampled by selecting a set of snapshots from the equilibrium dynamics of the classical isomorphism. We can then compare expectation values obtained for the classical model (in the $P=1$ limit of a single replica) with the corresponding approximations to the quantum case ($P>1$), in order to evaluate how equilibrium averages are transformed from one regime to the other. In the following analysis we study traces of observables only dependent on the positions and not the momenta. The numerical procedure for the calculation of the latter is not as straightforward as for the former, due to the fact that the configurations are the natural basis used in the expansion in \cref{def_Z_T} (see \cref{exp_val_iso} for a discussion). \\
A detailed description of the PIMD techniques used in the numerical sampling can be found in \cref{app:class_iso} and \cref{app:PIMD}. For the purpose of the following discussion, it suffices to mention that we attached the classical isomorphism of the quantum system to a massive thermostat. In particular, we connected each of the $NP$ momenta to $M=5$ bath particles, arranged in Nos\'e-Hoover chains \cite{martyna1996}.  We then collected $N_{\text{samp}}=5000$ uncorrelated phase space points from a long molecular dynamics trajectory, and we computed expectation values by averaging over the samples. The reliability of the integrator (discussed in detail in \cref{app:PIMD}) has been assessed through the conservation of the total energy (see \cref{def_H'cl} for the definition), which is the unique constant of motion for the system coupling the classical isomorphism and the thermostats. This energy remains constant with six-digits precision in each of the computed PIMD trajectories. In all the simulations  we fixed the number of particles to $N=8$. This rather small number of physical d.o.f.~allowed us to study the effect of the number of instances $P$. Several aspects can influence whether a specific finite approximation for the value of $P$ suffices to effectively reconstruct the quantum statistics. For example, an increasing number of replicas will be in general needed at lower temperatures, where the relevance of the quantum effects increases. In order to avoid any \textit{a priori} assumption on the choice of $P$, we computed the statistics for increasing values $P=1, 16, 32$ and $64$. We then accepted the statistics once convergence was reached. For the purposes of the present study, already the simplest quantum approximation $P=16$ turned out to converge satisfactorily.

\subsection{Configurational distributions}\label{subsec:distrib_q}
\begin{figure}[H]
	\centering
	\begin{minipage}{.45\textwidth}
		\centering
		\includegraphics[scale=0.55]{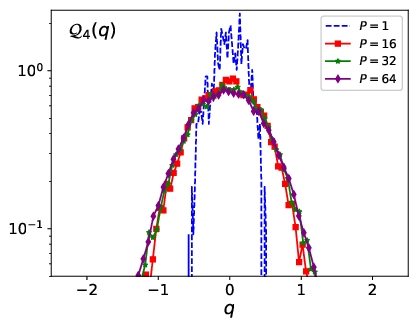}
		\caption{Probability to find a displacement $q$ from the ground state position at $T=0.01$, $\alpha=5$, for $j=4$ (center of chain)}\label{fig:Q_j_cent1}
	\end{minipage}
	\hfill
	\begin{minipage}{.45\textwidth}
		\centering
		\includegraphics[scale=0.55]{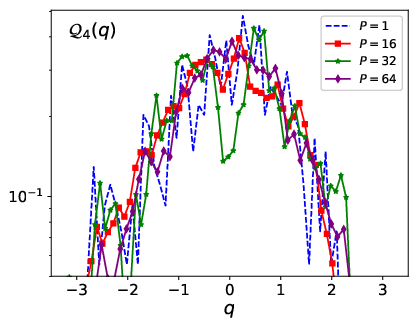}
		\caption{Probability to find a displacement $q$ from the ground state position at $T=5$, $\alpha=5$, for $j=4$ (center of chain)}\label{fig:Q_j_cent2}
	\end{minipage}
\end{figure}

\begin{figure}[H] 
	\centering
	\begin{minipage}{.45\textwidth}
		\centering
		\includegraphics[scale=0.55]{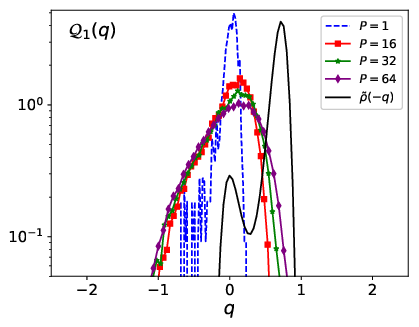}
		\caption{Probability to find a displacement $q$ from the ground state position at $T=0.01$, $\alpha=5$, for $j=1$ (left boundary)}\label{fig:Q_j_1}
	\end{minipage}
	\hfill
	\begin{minipage}{.45\textwidth}
		\centering
		\includegraphics[scale=0.55]{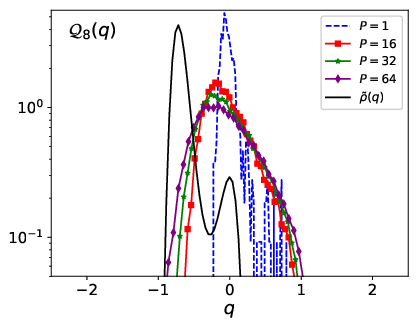}
		\caption{Probability to find a displacement $q$ from the ground state position at $T=0.01$, $\alpha=5$, for $j=8$ (right boundary)}\label{fig:Q_j_2}
	\end{minipage}
\end{figure}

In this section we discuss the formalism for the analysis of the configurations of a statistical mixture in Canonical equilibrium. The methods are then directly applied to the quantum FPU chain. The choice of observables that we compute in the following has been motivated by classical liquid state theory\cite{hansen_mdonald1986}, where n-particle densities are a basic ingredient.\\
We are in general interested in studying the configurations of arbitrary subsets of d.o.f. 
\begin{equation*}
\mathcal J = \{j_{1}, \cdots, j_n\} \subseteq \{1,\cdots, N\}
\end{equation*} 
This can be accomplished by analyzing the thermal traces
\begin{align}
&\mathcal Q_{\mathcal J}(q_{j_1},\cdots, q_{j_n}) \equiv\left. \frac 1 {Z_T} \tr\left\{e^{-\hat H/T}\prod_{m=1}^n\hat\delta(\hat q_{j_m}-\hat I q_{j_m})\right\} \right|_{\hat q_0=\hat q_{N+1}=\hat 0}= \label{def_multiv_Q0} \\
&\numeq{\ref{ave_rho_cl}} \lim_{P\to+\infty} \frac 1 {Z_{T,P}}\int_{\mathbb R^{2NP}} \mathrm d \mathbf q^1\cdots \mathrm  d \mathbf q^P \mathrm d \mathbf p^1\cdots \mathrm d \mathbf p^P \; \left.\exp\Le\{-\frac 1 T\mathcal H^{\mathrm{cl}}(\mathbf q^1\cdots \mathbf q^P)\Ri\}
\frac 1 P \sum_{k=1}^P \prod_{m=1}^n \delta (q_{j_m}^k- q_{j_m})\right|_{\substack{\mathbf q^{P+1}= \mathbf q^{1}\\ q_0^l=q_{N+1}^l=0,\;\;l=1\cdots P}} 
= \label{def_multiv_Qhalf}\\
&\equiv\lim_{P\to+\infty}\frac 1 P\sum_{k=1}^P \mathcal Q_{\mathcal J}^k(q_{j_1},\cdots, q_{j_n}) \label{def_multiv_Q}
\end{align}
In \cref{def_multiv_Q} we introduced the average over the canonical distribution of the classical isomorphism
\begin{equation}\label{def_QJk}
\mathcal Q_{\mathcal J}^k(q_{j_1},\cdots, q_{j_n}) \equiv \left\langle \prod_{m=1}^n \delta (q_{j_m}^k- q_{j_m}) \right\rangle_{T,P}
\end{equation}
whose dependence on $T$ and $P$ on the l.h.s. is implied for simplicity. \cref{def_multiv_Q0} is well defined and finite, as it involves the space integral of a trace class operator~\cite{gosson2017}, the Boltzmann density, over a $(N-n)$-dimensional hypersurface of the configuration space. We notice that all the distributions $\mathcal Q_{\mathcal J}^k$, $k=1\cdots P$ are equivalent, as the canonical distribution of the classical isomorphism in \cref{def_multiv_Qhalf} is invariant under permutation of the $P$ instances. However, with the simulation data it is still convenient to compute each of these $P$ distributions and to average over the replicas, in order to gain a $P$ factor in the statistics. (It is shown in \crefrange{fig:F_k_id1}{fig:F_k_id3} that the numerical results support this symmetry.) We list in the following a few relevant properties of these distributions, that can be directly inferred from their definition.
The computation of the multivariate distributions \cref{def_multiv_Q0} allows for a convenient dimensional reduction in the calculation of thermal traces of $n$-body operators, for $n<N$:
\begin{equation}\label{f_qJ_ave}
\lla \hat f(\hat q_{j_1},\cdots, \hat q_{j_n}) \rra_T =  \int_{\mathbb R^n} \mathrm d q_{j_1}\cdots \mathrm d q_{j_n}\; \mathcal Q_{\mathcal J}(q_{j_1},\cdots, q_{j_n}) f(q_{j_1},\cdots, q_{j_n}) 
\end{equation} 
Distributions of $n-1$ d.o.f. can be extracted from marginals in higher dimensions. For example, given $j_i \in \mathcal J$, we can determine
\begin{equation}\label{QJi}
\mathcal Q_{\mathcal J \backslash j_i}(q_1, q_{j_{i-1}}, q_{j_{i+1}}, q_{j_n}) = \int_{\mathbb R} \mathrm d q_{j_i} \; \mathcal Q_{\mathcal J} (q_1, \cdots, q_{j_n})
\end{equation}
Normalization is inherited from higher dimensions, e.g. $\mathcal Q_{\mathcal J \backslash j_i}$ in \cref{QJi} is normalized to one, provided the same holds for $\mathcal Q_{\mathcal J}$. The $n$-point distributions fulfill a symmetry condition w.r.t.~the center of the lattice, both for the classical and quantum statistics:
\begin{equation}\label{symm_center}
\mathcal Q_{\mathcal J}(q_{j_1}, \cdots, q_{j_n})=\mathcal Q_{\mathcal J^*}(-q_{j^*_1}, \cdots, -q_{j^*_n}),\hspace{5mm} \mathcal J^* = \{j^*_1, \cdots, j^*_n\}, \hspace{5mm} j^*_l = j_{N+1-l} \hspace{5mm}  l = 1\cdots n
\end{equation} 
In the present work we analyze these distributions in the domain of the configurations. However, it could be experimentally relevant to study the corresponding scattering patterns in reciprocal space. In particular, we could define a static structure factor for the subsystem $\mathcal J$ as
\begin{equation}\label{S_j}
S_{\mathcal J}(\bs \kappa) \equiv \int_{\mathbb R^n} \mathrm d q_{j_1}\cdots \mathrm d q_{j_n}\; \prod_{m=1}^n e^{i\kappa_{j_n} q_{j_n}} \mathcal Q_{\mathcal J}(q_{j_1}, \cdots, q_{j_n})
\end{equation}
where $\bs \kappa\in \mathbb R^n$ denotes the wavevector of the scattered image of a radiation which couples exclusively to $\mathcal J$. While \cref{S_j} involves a continuous Fourier transform, we show in \cref{app:harm_lim} how the discrete counterpart of the expression above allows to treat analytically these distributions in cases where the two-body potential can be identified, either approximately or exactly, with a harmonic interaction.\\
By taking single-particle subsystems $\mathcal J \equiv \{j\}$, we can specify \cref{def_multiv_Q0} as
 \begin{align}\label{def_Qj}
 &\mathcal Q_j(q) =\left. \frac 1 {Z_T} \tr\left\{e^{-\hat H/T}\hat\delta(\hat q_j- \hat I q)\right\}\right|_{\hat q_0=\hat q_{N+1}=\hat 0}= 
 \lim_{P\to+\infty}\frac 1 P\sum_{k=1}^P \mathcal Q_j^k(q) 
 \end{align}
The average of these  $N$ distributions, defined as
\begin{equation}\label{def_g_r}
g(q) \equiv\frac 1 N \sum_{j=1}^N \mathcal Q_j(q)
\end{equation}
resembles the total density of configurations, a standard object in Liquid-State Physics \cite{hans_mc_ch4}. 
\\ \\
In \cref{fig:Q_j_cent1} and \cref{fig:Q_j_cent2} we present the numerical results of \cref{def_Qj} for one of the central d.o.f. of the chain, $j=4$, for two different temperatures $T=0.01$ and $T=5$. Each graph includes the four sampled values of $P$.
The supports of the quantum distributions ($P>1$) at lower temperature (\cref{fig:Q_j_cent1}) are larger than the classical correspondents. Similar behavior has been seen in studies on the radial distribution functions of more realistic systems, such as water dimers \cite{czako2010} and trimers \cite{czako2010_bis}. These works highlight the importance of the zero-point energy in order to assess the enhanced delocalization from classical to quantum statistics at low temperature. This effect, expected to occur in the harmonic limit $\alpha=0$ (see \cref{sigma_q} in the Appendix), is preserved in the strongly nonlinear regime considered here. In general, the phenomenology stems as a direct consequence from the indetermination principle \cite{chandler1981}, which is exactly fulfilled in the $P\to+\infty$ limit of PIMD. In \cref{fig:Q_j_cent2} $\mathcal Q_4(q)$ is shown for a temperature $50$ times larger than in \cref{fig:Q_j_cent1}. In this case we can see a convergence of the classical and quantum results, as expected in the high energy limit discussed in \cref{app:class_iso}.  
\cref{fig:Q_j_1} and \cref{fig:Q_j_2} show the same distribution as above at low temperature $T=0.01$, for the d.o.f. at the left and right boundaries of the chain $j=1$ and $j=N$. 
The analytical (solid) curves in \cref{fig:Q_j_1} and \cref{fig:Q_j_2} refer to the distribution 
\begin{equation}\label{def_tilde_rho}
\tilde\rho(q)=\frac 1 {\tilde Z_T} e^{-V(q)/T}, \hspace{10mm} \tilde Z_T = \int_{\mathbb R}\mathrm d q \; e^{e^{-V(q)/T}}
\end{equation}
that would occur for the first and last moving particles, provided they were uniquely subjected to the potential of the closest boundaries. The choice of the signs can be understood via the following argument. From the perspective of the first d.o.f. in the configuration $q$, the source of the potential is located at a position $-q$, and vice-versa for $j=8$. We can notice a systematic shift of the effective classical numerical distribution w.r.t. this limit case. This behavior is caused by the other $N-1$ d.o.f., which have been integrated out in \cref{def_Qj}. Their action can be interpreted as an effective screening, which mitigates the strength of the repulsion of the walls. Finally, the mirror symmetry exhibited by the distributions on the two extreme ends of the chain agrees with \cref{symm_center}. Additional arguments that support this interpretation are discussed in \cref{app:force_field}, in the context of the numerical sampling of the force field. \\ \\ 
We can obtain a global picture of the fluctuations of the particles for different temperatures from the moments of the distributions of the configurations. From \cref{f_qJ_ave}, these averages can be easily computed. We consider in the following the central d.o.f. $j=\floor*{N/2}=4$, where the action of the boundaries is minimized. The results for the second moment
\begin{equation*}
\lla (q_4)^2 \rra_{T,P} = \int_{\mathbb R} \mathrm d  q_4 \; \mathcal Q_4( q_4) (q_4)^2 
\end{equation*}
are given in \cref{fig:q2_1} and \cref{fig:q2_2}:  
\begin{figure}[H]
	\centering
	\begin{minipage}{.45\textwidth}
		\vspace*{8mm}
		\includegraphics[scale=0.55]{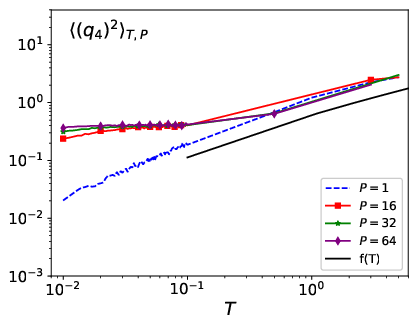}
		\caption{Fluctuation in configurational d.o.f. $j=4$ for $\alpha=1$, symbols denote simulation results, the continuous line corresponds to \cref{eq:fofT}}\label{fig:q2_1}
	\end{minipage}
	\hfill
	\begin{minipage}{.45\textwidth}
		\includegraphics[scale=0.55]{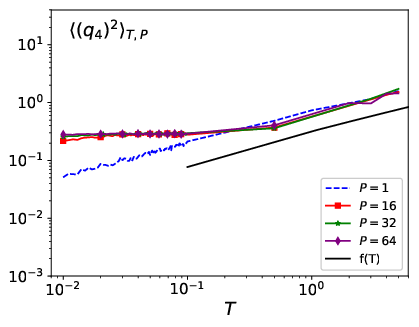}
		\caption{As \cref{fig:q2_1}, but for $\alpha=5 $}\label{fig:q2_2}
	\end{minipage}
\end{figure}
The average fluctuations for different values of $P$ converge to a unique curve above a thermal threshold, in agreement with the expected collapse to the classical distributions at high temperature. The quantum moments at low temperatures are systematically higher that the classical ones, supporting the arguments of a weaker quantum localization in this regime. The broadening of the statistics due to quantum dispersion observed in the nonlinear model is in line with the discussion of the harmonic limit in \cref{app:harm_pot}.
\\
The power law scaling $f(T)$ at high temperature in \cref{fig:q2_1} and \cref{fig:q2_2} can be estimated under the assumption that the central particle is uniquely subjected to a symmetric potential from the left and right first neighbors, and defined by
\begin{equation*}
V_4(q)\equiv V(q)+V(-q) = q^2+\frac\alpha 2 q^4
\end{equation*}
with a related one dimensional partiton function
\begin{equation*}
Z^{4}_T = \int_{\mathbb R}\mathrm d q\; e^{-V^4(q)/T} = \frac 1 {\sqrt{2\alpha}}e^{\frac 1 {4\alpha T}}K_{1/4}\left(\frac 1{4 \alpha T}\right)
\end{equation*}
where $K_\nu(z)$ denotes the modified Bessel function of the second kind. The second moment of the Boltzmann distribution generated by this potential can be computed analytically: 
\begin{align}
f(T) &\equiv \frac 1 {Z^{4}_T} \int \mathrm d q\; e^{-V_4(q)/T} q^2 = \\
&= \frac \pi{2\sqrt 2\alpha K_{1/4}\left(\frac 1{4 \alpha T}\right)}\left[-I_{-1/4}\left(\frac 1{4 \alpha T}\right)+(1+2\alpha T)I_{1/4}\left(\frac 1{4 \alpha T}\right)-I_{3/4}\left(\frac 1{4 \alpha T}\right)+I_{5/4}\left(\frac 1{4 \alpha T}\right)\right] \label{eq:fofT}
\end{align}
where $I_\nu(z)$ denotes the modified Bessel function of the first kind.
\\ \\
The analysis of the present chapter allowed us to identify numerically the effects of quantum dispersion on the statistics of the system in thermal equilibrium. In the following section we discuss the consequences of the this analysis on the dynamics of time-dependent correlation functions.  
\section{Dynamics}\label{dynamics}
\begin{figure}[H] 
	\centering
	\begin{minipage}{.45\textwidth}
		\vspace*{7.5mm}
		\centering
		\includegraphics[scale=0.55]{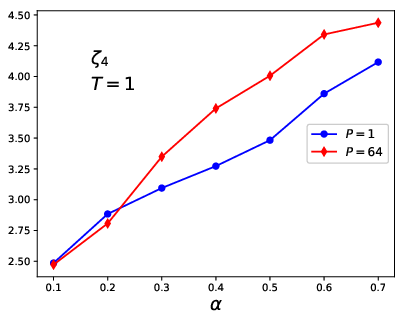}
		\caption{Values of $\zeta_4$ for different anharmonic parameters $\alpha$, from the classical ($P=1$) and quantum $(P=16)$ statistics}\label{fig:omega4}
	\end{minipage}
	\hfill
	\begin{minipage}{.45\textwidth}
		\centering
		\includegraphics[scale=0.55]{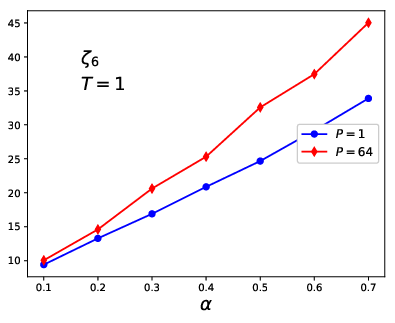}
			\caption{Same as \cref{fig:omega4} for $\zeta_6$}\label{fig:omega6}
	\end{minipage}
\end{figure}

\begin{figure}[H] 
 	\centering
 	\begin{minipage}{.45\textwidth}
 		\vspace*{7.5mm}
 		\centering
 		\includegraphics[scale=0.55]{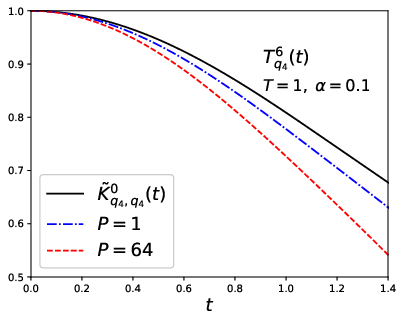}
 		\caption{Decay of $\lla q_{j,P} q_{j,P}(t)\rra_{TP}$ in the sixth order series expansion for the classical ($P=1$) and quantum $(P=16)$ dynamics; $T=1$ and $\alpha=0.1$}\label{fig:T6_P1}
 	\end{minipage}
 	\hfill
 	\begin{minipage}{.45\textwidth}
 		\centering
 		\includegraphics[scale=0.55]{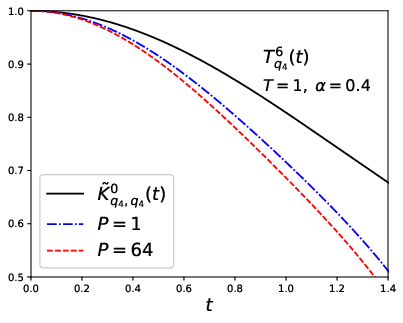}
 		\caption{Same as \cref{fig:omega4}, for  $\alpha=0.4$}\label{fig:T6_P64}
 	\end{minipage}
 \end{figure}

In \cref{sect:statistics_theo} we showed that the broadening of the spatial distributions from classical to quantum statistics is preserved under nonlinearities in the interaction. In the present section we analyze how this affects the evolution of the system. The Ring Polymer Molecular Dynamics (RPMD) approach \cite{craig2004}, \cite{habershon2013}, allows to compute the short-times dynamics of Kubo-transformed correlation functions  
\begin{equation}\label{def_KAB}
K_{\hat A, \hat B}(t)\equiv\frac 1{Z_\beta\beta} \int_0^{1/T} \mathrm d \lambda \; \tr \left\{e^{-(1/T-\lambda)\hat H_0} \hat A(\mathbf {\hat q}) e^{(-1/T+it) \hat H_0} \hat B(\mathbf {\hat q}) e^{-i\hat H_0 t}\right\}
\end{equation}
for two observables $\hat A$ and $\hat B$, which depend exclusively on the configurations. The method is based on the mapping of \cref{def_KAB} to the classical correlation 
\begin{equation}\label{tildeK_q}
\tilde K_{A,B}(t) \equiv\frac 1{Z_{TP}}\int_{\mathbb R^{2NP}}\mathrm d \mathbf q^1\cdots  \mathrm d \mathbf q^P \mathrm d \mathbf {p}^1\cdots  \mathrm d \mathbf {p}^P \; e^{-\frac 1 {TP}\mathcal H_P^{\mathrm{cl}}(\mathbf { q}^1, \cdots , \mathbf {q}^P, \mathbf {p}^1, \cdots, \mathbf {p}^P)} A_{P}(\mathbf q)\;  B_{P}(\mathbf q(t))
\end{equation}
where we defined
\begin{equation} \label{def_AP}
A_P(\mathbf q) \equiv A_P(\mathbf q^1,\cdots, \mathbf q^P) \equiv \frac 1 P\sum_{l=1}^P A(\mathbf{ q}^l)
\end{equation}
and 
\begin{equation}\label{def_HP_cl}
\mathcal H_P^{\mathrm{cl}}(\mathbf { q}^1, \cdots , \mathbf {q}^P, \mathbf {p}^1, \cdots, \mathbf {p}^P) = \sum_{j=0}^N\sum_{k=1}^P\left[\frac{(p_j^k)^2}{2} + \frac{P^2T^2}{2}(q_j^{k+1}-q_j^{k})^2+V(\mathbf q^k)\right]
\end{equation}
The dynamics in \cref{tildeK_q} is propagated in time via Newton's equations
\begin{align}
\dot q^k_j &= p_k^j  \label{RPMD1} \\
\dot p^k_j &= -P^2 T^2 \left(2q^k_j-q^{k-1}_j-q^{k+1}_j\right)-\frac{\partial V(\mathbf q^k)}{\partial q_j^k} \label{RPMD2}
\end{align} 
with periodic boundary conditions for the ring-polymer 
\begin{equation*}
q_0^k= q_{N+1}^k=0 \;\;\;\forall\; \;k=1\cdots N \hspace{10mm} q_j^{N+1}=q_j^{1} \;\;\;\forall\;\; j=1\cdots N
\end{equation*} 
The classical dynamics is recovered in the numerically exact limit of $P=1$ bead. Note that the r.h.s. of \cref{RPMD2} is $P$ times larger than the corresponding \cref{eom_path_int_std2}, which we used to propagate the stationary dynamics in the PIMD sampling scheme. This conventional change of notation is consistent with the original work introducing RPMD \cite{craig2004} (see \cite{tuckerman2010} for a discussion in this regard). In the potential energy term, this rescaling is accounted for by the additional factor $1/P$ in the canonical distribution in \cref{tildeK_q} (compare with the PIMD counterpart in \cref{def_rho_cl}). Therefore, the only adjustment needed to construct the stationary correlation \cref{tildeK_q} is to sample the momenta according to a Maxwell distribution at temperature $PT$. \\
In the following we study the decay of correlators of linear operators of the configurations. This choice allows to reliably predict real-time correlations till sixth order in time, while the accuracy would be smaller for nonlinear operators \cite{welsch2016}. The RPMD scheme is optimized for weakly nonlinear interactions; in our system this condition is satisfied, provided the magnitude of the nonlinear parameter is smaller than the quadratic coefficient, i.e. for $ \alpha  < 3/2$. 
Additionally, the precision of the method improves at high temperatures $T$ and high number of replicas $P$. Given that the two parameters appear in \cref{tildeK_q} and \cref{RPMD2} only via the product $PT$, we impose $PT\gg 1$ for $P>1$. This is satisfied by our choice of $P=64$ beads and $T=1$. The real-time dynamics of the classical isomorphism yields the exact result in the harmonic limit $\alpha=0$ (see \cref{app:harmonic_lim_RP}), hence we can keep this regime as a benchmark for the results from the nonlinear system.
We consider in the following the position autocorrelation function
\begin{align}\label{Kq_exp}
\lla q_{j,P} \;q_{j,P}(t)\rra_{TP}=\sum_{n=0}^{+\infty}(-1)^n\frac{t^{2n}}{(2n)!}\lla \llv \left(i\mathcal L^{(RP)}\right)^nq_{j,P} \rrv^2 \rra_{TP}\equiv \sum_{n=0}^{+\infty}(-1)^n\frac{t^{2n}}{(2n)!}\zeta_{2n}
\end{align}
where we introduced the ring-polymer Liouville operator 
\begin{equation}\label{def_iLRP}
i\mathcal L^{(RP)} \equiv \sum_{j=1}^N\sum_{k=1}^P\left[p_j^k\frac{\partial}{\partial q_j^k} -\frac{\partial V(\mathbf q^k)}{\partial q_j^k}\frac{\partial}{\partial p_j^k}-P^2T^2\left(2q_j^k-q_j^{k-1}-q_j^{k+1}\right)\frac{\partial}{\partial p_j^k}\right]
\end{equation}
A connection between the statistical analysis of the previous chapter and the present dynamical description can be drawn by analyzing the details of the first static correlators $\zeta_{2n}$ in \cref{Kq_exp}:
\begin{align}
&\zeta_0 = \lla \left (q_{j,P}\right)^2\rra_{TP}=\frac 1 {P^2}\sum_{k,k'=1}^P\lla q_j^k q_j^{k'}\rra_{TP} \label{omega0}\\
&\zeta_2  = \lla \left (p_{j,P}\right)^2\rra_{TP}=\frac 1{P^2}\sum_{k,k'=1}^P \lla p_j^k p_j^{k'} \rra_{TP}= \frac 1{P} \lla \left(p_j^1\right)^2 \rra_{TP} =T \label{omega2}\\
&\zeta_4  = \lla \left (F_{j,P}\right)^2\rra_{TP} = \frac 1 {P^2}\sum_{k,k' =1}^P\lla F_{j}^kF_{j}^{k'}\rra_{TP}=\frac 1 {P}\sum_{k =1}^P\lla F_{j}^kF_{j}^{1}\rra_{TP} \label{omega4}\\
&\zeta_6 = \lla  \left(i\mathcal L^{(RP)}F_{j,P}\right)^2\rra_{TP}=
\lla\left[\sum_{j'=1}^N\sum_{k=1}^P\left(p_{j'}^{k}\frac{\p F_{j,P}}{\partial q_{j'}^{k}}\right)\right]^2\rra_{T,P}=\sum_{j',j''=1}^N\sum_{k,k'=1}^P\lla p_{j'}^{k}p_{j''}^{k'}\rra_{T,P}\lla \frac{\partial F_{j,P}}{\partial q_{j'}^{k}} \frac{\partial F_{j,P}}{\partial q_{j''}^{k'}}\rra_{TP}= \nn\\
&=TP\sum_{j' =1}^N\sum_{k =1}^P \lla \left(\frac{\partial F_{j,P}}{\partial q_{j'}^{k}} \right)^2\rra_{TP}=TP\sum_{j' =1}^N\sum_{k=1}^P \lla \left(\frac 1 P\sum_{k'=1}^P\frac{\partial F_{j}^{k'}}{\partial q_{j'}^{k}} \right)^2\rra_{T,P}=\frac TP  \sum_{k=1}^P\lla \left(\frac{\partial F_{j}^{k}}{\partial q_{j}^{k}} \right)^2+\left(\frac{\partial F_{j}^{k}}{\partial q_{j-1}^{k}} \right)^2+\left(\frac{\partial F_{j}^{k}}{\partial q_{j+1}^{k}} \right)^2\rra_{TP}=\label{pp_delta}\\
&= T\lla \left(\frac{\partial F_{j}^{1}}{\partial q_{j}^{1}} \right)^2+\left(\frac{\partial F_{j}^{1}}{\partial q_{j-1}^{1}} \right)^2+\left(\frac{\partial F_{j}^{1}}{\partial q_{j+1}^{1}} \right)^2\rra_{TP} \label{omega6}
\end{align}
where 
\begin{align*}
F_j^k 
&= -\frac{\partial V(\mathbf q^k)}{\partial q_j^k} = - V'(q_j^k-q_{j-1}^k)+ V'(q_{j+1}^k-q_{j}^k)  \\
\frac{\partial F_{j}^{k}}{\partial q_{l}^{k}}& = 
\begin{cases}
 V''(q_j^k-q_{j-1}^k)&l=j-1\\
-V''(q_j^k-q_{j-1}^k)- V''(q_{j+1}^k-q_{j}^k)&l=j\\
V''(q_{j+1}^k-q_{j}^k) &l=j+1
\end{cases}
\end{align*}
In \cref{fig:omega4} and \cref{fig:omega6} we compare the magnitude of the classical and quantum values of $\zeta_4$ and $\zeta_6$ for different anharmonicities $\alpha$. 
The classical and quantum curves tend to converge as $\alpha\to 0^+$ as expected, yielding the same harmonic limit in \cref{sol_Keta_RP} and \cref{sol_Kq_RP}. The series coefficients of the quantum correlation are in general larger than the classical ones, apart from a small crossover of $\zeta_4$ at $\alpha=0.2$. This signals that the decay of the quantum correlations at short times occurs on a faster time-scale than the classical ones. To understand this last point, let us consider the sixth order expansion of \cref{Kq_exp}:
\begin{equation}\label{def_T6}
T_{6}(t)\equiv\sum_{l=0}^6(-1)^l \frac{\zeta_{2l} }{(2l)!}t^{2l}
\end{equation}
The maximual time $t_\epsilon$ in which $T^6$ is able to effectively approximate the full time correlation can be estimated for small anharmonicites from the integrable limit benchmark, by imposing
\begin{equation*}
t_\epsilon\equiv \min_j\left\{t\; :\llv \cos (\omega_j t)- T^j_{6}(\omega_j t)\rrv \le \epsilon\right\} 
\end{equation*}
For $\epsilon=0.1$, we can determine numerically $t_\epsilon \simeq 1.4$. $T_6(t)$ with coefficients computed via \crefrange{omega0}{omega6} is shown in \cref{fig:T6_P1} and \cref{fig:T6_P64} for $\alpha=0.1$ and $\alpha=0.4$.
In \cref{fig:T6_P1} and \cref{fig:T6_P1} we see that larger magnitude of the Taylor coefficients in the quantum w.r.t. classical dynamics is associated to a speed up in the decorrelation at short times. \\ \\
A connection between the dynamical and statistical analyses can be drawn by rewriting the coefficient $\zeta_6$ in \cref{omega6} in terms of the coarse-grained distributions in \cref{def_multiv_Q0}, via \cref{f_qJ_ave}:
\begin{align}
\zeta_6 &= \int_{\mathbb R^3}\mathrm d q_{j-1}\mathrm d q_{j}\mathrm d q_{j+1}\; \mathcal Q_{j-1, j, j+1}(q_{j-1}, q_j, q_{j+1}) \left[   \left(\frac{\partial F_{j}}{\partial q_{j}} \right)^2+\left(\frac{\partial F_{j}}{\partial q_{j-1}} \right)^2+\left(\frac{\partial F_{j}}{\partial q_{j+1}} \right)^2   \right] \label{omega6_stat}
\end{align}
We define an effective compact support of the spatial distributions
\begin{equation*}
\mathrm{supp}_\epsilon\{\mathcal Q_{\mathcal J}\}\equiv \{(q_{j_1},\cdots, q_{j_1}):\;\;\mathcal Q_{\mathcal J}(q_{j_1},\cdots, q_{j_1}) \ge \epsilon\}, \hspace{10mm} \epsilon \ll  1
\end{equation*}
From the results of the previous section we know that the quantum distributions have a larger support than the classical ones: 
\begin{equation*}
\mathrm{supp}_\epsilon\left\{\mathcal Q^{\mathrm{qu}}_{j-1, j, j+1}\right\}= \prod_{i\in\{j-1,j,j+1\}}
\mathrm{supp}_\epsilon\left\{\mathcal Q^{\mathrm{qu}}_{i}\right\}\ge\mathrm{supp}_\epsilon\left\{\mathcal Q^{\mathrm{cl}}_{j-1, j, j+1}\right\}
\end{equation*}
Given that \cref{omega6_stat} involves the average of positive function of the configurations, the fact that $\zeta_6^{\mathrm{qu}}>\zeta_6^{\mathrm{cl}}$ suggests that the tails of the quantum distributions play a significant role for the decay at short times of the auto-correlation function \cref{Kq_exp}. These low-probability configurations correspond to large excursions from the equilibrium averages into non classical regions. The deviations from classical paths involve a faster loss of correlation in dynamics at short times, in agreement with the trends observed in \cref{fig:T6_P1} and \cref{fig:T6_P64}.
\section{Conclusions}
We presented a numerical study comparing the classical and quantum FPU chains, in thermal equilibrium. A statistical analysis shows that the quantum system exhibits higher thermal fluctuations than the classical one; the effect increases for lower temperatures. The present observations allow us to conclude that the thermal fluctuations in the strongly nonlinear regime follow the same trend of the harmonic limit. We have analyzed the impact of these observations on a dynamical framework. The large quantum fluctuations have a direct impact on the magnitude of the first orders of the spatial auto-correlation functions. This translates in a speed up of their short-times decay.
\\
It would be interesting to extend the analysis at longer times, via a numerical approach that takes full advantage of the weakly nonlinear nature of the system. An analysis at larger time scales would allow us to assess whether the discrepancy observed in the first dynamical coefficients saturates or increases at higher orders. This would allow to assess how interesting phenomena observed in classical FPU chains, from anomalous heat conduction \cite{mellet2015}, to intermittencies \cite{danieli2017}, are affected by quantum dispersion. Additionally, the construction of isomorphisms analogous to the one discussed in the present work, could allow to translate results from classical systems in two dimensions, as \cite{wang2019} and \cite{benettin2008}, to quantum-mechanical one dimensional models.
\section{Acknowledgments}
We thank Hans-Peter Breuer, Andeas Buchleitner, Arshia Atashpendar, Hugues Meyer and Andreas H\"artel for proficient discussions. We also thank an anonymous reviewer for pointing out a critical mistake. The simulations have been performed on the NEMO computing cluster facility with support by the state of Baden-Württemberg through bwHPC and the German Research Foundation (DFG) through grant no INST 39/963-1 FUGG.
\begin{appendices}
\crefalias{section}{appsec}
\section{The Quantum FPU Hamiltonian}\label{app:deriv_H}
\subsection{Derivation from a physical interaction}
In this appendix we show how the classical FPU Hamiltonian in \cref{def_H} can be seen as the fourth-order expansion of a physical two-body potential, following an approach close to \cite{stoppato2016}. Let us consider at first a quantum mechanical system identified by the Hamiltonian 
\begin{align}
 \hat K(\mb{\hat x}, \mb{\hat y})&=\sum_{j=0}^N\Le[\frac{\hat y_j^2}{2m}+\hat \phi(\hat x_{j+1}-\hat x_j)\Ri]\equiv \hat {\mathcal T}(\mb{\hat y})+\hat {\mathcal V}(\mb{\hat x}) \label{def_K}\\
\hat y_j &= -i\hbar\frac{\partial}{\partial \hat x_j}\label{def_CCR_K}
\end{align}
where $\hat x_j$'s and $\hat y_j$'s have respectively the dimensions of configurations and momenta. $\hat\phi(\hat r)$ denotes an analytical potential depending on the distance between nearest neighbors. The total Hilbert space $\mathcal C$ can be factorized as as the tensor product of the single-particle spaces $\mathcal H_i$ \cite{zanardi2004}: 
\begin{equation}\label{def_C}
\mathcal C =\otimes_{i=1}^N \mathcal H_i, \hspace{10mm} \hat x_i\in  \mathcal H_i
\end{equation}
We choose Dirichlet boundary conditions 
\begin{align}\label{bc_x}
\hat x_0=\hat 0 \hspace{10mm} \hat x_{N+1}=\hat I L
\end{align}
where $L\in \mathbb R^+$ denotes the total length of the chain and $\hat I$ is the identity operator. Hamilton's equations of the corresponding classical system are
\begin{equation}\label{hamilt_cl}
\ddot y_j/m = \phi'(x_{j+1}-x_j)-\phi'(x_j-x_{j-1})\hspace{10mm}j=1\cdots N
\end{equation}
Irrespectively of the specific choice of $\phi$, the configuration in which the forces on the left and right on each particle are equal in magnitude and opposite in sign corresponds to an equilibrium point for \cref{hamilt_cl}. This occurs when the distances between neighbours are constant and equal to 
\begin{equation}\label{def_a}
x_{j+1}-x_{j} = d=L/(N+1) \hspace{10mm} \forall\;\; j=0\cdots N
\end{equation}
We are interested the action of the first-order nonlinearities on the dynamics. In this perspective, it seems convenient to expand the analytic potential in series. As a center of the expansion, we can choose the equilibrium point defined in \cref{def_a}. A natural set of coordinates is then identified by the displacements from the classical minimum: 
\begin{equation}\label{def_q_p}
\hat x_j \equiv \alpha  (\hat q_j+\hat I j) d, \hspace{10mm} \hat y_j \equiv\beta  \hat p_j \frac{md}\tau,  \hspace{10mm} s=t\tau
\end{equation}
$s$ and $t$ denote respectively the physical time conjugated to $\hat K$ and the dimensionless time in the new coordinate system. $\tau$ is a yet undefined temporal scaling factor. The b.c. in the new coordinates are written as
\begin{equation}\label{bc_q}
\hat q_0=\hat q_{N+1}=\hat 0
\end{equation}
The CCR are fixed via the following prescription in the configurational representation:
\begin{equation}\label{def_CCR_H}
\hat p_j\equiv -i\frac{\partial}{\partial \hat q_j}
\end{equation}
The correspondence between \cref{def_K} and \cref{def_H} $\hat H$ is fixed by expanding the analytic potential to fourth order:
\begin{align}
& \sum_{j=0}^N\frac{\hat y_j^2}{2m}+\phi\Le(\alpha (\hat q_{j+1}-\hat q_j+\hat I)d\Ri)=\nn\\
&\sum_{j=0}^N\frac{\hat y_j^2}{2m}+\sum_{n=2}^4\phi^{(n)}(\alpha d\hat I)\frac {(\alpha d)^n}{n!}(\hat q_{j+1}-\hat q_j)^n+\mathcal O((\alpha d)^5) \equiv   \hat H \f{\gamma md^2}{\tau^2}+\mathcal O((\alpha d)^5) \label{equiv_4}
\end{align} 
where we introduced a dimensionless scaling factor $\gamma$ for the energy. In \cref{equiv_4} we omitted on purpose the zeroth-order contribution $\phi(\alpha d\hat I)$, as it does not play any role in the dynamics. The linear term is identically zero for any value of $\hat\phi^{(1)}(\alpha d\hat I)$:
\begin{equation*}
\hat\phi^{(1)}(\alpha\hat d)\sum_{j=0}^N(\hat q_{j+1}-\hat q_j)= \hat\phi^{(1)}(\alpha\hat d)\Le(\hat q_{N+1}-\hat q_0\Ri)=0
\end{equation*}
where we inserted the b.c. from \cref{bc_q}. 
The transformation is completely defined once we find a closed expression for the yet undefined constants $\alpha, \beta, \gamma$ and $\tau$ in terms of the parameters of the original system. We therefore need to define a set of four equations in terms of these unknown quantities. From \cref{def_CCR_K}, \cref{def_q_p} and \cref{def_CCR_H} it follows 
\begin{equation}\label{constr1}
\frac{\alpha\beta}\tau=\frac \hbar{md^2}
\end{equation} 
Through \cref{constr1} we can write the dimensionless correspondents of the CCR:
\begin{equation*}
[\hat x_j, \hat y_k] = i\hbar \delta_{jk}\;\; \implies \;\; [\hat q_j, \hat p_k] = i   \delta_{jk}
\end{equation*}	 
A second constraint can be inferred from the Heisenberg equation of motion of the positions in the physical coordinates:
\begin{align}
\frac{\mr d \hat x_j}{\mr d s}&=\f i \hbar [\hat K, \hat x_j
]=\nn
\f {\hat y_j}{m}\nn  \\
\frac{\alpha d}{\tau}\frac{\mr d \hat q_j}{\mr d t}&=\frac i \hbar\f{\gamma m d^2}{\tau^2}  [\hat H,\hat q_j]\alpha d\implies \frac{\gamma m d^2}{\hbar\tau} = 1 \implies\frac \gamma \tau = \frac{\hbar}{m d^2} \label{constr2}
\end{align}
The last two constraints can be fixed by imposing the normalization of the kinetic energies and the harmonic terms in \cref{def_K} and \cref{def_H}, via \cref{equiv_4}. This yields
\begin{align}
\gamma \f{md^2}{\tau^2} &= \frac{m\beta^2 d^2}{\tau^2} \implies \f \gamma{\beta^2}= 1 \label{constr3}\\
\gamma \f{md^2}{\tau^2} &= \alpha^2d^2 \hat\phi^{(2)}(\alpha d\hat I) \implies \frac\gamma{\alpha^2\tau^2} = \frac{\phi^{(2)}(\alpha d\hat I)}{m} \label{constr4}
\end{align}
In order to find an explicit solution for the parameters of the system, it would be convenient to require a homogeneous scaling law for the physical potential:
\begin{equation}\label{scaling_phi}
\hat \phi(\alpha d\hat I)\equiv f(\alpha)\hat \phi( d\hat I)
\end{equation}
The second derivative in \cref{constr4} would preserve the homogeneity in the scaling, simplifying the task of extracting the parameters of the system; in particular:
\begin{equation}\label{scaling_phi2}
\hat \phi^{(2)}(\alpha d\hat I) = \Le.\nabla^2_{\hat r}\hat \phi(\hat r)\Ri|_{\hat r=\alpha d\hat I}=(\alpha d)^{-2}\phi^{(2)}(d\hat I)f(\alpha)
\end{equation} 
\cref{scaling_phi} fixes an implicit constraint on the available choices of $\hat \phi$. The condition would be, however, satisfied by some realistic interaction, e.g. by an exponential Morse-like potential. This would be directly connected to other relevant dynamical systems as the Toda lattice \cite{yoshida1988}. As we do not need to take an explicitly choice for the physical potential, this does not represent an effective limitation for the subsequent analysis. The system of \crefrange{constr1}{constr4} with the additional scaling law in \cref{scaling_phi} is then complete, and depending on the functional form for $f(\alpha)$ it can be solved, either exactly or with numerical means. For the purpose of the present work, from \crefrange{constr1}{constr3}, we can extract the condition 
\begin{equation}\label{alpha_eq_beta}
\alpha=\beta
\end{equation}
The assumptions of $\beta>0$ and $\gamma >0$ have been used to fix \cref{alpha_eq_beta}. These ensure respectively a confining interaction at fourth order, and that the sign between the two Hamiltonians is preserved. The constraint in \cref{alpha_eq_beta} may seem as a strong limitation in terms of the allowed choices for the two-body interaction. It would still be legitimate to drop this constraint, and define in the first place a dynamical system ruled by \cref{def_H}, without a direct connection with \cref{def_K}. In the present work we still chose to fulfill the constraint \cref{alpha_eq_beta}.
As a side remark, we note that the ``fourth-order'' correspondence between the two dynamical systems has been written in a closed form depending on the analytic potential only through the value of its second derivative. This is a direct consequence of the fact that two constraints in \cref{constr1} and \cref{constr2} are bounded by the preservation the CCR. (This highlights that there exists an infinite set of analytical potentials $\hat \phi$ producing the same FPU Hamiltonian \cref{def_H}, provided the first terms of the expansion are the same.) \\
Following \cite{anderson1994}, we can construct the operator $\hat C$ associated to the canonical transformation, defined s.t.
\begin{equation*}
\frac 1{\hbar}[\hat C\hat x_j\hat C^{-1}, \hat C\hat y_k\hat C^{-1}] = i\delta_{j,k} \equiv \frac{\alpha \beta m d^2}{\hbar\tau} [\hat q_j, \hat p_k] \numeq{\ref{constr1}} [\hat q_j, \hat p_k]
\end{equation*}
In our case, it is given by the translations of the configurations
\begin{equation*}
\hat C = \hat C(\mb{\hat y}) = 
\prod_{j=1}^N e^{ -id\alpha j  \hat y_j/\hbar}
\end{equation*}
acting as 
\begin{align}
\hat C \hat x_j  \hat C^{-1} &=  e^{ -id\alpha j  \hat y_j/\hbar} \hat x_j e^{id\alpha j  \hat y_j/\hbar} = \Le([e^{ -i\hat d\alpha j  \hat y_j/\hbar}, \hat x_j ]+\hat x_je^{-i d\alpha j  \hat y_j/\hbar}\Ri)e^{id\alpha j  \hat y_j/\hbar}= \nn \\
&= \Le(-i\hbar\frac{\partial}{\partial \hat y_j}e^{ id\alpha j  \hat y_j/\hbar}\Ri)e^{-i d\alpha j  \hat y_j/\hbar} +\hat x_j = \hat x_j -\hat I d\alpha j \numeq{\ref{def_q_p}}\alpha\hat q_j\label{comm_prop1} \\
\hat C \hat y_j  \hat C^{-1} &= y_j \nn
\end{align}
In \cref{comm_prop1} we used the relation valid for any analytic operator of the momenta $\hat F(\hat y_j)$
\begin{align*}
[\hat F(\hat y_j), \hat x_j] &=\sum_{n=0}^{+\infty}\frac{\hat F^{(n)}(0)}{n!}[\hat y_j^n, \hat x_j] = \sum_{n=0}^{+\infty}\frac{\hat F^{(n)}(0)}{n!}\Le(\hat y_j[\hat y_j^{n-1}, \hat x_j]+[\hat y_j, \hat x_j]\hat y_j^{n-1}\Ri)=\cdots= \sum_{n=0}^{+\infty}\frac{\hat F^{(n)}(0)}{n!} \Le(-i\hbar n \hat y_j^{n-1}\Ri) = \\
&=-i\hbar \frac{\p \hat F(\hat y_j)}{\p \hat y_j}
\end{align*}
\subsection{Canonical ensemble}\label{app:canonical}
The NVT ensemble for the FPU Hamiltonian can be constructed from the correspondent ensemble for the Hamiltonian $\hat K$ in \cref{def_K}
\begin{equation}\label{def_Xi_theta}
\Xi_\theta \equiv \tr\{ e^{-\hat K/(\kappa_B \theta)}\}=\int_{D(L)} \mr d \mb x\;\bra {\mathbf {x}} e^{-\hat K/(\kappa_B \theta)}  \ket {\mathbf x}
\end{equation}
where $\theta$ and $\kappa_B$ denote respectively the physical temperature and the Boltzmann constant. The trace in \cref{def_Xi_theta} has been taken over the positions eigenstates which, as a consequence of \cref{def_C}, can be factorized as
\begin{equation*}
\ket {\mathbf  x} = \ket{x_1}\otimes \cdots \otimes \ket{x_N}
\end{equation*} 
As in the present work we consider distinguishable particles, no  symmetrization/antistymmetrization is required for the physical states in the traces. The integration domain in \cref{def_Xi_theta} is
\begin{equation}
D(L)= \{\mathbb R^N, \;\; \hat q_0=\hat 0,\hat q_{N+1}=\hat I L\}
\end{equation}
Let us remark that the prescription of fixed length in the Canonical ensemble is not identically satisfied by the b.c. in \cref{bc_x}. In particular, the size of the system could exceed the length $L$, as the $N$ moving particles are allowed to cross the boundaries. However, as we stick to $\beta>0$ in the two body potential in \cref{def_H}, the system is globally confining; the total length is approximately conserved and equal to $L$, apart from fluctuations. This allows us to identify $D(L) \simeq L^N$.
\\\\
The dimensionless temperature $T$, in the Canonical ensemble of the FPU Hamiltoninan in \cref{def_Z_T}, can be computed by identifying the thermal average of the kinetic energies in \cref{def_H} and \cref{def_K}: 
\begin{equation*}
\lla \hat{ \mathcal T}\rra_\theta = \frac{N}2\kappa_B \theta \equiv \frac N 2 T \frac{\gamma md^2}{\tau^2} = \lla \hat T\rra_T \frac{\gamma md^2}{\tau^2}
\end{equation*}
This yields
\begin{equation}\label{def_T}
	T \equiv \frac{\kappa_B \theta\tau^2}{\gamma m d^2} 
\end{equation}
\section{Construction of the classical isomorphism}\label{app:class_iso}
\subsection{Partition function}\label{part_func_iso}
In this section we construct the classical isomorphism used to determine the quantum statistics. The underlying idea is that the path integral in \cref{def_Z_T} can be discretized via the application of Trotter theorem on the Canonical density operator. This is transformed into a countable product of infinitesimal commuting exponential operators, which finally leads to the statistics of an effective classical model. In particular, the symmetrized version of the Trotter theorem on the quantum Gibbs-Boltzmann distribution allows us to write:
\begin{equation}\label{exp_trotter}
e^{-\hat H(\hat {\mathbf q}, \hat {\mathbf p})/T} = \lim_{P\to+\infty}\Le[e^{- { \hat V(\hat {\mathbf q})}/(2PT)}e^{-\hat { T}(\hat {\mathbf p})/(PT)}e^{- { \hat V(\hat {\mathbf q}})/(2PT) }\Ri]^P
\end{equation}
The matrix element of \cref{exp_trotter} on the positional basis is
\begin{align} 
\bra {\mathbf q'} e^{-\hat H(\hat {\mathbf q}, \hat {\mathbf p})/T}\ket {\mathbf q} &=  \lim_{P\to+\infty} \bra {\mathbf q'} \Le[e^{- { \hat V(\hat {\mathbf q})}/(2PT)}e^{-\hat { T}(\hat {\mathbf p})/(PT)}e^{- { \hat V(\hat {\mathbf q}})/(2PT)}\Ri]^P \ket{\mathbf q}= \\
&=\lim_{P\to+\infty} \left.\int_{\mathbb R^{N(P-1)}}\mathrm d \mathbf q^2\cdots \mathrm d\mathbf q^P \; \prod_{k=1}^P \bra{\mathbf q^{k+1}} \hat \Omega_P \ket{\mathbf q^{k}} \right|_{\substack{\mathbf q^{P+1}= \mathbf q'\\\mathbf q^{1}= \mathbf q}}= \nn \\
&=\lim_{P\to+\infty} \left.\int_{\mathbb R^{N(P-1)}}\mathrm d \mathbf q^2\cdots \mathrm d\mathbf q^P \; \prod_{k=1}^P e^{- {  V( {\mathbf q}^{k+1}})/(2PT)} \bra{\mathbf q^{k+1}} \hat e^{-\hat { T}(\hat {\mathbf p})/(PT)} \ket{\mathbf q^{k}}e^{- {  V( {\mathbf q}^{k}})/(2PT)} \right|_{\substack{\mathbf q^{P+1}= \mathbf q'\\\mathbf q^{1}= \mathbf q}} \label{matr_elem_gib}
\end{align}
where we inserted $P-1$ times the decomposition of the identity on the configurations, and we defined the product of infinitesimal operators
\begin{equation}
\hat \Omega_P \equiv e^{- { \hat V(\hat {\mathbf q})}/(2PT)}e^{-\hat { T}(\hat {\mathbf p})/(PT)}e^{- { \hat V(\hat {\mathbf q}})/(2PT)}
\end{equation}
To evaluate the matrix element in \cref{matr_elem_gib} it is convenient to add an additional decomposition on the basis of the momenta:
\begin{align}
\bra{\mathbf q^{k+1}} \hat e^{-\hat { T}(\hat {\mathbf p})/(PT)} \ket{\mathbf q^{k}} = \int_{\mathbb R^N} \mathrm d \mathbf p \; \braket{\mathbf q^{k+1}|  \mathbf p}e^{-T(\mathbf p)/(TP)}\braket{\mathbf p|  \mathbf q^k} \label{matr_elem_gib2}
\end{align} 
The scalar products in \cref{matr_elem_gib2} are defined through the solution of the differential equation
\begin{align}
\bra{\mathbf q}\hat{\mb p}\ket{\mb p}={\mb p}\braket{\mathbf q |\mathbf p}
\nn \implies-i\f{\p}{\p{\mathbf q}}\braket{\mathbf q|\mathbf p}={\mathbf p}\braket{\mathbf q|\mathbf p} \label{free_wa}
\end{align}
that is 
\begin{equation*}
\braket{\mathbf q|\mathbf{p}}=Ce^{i\mathbf{q\cdot p}}
\end{equation*}
The constant $C\in \mathbb C$ is fixed by imposing the normalization of the momenta eigenstates:  
\begin{align}
\braket{\mb p|\mb{p'}}&=\int_{\mathbb R^{N}} \mr d \mb q\;\braket{\mb p|\mb q}\braket{\mathbf q|\mathbf{p'}}=\llv C \rrv^2\int_{\mathbb R^{N}}\mr d\mathbf q\; e^{i\cdot\mathbf q\cdot(\mb {p'-p})}= \nn \\
&= \llv C \rrv^2\Le[\prod_{j=1}^N\int_{\mathbb R}\mr d q_j e^{i q_j(p_j'-p_j)}\Ri]=\llv C\rrv^2(2\pi)^N \delta(\mathbf{p-p'}) \implies \llv C \rrv \equiv \f 1{(2\pi)^{N/2}} \nn \\
&\implies \braket{\mathbf q|\mathbf{p}}=\f{1}{(2\pi)^{N/2}}e^{i\mathbf{q\cdot p}} \label{plane_w}
\end{align}
where we fixed the arbitrary phase of $C$ to zero. 
\cref{matr_elem_gib} is rewritten through \cref{matr_elem_gib2} and \cref{plane_w} as 
\begin{align}
& \bra {\mathbf q'} e^{-\hat H(\hat {\mathbf q}, \hat {\mathbf p})/T}\ket {\mathbf q} = \nn \\
&=\lim_{P\to+\infty}\frac 1{(2\pi)^{NP}} \int_{\mathbb R^{N(P-1)}}\mathrm d \mathbf q^2\cdots \mathrm d\mathbf q^P \; \int_{\mathbb R^{NP}}\mathrm d \mathbf p^1\cdots \mathrm d\mathbf p^P  \prod_{k=1}^P e^{- {  V( {\mathbf q}^{k+1}})/(2PT)} \times \nn \\
&\times\left. e^{i\mathbf q^{k+1}\cdot  \mathbf p^k}e^{-T(\mathbf p^k)/(TP)} e^{-i\mathbf q^{k}\cdot  \mathbf p^k} e^{- {  V( {\mathbf q}^{k}})/(2PT)} \right|_{\substack{\mathbf q^{P+1}= \mathbf q'\\\mathbf q^{1}= \mathbf q}} = \nn\\
&=\lim_{P\to+\infty}\frac {e^{-\frac {TP}2\left \lvert \mathbf q^k-\mathbf q^{k+1}\right\rvert^2}}{(2\pi)^{NP}} \int_{\mathbb R^{N(P-1)}}\mathrm d \mathbf q^2\cdots \mathrm d\mathbf q^P \; \int_{\mathbb R^{NP}}\mathrm d \mathbf p^1\cdots \mathrm d\mathbf p^P  \prod_{k=1}^P e^{- {  V( {\mathbf q}^{k+1}})/(2PT)} \times \nn \\
&\times\left.e^{-\frac 1{2TP}\left(\mathbf p^k + iTP (\mathbf q^k-\mathbf q^{k+1})\right)^2} e^{- {  V( {\mathbf q}^{k}})/(2PT)} \right|_{\substack{\mathbf q^{P+1}= \mathbf q'\\\mathbf q^{1}= \mathbf q}}  \label{nondiag_mat_el}
\end{align}
The partition function in \cref{def_Z_T} is finally obtained by tracing over the diagonal elements $\mathbf q'=\mathbf q$, i.e. $\mathbf q^1=\mathbf q^{P+1}$, with the appropriate b.c. of the FPU Hamiltonian:
\begin{align}
&Z_T =\left. \int_{\mathbb R^N} \mathrm d \mathbf q \; \bra {\mathbf q} e^{-\hat H(\hat {\mathbf q}, \hat {\mathbf p})/T}\ket {\mathbf q} \right|_{ q_0= q_{N+1}= 0}= \\
&= \lim_{P\to+\infty}\left(\frac{\sqrt P}{2\pi}\right)^{NP}\int_{\mathbb R^{2NP}} \mathrm d \mathbf q^1\cdots \mathrm  d \mathbf q^P \mathrm d \mathbf p^1\cdots \mathrm d \mathbf p^P \; \left.\exp\Le\{-\frac 1 T\sum_{k=1}^P \Le[ \frac 1 2 (\mathbf p^k)^2+ \frac{T^2P}2 \left \lvert \mathbf q^{k+1}-\mathbf q^k \right \rvert^2 +\frac 1 P V(\mathbf q^k) \Ri]\Ri\}\right|_{\substack{\mathbf q^{P+1}= \mathbf q^{1}\\ q_0^l=q_{N+1}^ l=0,\;\;l=1\cdots P}} \label{Z_T_iso} = \\
&\equiv   \lim_{P\to+\infty} Z_{T,P}
\end{align}
The summation of the potential in \cref{nondiag_mat_el} has been simplified by noticing:
\begin{equation*}
\left.\frac 1 2 \sum_{k=1}^P \left[ V(\mathbf q^{k+1})+V(\mathbf q^{k})\right] \right|_{\substack{\mathbf q^{P+1}= \mathbf q^{1}}} = \sum_{k=1}^P V(\mathbf q^{k})
\end{equation*}
This allows us to identify the canonical distribution of the classical isomorphism:
\begin{align}
\rho_T^{\text{cl}}(\mathbf q^1,\cdots, \mathbf q^P, \mathbf p^1, \cdots, \mathbf p^P)&=\frac 1 {Z_T}\exp\left\{-\frac 1 T\mathcal H^{\mathrm{cl}}(\mathbf q^1,\cdots, \mathbf q^P, \mathbf p^1, \cdots, \mathbf p^P)\right\} \label{def_rho_cl} \\
\mathcal H^{\mathrm{cl}}(\mathbf q^1,\cdots, \mathbf q^P,& \mathbf p^1, \cdots, \mathbf p^P) \equiv \sum_{k=1}^P \Le[ \frac 1 2 (\mathbf p^k)^2+ \frac{T^2P}2 \left \lvert \mathbf q^{k+1}-\mathbf q^k \right \rvert^2 +\frac 1 P V(\mathbf q^k)\Ri] \label{def_H_cl}
\end{align} 
We can notice that the coupling term between adjacent replicas diverges at $T\to+\infty$. This implies that the interaction between the different instances becomes rigid at high temperature: all the copies of the system collapse to a single, classical replica.
\subsection{Expectation values on the classical isomorphism}\label{exp_val_iso}
The formalism presented in \cref{part_func_iso} for the partition function of the canonical ensemble can be conveniently extended to the calculation of thermal averages of observables depending on the configurations. In particular:
\begin{align}
&\lla \hat A( \mathbf{\hat q})\rra_{T} = \frac 1{Z_T} \int \mathrm d \mathbf q\; \bra{\mathbf q}e^{-\hat H/T} \hat A(\mathbf{\hat q}) \ket{\mathbf q}=\frac 1{Z_T}\int \mathrm d \mathbf q\; \bra{\mathbf q}e^{-\hat H/T}  \ket{\mathbf q}  A(\mathbf{ q}) = \label{ave_A_q} \\
&\numeq{\ref{Z_T_iso}}  \lim_{P\to+\infty}\frac{1 }{Z_{T,P}}\int \mathrm d \mathbf q^1\cdots \mathrm  d \mathbf q^P \mathrm d \mathbf p^1\cdots \mathrm d \mathbf p^P \; \left.\exp\Le\{-\frac 1 T\sum_{k=1}^P \Le[ \frac 1 2 (\mathbf p^k)^2+ \frac{T^2P}2 \left \lvert \mathbf q^{k+1}-\mathbf q^k \right \rvert^2 +\frac 1 P V(\mathbf q^k) \Ri]\Ri\}  A(\mathbf{ q}^1)\right|_{\substack{\mathbf q^{P+1}= \mathbf q^{1}\\ q_0^l=q_{N+1}^ l=0,\;\;l=1\cdots P}} = \label{ave_A_q1} \\
&= \lim_{P\to+\infty}\frac{1}{Z_{T,P}}\int \mathrm d \mathbf q^1\cdots \mathrm  d \mathbf q^P \mathrm d \mathbf p^1\cdots \mathrm d \mathbf p^P \; \left.\exp\Le\{-\frac 1 T\mathcal H^{\mathrm{cl}}(\mathbf q^1,\cdots, \mathbf q^P, \mathbf p^1, \cdots, \mathbf p^P)\Ri\} A_P(\mathbf q^1,\cdots, \mathbf q^P) \right|_{\substack{\mathbf q^{P+1}= \mathbf q^{1}\\ q_0^l=q_{N+1}^ l=0,\;\;l=1\cdots P}} \label{ave_rho_cl}
\end{align}
where the estimator $A_P$ in \cref{ave_rho_cl} has been defined in \cref{def_AP}. 
\cref{ave_rho_cl} follows as the $P$ realizations of the dynamics are identical. Instead of choosing an arbitrary replica (as $k=1$ in \cref{ave_A_q1}) it can be in general convenient to compute the statistics for all $P$ instances, and average between them in order to increases the statistics. 
The evaluation of the trace of operator depending on the momenta is not as straightforward as in the case of the configurations; in particular, such observables could not be anymore expressed in the natural basis in which the trace is expanded in \cref{ave_A_q}. We refer to \cite{tuckerman2010} for a discussion of the issue, while in the present work we will only consider the statistics of operators depending exclusively on the positions.
\section{Path integral molecular dynamics}\label{app:PIMD}
In this appendix we present the numerical scheme used to sample the equilibrium distribution of the classical isomorphism in \cref{def_rho_cl}. The approach is based on the propagation of an extended Molecular Dynamics scheme in equilibrium, driving the system towards a suitable constant-energy surface on phase space, ergodically (more on this later). We can argue that some enhancements are advisable w.r.t. a direct propagation of the Hamiltonian in \cref{def_H_cl} for a few potential issues. Firstly, a discrepancy between the timescales of the harmonic coupling of the replicas and the physical potential could hinder equilibration. Secondly, equilibration is slowed down by the small size of the system, including only $N=8$ particles. In the following we describe how to efficiently deal with those issues, following \cite{tuckerman2010, martyna1992, tuckerman1993, tuckerman2006}. 
\subsection{Staging variables}
The equations of motion of the Hamiltonian of the classical isomorphism in \cref{def_H_cl} are
\begin{align}
\dot q_j^k &= p_j^k \label{eom_path_int_std1} \\
\dot p_j^k &= -T^2P\left(2q_j^k-q_j^{k-1}-q_j^{k+1}\right)-\frac 1 P \frac{\partial V(\mathbf q^k)}{\partial q_j^k} \label{eom_path_int_std2}
\end{align}
As discussed in \cite{tuckerman2010} and \cite{hall1984}, critical issues could hinder the equilibration of the system via the simple propagation of the equations of motion (e.o.m.) in \cref{eom_path_int_std1} and \cref{eom_path_int_std2}. This could be seen by passing to the normal modes' coordinates for the harmonic coupling term in the potential of \cref{def_H_cl}. We would find a frequency spectrum ranging from $0$ to $4PT$ (see Section $1.7$ of \cite{tuckerman2010} for a derivation), where $P\gg 1$. The allowed timestep for the MD scheme would be bounded from above by the inverse of the highest frequency; this in turn would imply that the dynamics in the lower part of the spectrum, involving the largest timescales, would be poorly sampled. To solve this issue it is convenient to introduce a new set of configurations, the so-called \textit{staging variables}. These coordinates are constructed in order to uncouple the harmonic bound in \cref{def_H_cl}. A unique frequency is then assigned to each of the new oscillators. We will see in the discussion of \cref{Z_T_iso_in_u} that this can be done by fixing the fictitious masses of the ring polymer properly. \\
The the transformation to the staging variables $\mathbf u^k$ is defined s.t. \cite{tuckerman1993}:
\begin{align}
\mb u^1 &=\mb q^1 \nn \\
\mb u^k &=  \mb q^k -\f{(k-1)\mb q^{k+1}+\mb q^1}k \hspace{5mm} k = 2\cdots P\label{uk_qk}
\end{align}
with inverse
\begin{align}
\mb q^1 &=\mb u^1 \nn \\
\mb q^k &=  \mb u^k +\f{k-1}k\mb q^{k+1}+\f 1k\mb u^1 \hspace{5mm} k = 2\cdots P\label{qk_uk}
\end{align}
We can induce recursively a closed relation between the particles' displacements and the staging variables:
\begin{align}
\mb q^{P}&=  \mb u^P +\f{P-1}P\mb q^{1}+\f 1P\mb u^1=\mb u^1+\mb u^P \label{qu_recurs} \\
\mb q^{P-1}&=  \mb u^{P-1} +\f{P-2}{P-1}\mb q^{P}+\f 1{P-1}\mb u^1 = \mb u^{P-1} +\f{P-2}{P-1}(\mb u^{1}+\mb u^{P})+\f 1{P-1}\mb u^1 = \mb u^{P-1}+\f{P-2}{P-1}\mb u^P + \mb u^1 = \sum_{l=P-1}^{P}\f{P-2}{l-1}\mb{u}^l + \mb u^1 \nn \\
\mb q^{P-2}&= \mb u^{P-2}+\f{P-3}{P-2}\Le[\sum_{l=P-1}^P\f{P-2}{l-1}\mb u^l +\mb u^1\Ri]+\f 1 {P-2}\mb u^1 = \sum_{l=P-2}^P\f{P-3}{l-1}\mb u^l +\mb u^1 \nn
\end{align}
In \cref{qu_recurs} we used the periodic b.c. $\mathbf q^{P+1}=\mathbf q^1$. We finally infer
\begin{align}\label{inv_tr}
\mb q^1 &= \mb u^1 \nn \\
\mb q^{k} &= \sum_{l=k}^P\f{k-1}{l-1}\mb u^l +\mb u^1 \hspace{8mm} k =2\cdots P
\end{align}
It can be proven that the harmonic term decouples in the new variables \cite{tuckerman2010}:
\begin{equation}
\sum_{k=1}^P(\mathbf q_{k+1}-\mathbf q_k)^2 =\sum_{k=2}^P\frac{k}{k-1}\mathbf u_k^2= \sum_{k=2}^P\sum_{j=1}^N\frac{k}{k-1}(u_j^k)^2\label{quadr_u}
\end{equation}
We can write the $NP$ Cartesian components of the configurations and staging variables in \textit{row-major} order, defined s.t. $\forall \; k=1\cdots P, \; \; j= 1\cdots N$. Through the notation introduced in \cref{row_maj}, the Jacobian of the transformation is conveniently defined as
\begin{equation*}
\mathbf {J} \equiv\left\{ \frac{\partial q_i}{\partial u_l} \right\}_{i,l=1,\cdots, NP}
\end{equation*}
its components are
\begin{align*}
J_{m,n} \equiv \f{\partial q_j^m}{\partial u_j^n}=
\begin{cases}
\delta_{mn},& m=1 \\
=\f{\p }{\p u_j^n}\Le(u_j^1+\sum_{l=m}^P\f{m-1}{l-1}u_j^l\Ri), &m=2\cdots P
\end{cases}=
\begin{cases}
\delta_{mn},& m=1 \\
\theta(n-m)\f{m-1}{n-1} &m=2\cdots P
\end{cases}
\end{align*}
i.e. it is a triangular matrix, hence its determinant is given by the product of the diagonal term, which are constant and equal to one. The partition function in \cref{Z_T_iso} is rewritten as 
\begin{align}
Z_T&=\lim_{P\to+\infty} P^{NP/2}\int_{\mathbb R^{NP}}\mr d \mathbf p^1\cdots\mr d \mathbf p^P \;\exp\Le\{-\frac 1 T\sum_{k=1}^P\f{(\mathbf p^k)^2}{2\mu'_k}\Ri\}\times \nn\\
&\times\int_{\mathbb R^{NP}}\mr d\mathbf u^1 \cdots\mr d\mathbf u^P\exp\Le\{-\frac 1 T\sum_{k=1}^P\Le[\f{1}{ P} V(\mathbf q^{k}(\mathbf u)) +\frac{\mu_kT^2P}{2}(\mathbf u^k)^2\Ri]\Ri\} \label{Z_T_iso_in_u}  \\
&\mu_k \equiv \begin{cases}
0,& k=1 \\
\frac{k}{k-1},& k=2,\cdots, P 
\end{cases} \label{def_mu} \hspace{10mm}
\mu'_k \equiv \begin{cases}
1,& k=1 \\
\mu_k,& k=2,\cdots, P 
\end{cases}
\end{align}
where $\mathbf u\equiv(\mathbf u_1\cdots \mathbf u_P)\in\mathbb R^{NP}$. The fictitious kinetic masses $\mu_k'$ in \cref{Z_T_iso_in_u} have been adapted to the ones of the harmonic coupling between the replicas, without any effective consequence on the statistics:
\begin{align*}
&\int_{\mathbb R^{NP}}\mr d \mathbf p^1\cdots\mr d \mathbf p^P \;\exp\Le\{-\frac 1 T\sum_{k=1}^P\f{(\mathbf p^k)^2}{2\mu'_k}\Ri\}= \prod_{k=1}^P(2\pi \mu_k'T)^{N/2}=(2\pi T)^{NP/2}\prod_{k=1}^P {\mu_k'}^{N/2}= \\
&=(2\pi T)^{NP/2} \left(1 \cdot 2\cdots \frac{k}{k-1}\frac{k+1}{k}\cdots \frac P{P-1}\right)^{N/2}= (2\pi TP)^{NP/2}= \\
&=(P)^{NP/2}\int_{\mathbb R^{NP}}\mr d \mathbf p^1\cdots\mr d \mathbf p^P \;\exp\Le\{-\frac 1 T\sum_{k=1}^P\f{(\mathbf p^k)^2}{2}\Ri\}
\end{align*}
Let us remark that the prefactor $P^{NP/2}$ in \cref{Z_T_iso} has been absorbed in the new masses $\mu'_k$. 
The classical-like Hamiltonian 
\begin{align}\label{def_tilde_H_cl}
\tilde {\mc H}_{cl}(\mb u,\mb p) &= \sum_{k=1}^P\Le[\frac{(\mb p^k)^2}{2{\mu_k'}}+\f{1}{ P} V(\mb q^{k}(\mb u)) +\frac {\mu_kPT^2}{2}(\mb u^k)^2\Ri] 
\end{align}
generates for each of the $NP$ Cartesian components of the coordinates $(u_j^k, p_j^k)$ the e.o.m.'s 
\begin{align}
\dot u_{j}^k &=\f{\p \tilde {\mc H}_{cl}}{\p p_j^k} = \f{\p }{\p p_j^k}\Le(\sum_{j=1}^N\sum_{k=1}^P\f{(p_j^k)^2}{2\mu_k'}\Ri) = p_j^k/{\mu_k'} \nn \\
\dot p_j^k&= -\mu_k P T^2u_j^k-\f 1 P\sum_{l=1}^P\f{\p V(\mb q^l(\mb u))}{\p u_j^k}=-\mu_kPT^2u_j^k-\f 1 P\f{\p V(\mb q(\mb u))}{\p u_j^k} \label{eom_up}
\end{align}
where in the last line we defined
\begin{equation*}
V(\mb q(\mb u))\equiv\sum_{l=1}^P V(\mb q^l(\mb u))
\end{equation*}
We can fix a recursion relation for the calculation of the forces in the staging variables, in order to relate them to the forces w.r.t. the primitive variables, which can be directly computed in a simulation. For $k=1$ we can write
\begin{align*}
\f 1 P\f{\p V(\mb q(\mb u))}{\p u_j^1}&=\f 1 P \sum_{l=1}^P\f{\p V(\mb q(\mb u))}{\p q_j^l}\f{\p q_j^l(\mb u_j)}{\p u_j^1}\numeq{\cref{inv_tr}}
\f 1 P \sum_{l=1}^P\f{\p V(\mb q(\mb u))}{\p q_j^l}\f{\p u_j^1}{\p u_j^1}
=\f 1 P \sum_{l=1}^P\f{\p V(\mb q(\mb u))}{\p q_j^l}
\end{align*}
while for $k=2\cdots P$
\begin{align}
\f 1 P\f{\p V(\mb q(\mb u))}{\p u_j^k}&=\f 1 P\sum_{l=1}^P\f{\p V(\mb q(\mb u))}{\p q_j^l}\f{\p q_j^l(\mb u_j)}{\p u_j^k} \numeq{\cref{inv_tr}} \f 1 P \sum_{l=1}^P\f{\p V(\mb q(\mb u))}{\p q_j^l}\f{\p}{\p u_j^k}\Le( \sum_{m=l}^P\f{l-1}{m-1} u^m_j + u^1_j \Ri)=\label{rec_F1}\\
&= \f 1 P \sum_{l=1}^P\f{\p V(\mb q(\mb u))}{\p q_j^l}\f{\p}{\p u_j^k}\Le( \theta(k-l)\f{l-1}{k-1} u^k_j  \Ri)= \f 1 P\sum_{l=1}^k\f{\p V(\mb q(\mb u))}{\p q_j^l}\f{l-1}{k-1}=\nn\\
&=\f{k-2}{k-1}\Le[\f 1 P\sum_{l=2}^{k-1}\f{\p V(\mb q(\mb u))}{\p q_j^l}\f{l-1}{k-2}\Ri]+\f 1 P\f{\p V(\mb q(\mb u))}{\p q_j^k}=\f{k-2}{k-1}\f 1 P\f{\p V(\mb q(\mb u))}{\p u_j^{k-1}}+\f 1 P\f{\p V(\mb q(\mb u))}{\p q_j^k} \label{rec_F2}
\end{align}
where 
\begin{equation*}
\theta(x) = \begin{cases}
1,& x\ge 0 \\
0,& x<0 
\end{cases}
\end{equation*}
In \cref{rec_F2} we inserted recursively the expression from \cref{rec_F1}. In the specific case of the FPU potential in \cref{def_H} we have $\forall~k=1\cdots P$:
\begin{align}
\f{\p V(\mb q(\mb u))}{\p q_j^k}=\f{\p}{\p q_j^k}\sum_{l=1}^PV(\mb q^l(\mb u))=\f{\p}{\p q_j^k}\sum_{l=1}^P\sum_{m=1}^NV(q_{m+1}^l-q_{m}^l)=
\begin{cases}
V'(q_1^k)-V'(q_2^k-q_1^k),& j=1 \\
V'(q_j^k-q_{j-1}^k)-V'(q_{j+1}^k-q_j^k),&j=2,\cdots, N-1 \\
V'(q_N^k-q_{N-1}^k)-V'(-q_j^{N}),&j=N 
\end{cases}\label{dV_qjk}
\end{align}
where in the last expression we inserted the b.c. $q_0^k=q_{N+1}^k=0, \;\;\forall k=1\cdots P$.
\subsection{Nos\'e-Hoover chain}\label{app:NHC}
In this appendix we report the details of the numerical scheme used for the equilibrium sampling for the quantum $FPU$ system. The method is also applicable to the classical statistics, in the limit of $P=1$ replica. Each of the $NP$ momenta of the system is connected to a sequence of $M$ thermostats, constructing a so-called \textit{Nos\'e-Hoover chain}. The coupling of those additional degrees of freedom to the physical ones constitutes a non-Hamiltonian dynamical system, spanning phase space according to the thermal distribution of the classical isomorphism in \cref{def_rho_cl}. We refer again to \cite{tuckerman2010, martyna1992, tuckerman1993, tuckerman2006} for an exhaustive explanation of the procedure described in the following. As a notation, let us label with $\{(\eta_{jk}^\gamma, p_{\eta_{jk}^\gamma})\}_{\gamma=1}^M$ the particles of the thermostat attached to the d.o.f. $(p_j^k, u_j^k)$. We will see that the following coupled equations of motions yield to an artificial dynamics generating the correct canonical distribution:
\begin{align}
{\dot { u}^k_j}&= p^k_j/\mu_k'  \label{ref_eom_NHC} \\
{\dot {p}^k_j}&=-\mu_k PT^2u^k_j-\f 1 P \f{\p V(\mb q(\mb u))}{\p u_j^k}-\f{p_{\eta_{jk}^1}}{Q_{jk}^1}p_j^k  \\ 
\dot \eta_{jk}^\gamma &= \f{p_{\eta_{jk}^\gamma}}{Q_{jk}^\gamma} &\gamma=1\cdots M\label{NHC_latt0}\\
\dot p_{\eta_{jk}^1} &= \f{({p}^k_j)^2}{\mu'_k}-T-\f{p_{\eta_{jk}^2}}{Q_{jk}^2}p_{\eta_{jk}^1} \\
\dot p_{\eta_{jk}^\gamma} &= \Le[\f{(p_{\eta_{jk}^{\gamma-1}})^2}{Q_{jk}^{\gamma-1}}-T\Ri] -\f{p_{\eta_{jk}^{\gamma+1}}}{Q_{jk}^{\gamma+1}}p_{\eta_{jk}^\gamma}&\gamma=2\cdots M-1 \\
\dot p_{\eta_{jk}^M} &= \Le[\f{(p_{\eta_{jk}^{M-1}})^2}{Q_{jk}^{M-1}}-T\Ri] \label{NHC_latt} \hspace{10mm}  &j=1\cdots~N,~~k=1\cdots~P
\end{align}
The parameters $Q_{jk}^\gamma$ can be interpreted as masses, tuning the timescale of the evolution of the bath variables \cite{tuckerman2006}. 
The term $-p_{\eta_{jk}^1}/Q_{jk}^1 p_j^k$ acts as a damp or boost of the $(j,k)$-th momentum. From its dynamics in the second and fourth equations above, we can see that if the $({p}^k_j)^2/(2\mu'_k)>T/2$, it will reduce the momentum of the $(j,k)$-th degree of freedom, or vice versa.
It has been shown in \cite{martyna1992} that the optimal choice of the thermostats' masses is
\begin{align}
Q_{jk}^\gamma &=
\begin{cases} 
{\tilde \tau}^2 T, & k=1 \\
1/(PT), & k = 2\cdots P
\end{cases}\label{def_Q_jk}
\end{align}
where $\tilde \tau$ is a timescale associated to the dynamics of classical system. It turns out that an integration scheme based on a simple Taylor expansion would be insufficient to generate the correct canonical distribution \cite{tuckerman1999_comm}; a suitable numerical procedure is described in the following. Let us define a phase of the global system
\begin{equation}\label{def_x}
\mb x = ( u_1^1\cdots u_N^P, \bs\eta_1^1, \cdots \bs \eta_N^P,  p_1^1\cdots \cdots p_N^P, \mb p_{\eta_1^1}\cdots \cdots\mb p_{\eta_N^P})
\end{equation}
where 
\begin{align*}
\bs\eta_j^k= (\eta_{jk}^1,\cdots, \eta_{jk}^M) \\
\mb p_{\eta_j^k}= (p_{\eta_{jk}}^1,\cdots, p_{\eta_{jk}}^M)
\end{align*}
\cref{NHC_latt} is implicitly expressed as
\begin{equation*}
\dot {\mb x} = \bs \xi(\mb x) = i\mc L\mb x \hspace{8mm} i\mc L = \bs\xi(\mb x)\cdot \nabla_{\mb x}
\end{equation*}
We can then separate the Hamiltonian and the thermostats' part of the Liouville operator, according to
\begin{equation*}
i\mc L = i\mc L_1+i\mc L_2+i\mc L_{NHC}
\end{equation*}
where
\begin{align*}
i\mc L_1&=\sum_{j=1}^N\sum_{k=1}^P\f{p_j^k}{\mu'_k}\f{\p}{\p u_j^k} \hspace{8mm} i\mc L_2=\sum_{j=1}^N\sum_{k=1}^P\Le[-\mu_k PT^2 u^k_j-\f 1 P \f{\p V(\mb q(\mb u))}{\p u_j^k}\Ri]\f{\p}{\p p_j^k} \\
i\mc L_{NHC} &= \sum_{j=1}^N\sum_{k=1}^P\Bigg\{-\f{p_{\eta_{jk}^1}}{Q_{jk}^1}p_j^k\f{\p}{\p p_j^k}+\sum_{\gamma=1}^{M}\f{p_{\eta_{jk}^\gamma}}{Q_{jk}^\gamma}\f{\p}{\p\eta^\gamma_{jk}}+\Le( \f{({p}^k_j)^2}{\mu'_k}-T-\f{p_{\eta_{jk}^2}}{Q_{jk}^2}p_{\eta_{jk}^1}\Ri)\f{\p}{\p p_{\eta_{jk}^1}}+\\
&+\sum_{\gamma=2}^{M-1}\Le[\Le(\f{(p_{\eta_{jk}}^{\gamma-1})^2}{Q_{jk}^{\gamma-1}}-T-\f{p_{\eta_{jk}^{\gamma+1}}}{Q_{jk}^{\gamma+1}}p_{\eta_{jk}^1}\Ri)\f{\p}{\p p_{\eta_{jk}^\gamma}}\Ri]+\Le(\f{(p_{\eta_{jk}^{M-1}})^2}{Q_{jk}^{M-1}}-T\Ri)\f{\p}{\p p_{\eta_{jk}^M}}\Bigg\}=\\
&=\sum_{j=1}^N\sum_{k=1}^P\Bigg\{-\f{p_{\eta_{jk}^1}}{Q_{jk}^1}p_j^k\f{\p}{\p p_j^k}+\sum_{\gamma=1}^{M}\f{p_{\eta_{jk}^\gamma}}{Q_{jk}^\gamma}\f{\p}{\p\eta_{jk}^\gamma}+\sum_{\gamma=1}^{M-1}\Le( G_{jk}^\gamma-\f{p_{\eta_{jk}^{\gamma+1}}}{Q_{jk}^{\gamma+1}}p_{\eta_{jk}^\gamma}\Ri)\f{\p}{\p p_{\eta_{jk}^\gamma}}+G_{jk}^M\f{\p}{\p p_{\eta_{jk}^M}}\Bigg\}
\end{align*}
where we defined 
\begin{align*}
G_{jk}^1 &=\f{(p_{j}^k)^2}{\mu_k'}-T\\
G_{jk}^\gamma &= \f{(p_{\eta_{jk}}^{\gamma-1})^2}{Q_{jk}^{\gamma-1}}-T,\hspace{8mm}\gamma = 2,\cdots M
\end{align*}
The symmetric Trotter theorem allows us to decompose the total propagator into an infinite product of infinitesimal exponential operators:
\begin{align*}
&e^{i\mc L t} = e^{(i\mc L_1+i\mc L_2+i\mc L_{NHC}) t} =  \lim_{P\to+\infty} \Le[ e^{i\mc L_{NHC} t/(2P)} e^{i(\mc L_1+\mc L_2)t/P}e^{i\mc L_{NHC} t/(2P)}\Ri]^P= \lim_{\substack{P\to+\infty\\\Delta t\to 0^+}} \Le[ e^{i\mc L_{NHC} \Delta t/2} e^{i(\mc L_1+\mc L_2)\Delta t}e^{i\mc L_{NHC}\Delta t/2}\Ri]^P 
\end{align*}
where in the last identity we defined $\Delta t = t/P$. The Trotter decomposition at finite values of $P$ induces an error
\begin{equation*}
e^{i\mc L t} = \Le[ e^{i\mc L_{NHC} \Delta t/2} e^{i\mc L_2\Delta t/2}e^{i\mc L_1\Delta t }e^{i\mc L_2\Delta t/2}e^{i\mc L_{NHC}\Delta t/2}\Ri]^P +\mc O(P\Delta t^3)
\end{equation*} 
As $P =t/\Delta t$, the total error increases as $\mc O(\Delta t^4)$. The single timestep is instead decomposed into
\begin{equation}\label{propag_Dt}
e^{i\mc L \Delta t} = e^{i\mc L_{NHC} \Delta t/2} e^{i\mc L_2\Delta t/2}e^{i\mc L_1\Delta t }e^{i\mc L_2\Delta t/2}e^{i\mc L_{NHC}\Delta t/2} +\mc O(\Delta t^3)
\end{equation}
with a ``local'' error $\mc O(\Delta t^3)$. The operator $e^{i\mc L_{NHC}\Delta t}$ needs to be further factorized in order to be applied analytically. The dynamics of the thermostats' d.o.f. is in general faster than the one of the physical d.o.f., as the masses of the two scale respectively as $\mc O(P^{-1})$ (\cref{def_Q_jk}) and $\mc O(1)$ (\cref{def_mu}). To efficiently incorporate such timescales' separation, we can apply as high order decomposition of the propagator the so called \textit{Suzuki-Yoshida scheme} \cite{yoshida1990}. The method involves a primitive factorization of NHC propagator in $n_{sy}$ terms, with suitable weights $w_\alpha$:
\begin{equation}\label{SY_fact}
e^{i\mc L_{NHC} \Delta t/2} = \prod_{\alpha=1}^{n_{sy}} S(w_\alpha \Delta t/2)
\end{equation}
We can then apply a method called \textit{Reference System Propagator Algorithm} (RESPA) \cite{tuckerman2010} to the primitive factors. The method consists in a further decomposition of the thermostats' propagater dynamics in $n_R$ segments of size $\Delta t/n_R$. The choice of $n_R$ can be assessed \textit{a-posteri} from the conservation law of the NHC dynamics discussed later. We then have:
\begin{equation}\label{SY_RESPA}
e^{i\mc L_{NHC} \Delta t/2} = \prod_{\alpha=1}^{n_{sy}} [S(w_\alpha \Delta t/(2n_R))]^{n_R}
\end{equation}
The Yoshida-Suzuki weights $\{w_\alpha\}_{\alpha=1}^{n_{sy}}$ are determined numerically, as a solution of a set of algebraic equations. In our case we consider an expansion till sixth order, where $n_{sy}=7$. By defining $\delta_\alpha\equiv w_\alpha\Delta t/n_R$, we can progressively construct a primitive factorization of $S(w_\alpha \Delta t/(2n_R))$ by decomposing the primitive factors in \cref{SY_RESPA} via the Trotter theorem \cite{martyna1996}:
\begin{align}
&S(\delta_\alpha/2) = \exp\Le\{\f{\delta_\alpha}2\sum_{j=1}^N\sum_{k=1}^P\Le[-\f{p_{\eta_{jk}}^1}{Q_{jk}^1}p_j^k\f{\p}{\p p_j^k}+\sum_{\gamma=1}^{M}\f{p_{\eta_{jk}^\gamma}}{Q_{jk}^\gamma}\f{\p}{\p\eta^\gamma_{jk}}+\sum_{\gamma=1}^{M-1}\Le( G_{jk}^\gamma-\f{p_{\eta_{jk}^{\gamma+1}}}{Q_{jk}^{\gamma+1}}p_{\eta_{jk}^\gamma}\Ri)\f{\p}{\p p_{\eta_{jk}^\gamma}}+G_{jk}^M\f{\p}{\p p_{\eta_{jk}^M}}\Ri]\Ri\}=\nn\\
&= \exp\Le\{\f{\delta_\alpha}4\sum_{j=1}^N\sum_{k=1}^PG_{jk}^M\f{\p}{\p p_{\eta_{jk}^M}}\Ri\}\exp\Le\{\f{\delta_\alpha}2\sum_{j=1}^N\sum_{k=1}^P\Le[-\f{p_{\eta_{jk}}^1}{Q_{jk}^1}p_j^k\f{\p}{\p p_j^k}+\sum_{\gamma=1}^{M}\f{p_{\eta_{jk}^\gamma}}{Q_{jk}^\gamma}\f{\p}{\p\eta_{jk}^\gamma}+\sum_{\gamma=1}^{M-1}\Le( G_{jk}^\gamma-\f{p_{\eta_{jk}^{\gamma+1}}}{Q_{jk}^{\gamma+1}}p_{\eta_{jk}^\gamma}\Ri)\f{\p}{\p p_{\eta_{jk}^\gamma}}\Ri]\Ri\}\times \nn \\
&=\exp\Le\{\f{\delta_\alpha}4\sum_{j=1}^N\sum_{k=1}^PG_{jk}^M\f{\p}{\p p_{\eta_{jk}^M}}\Ri\}+\mc O(\Delta t^3) \label{S_da1}
\end{align}
We can then proceed by expanding with the same Trotter formula the central exponential in \cref{S_da1}, while keeping a global error $\mc O(\Delta t^3)$:
\begin{align}
&S(\delta_\alpha/2) =\exp\Le\{\f{\delta_\alpha}4\sum_{j=1}^N\sum_{k=1}^PG_{jk}^M\f{\p}{\p p_{\eta_{jk}^M}}\Ri\}
\exp\Le\{\f{\delta_\alpha}4\sum_{j=1}^N\sum_{k=1}^P\sum_{\gamma=1}^{M-1}\Le( G_{jk}^\gamma-\f{p_{\eta_{jk}^{\gamma+1}}}{Q_{jk}^{\gamma+1}}p_{\eta_{jk}^\gamma}\Ri)\f{\p}{\p p_{\eta_{jk}^\gamma}}\Ri\}\times \nn\\
&\times \exp\Le\{\f{\delta_\alpha}2\sum_{j=1}^N\sum_{k=1}^P\Le[-\f{p_{\eta_{jk}}^1}{Q_{jk}^1}p_j^k\f{\p}{\p p_j^k}+\sum_{\gamma=1}^{M}\f{p_{\eta_{jk}^\gamma}}{Q_{jk}^\gamma}\f{\p}{\p\eta_{jk}^\gamma}\Ri]\Ri\}
\exp\Le\{\f{\delta_\alpha}4\sum_{j=1}^N\sum_{k=1}^P\sum_{\gamma=1}^{M-1}\Le( G_{jk}^\gamma-\f{p_{\eta_{jk}^{\gamma+1}}}{Q_{jk}^{\gamma+1}}p_{\eta_{jk}^\gamma}\Ri)\f{\p}{\p p_{\eta_{jk}^\gamma}}\Ri\}\times \label{they_comm}\\
&\times\exp\Le\{\f{\delta_\alpha}4\sum_{j=1}^N\sum_{k=1}^PG_{jk}^M\f{\p}{\p p_{\eta_{jk}^M}}\Ri\} +\mc O(\Delta t^3) \nn
\end{align} 
We can notice that the contributions in $e^{i\mc L_{NHC}t}$ of different values of $\gamma$ are in general non commuting, due to the coupling between neighboring d.o.f. through the masses $G^\gamma_{jk}$. For example, let us define a general test function of the thermostats' momenta $f(\mb p_{\bs\eta})=f(\mb p_{\eta_{11}}, \cdots,\mb p_{\eta_{NP}})$, and let us consider the application of the same differential for two thermostat contributions $\gamma,\gamma'=\gamma+1$; then
\begin{align}
&\Le(G_{jk}^\gamma\f{\p}{\p p_{\eta_{jk}^\gamma}}\Ri)\Le(G_{jk}^{\gamma'}\f{\p}{\p p_{\eta_{jk}^{\gamma'}}}\Ri)f(\mb p_{\bs\eta})=G_{jk}^\gamma\f{\p G_{jk}^{\gamma'}}{\p p_{\eta_{jk}^\gamma}}\f{\p f(\mb p_{\bs\eta})}{\p p_{\eta_{jk}^{\gamma'}}}+G_{jk}^{\gamma}G_{jk}^{\gamma'}\f{\p^2 f(\mb p_{\bs\eta})}{\p p_{\eta_{jk}^\gamma}\p p_{\eta_{jk}^{\gamma'}}}= \nn \\
&=2G_{jk}^\gamma \f{p_{\eta_{jk}}^\gamma}{Q_{jk}^{\gamma}}\f{\p f(\mb p_{\bs\eta})}{\p p_{\eta_{jk}^{\gamma+1}}}+G_{jk}^{\gamma}G_{jk}^{\gamma+1}\f{\p^2 f(\mb p_{\bs\eta})}{\p p_{\eta_{jk}^\gamma}\p p_{\eta_{jk}^{\gamma+1}}} \nn
\end{align}
\begin{align}
&\Le(G_{jk}^{\gamma'}\f{\p}{\p p_{\eta_{jk}^{\gamma'}}}\Ri)\Le(G_{jk}^{\gamma}\f{\p}{\p p_{\eta_{jk}^{\gamma}}}\Ri)f(\mb p_{\bs\eta})=G_{jk}^{\gamma'}\f{\p G_{jk}^{\gamma}}{\p p_{\eta_{jk}^{\gamma'}}}\f{\p f(\mb p_{\bs\eta})}{\p p_{\eta_{jk}^{\gamma}}}+G_{jk}^{\gamma'}G_{jk}^{\gamma}\f{\p^2 f(\mb p_{\bs\eta})}{\p p_{\eta_{jk}^{\gamma'}}\p p_{\eta_{jk}^{\gamma}}}= G_{jk}^{\gamma'}G_{jk}^{\gamma}\f{\p^2 f(\mb p_{\bs\eta})}{\p p_{\eta_{jk}^{\gamma'}}\p p_{\eta_{jk}^{\gamma}}}
\end{align}
that is
\begin{equation}\label{non_comm_d}
\Le[G_{jk}^\gamma\f{\p}{\p p_{\eta_{jk}^\gamma}},G_{jk}^{\gamma'}\f{\p}{\p p_{\eta_{jk}^{\gamma'}}}\Ri]=2G_{jk}^\gamma \f{p_{\eta_{jk}}^\gamma}{Q_{jk}^{\gamma}}\f{\p f(\mb p_{\bs\eta})}{\p p_{\eta_{jk}^{\gamma+1}}}
\end{equation}
\cref{non_comm_d} implies that it is not possible to apply a simple factorization of the exponential of the sums in terms of products of exponentials. We can however notice that the $NHC$ dynamics acts is completely separable in the physical d.o.f. $(j,k)$. We can therefore apply an additional symmetric splitting for the decomposition of the sums over $\gamma$ in \cref{they_comm}, while factorizing the exponentials of the sums over $j$ and $k$ into products of exponentials. This yields
\begin{align}
S(\delta_\alpha/2) &=\prod_{j=1}^N\prod_{k=1}^P\exp\Le\{\f{\delta_\alpha}4G_{jk}^M\f{\p}{\p p_{\eta_{jk}^M}}\Ri\}
\prod_{\gamma=M-1}^{1}\exp\Le\{-\f{\delta_\alpha}8 \f{p_{\eta_{jk}^{\gamma+1}}}{Q_{jk}^{\gamma+1}}p_{\eta_{jk}^\gamma}\f{\p}{\p p_{\eta_{jk}^\gamma}}\Ri\}\exp\Le\{\f{\delta_\alpha}4  G_{jk}^\gamma\f{\p}{\p p_{\eta_{jk}^\gamma}}\Ri\}\times \nn\\
& \exp\Le\{-\f{\delta_\alpha}8 \f{p_{\eta_{jk}^{\gamma+1}}}{Q_{jk}^{\gamma+1}}p_{\eta_{jk}^\gamma}\f{\p}{\p p_{\eta_{jk}^\gamma}}\Ri\}\Le[\exp\Le\{-\f{\delta_\alpha}2\f{p_{\eta_{jk}}^1}{Q_{jk}^1}p_j^k\f{\p}{\p p_j^k}\Ri\}\prod_{\gamma'=1}^{M}\exp\Le\{\f{\delta_\alpha}2\f{p_{\eta_{jk}^{\gamma'}}}{Q_{jk}^{\gamma'}}\f{\p}{\p\eta_{jk}^{\gamma'}}\Ri\}\Ri]\times \label{they_comm2}\\
&\times\prod_{\gamma''=1}^{M-1}\exp\Le\{-\f{\delta_\alpha}8 \f{p_{\eta_{jk}^{\gamma''+1}}}{Q_{jk}^{\gamma''+1}}p_{\eta_{jk}^{\gamma''}}\f{\p}{\p p_{\eta_{jk}^{\gamma''}}}\Ri\}\exp\Le\{\f{\delta_\alpha}4  G_{jk}^{\gamma''}\f{\p}{\p p_{\eta_{jk}^{\gamma''}}}\Ri\}\Le\{-\f{\delta_\alpha}8 \f{p_{\eta_{jk}^{\gamma''+1}}}{Q_{jk}^{\gamma''+1}}p_{\eta_{jk}^{\gamma''}}\f{\p}{\p p_{\eta_{jk}^{\gamma''}}}\Ri\}\exp\Le\{\f{\delta_\alpha}4G_{jk}^M\f{\p}{\p p_{\eta_{jk}^M}}\Ri\} +\mc O(\delta_\alpha^3) \nn = \\
&\equiv\prod_{j=1}^N\prod_{k=1}^P S_{jk}(\delta_\alpha/2) +\mc O(\delta_\alpha^3)
\end{align}
The first exponential in \cref{they_comm} has been identically factorized in commuting operators in the square brackets of \cref{they_comm2}. 
\\ \\
The dynamics in \cref{NHC_latt} conserves the Hamiltonian
\begin{align}
\mc H'_{cl} &= \tilde{\mc H}_{cl}+\sum_{k=1}^P\sum_{j=1}^N\sum_{\gamma=1}^{M}\Le[\f{(p_{\eta_{jk}}^\gamma)^2}{2Q_{jk}^\gamma}+T\eta_{jk}^\gamma\Ri] \numeq{\cref{def_tilde_H_cl}} \sum_{j=1}^N\sum_{k=1}^P\Le[\f{(p_j^k)^2}{2{\mu_k'}}+\f{1}{ P} V(\mb q^{k}(\mb u)) +\f {\mu_kP T^2}2(u_j^k)^2   + \sum_{\gamma=1}^{M}\Le(\f{(p_{\eta_{jk}}^\gamma)^2}{2Q_{jk}^\gamma}+T\eta_{jk}^\gamma\Ri)\Ri] = \nn \\
&=\sum_{j=1}^N\sum_{k=1}^P\Le[\f{(p_j^k)^2}{2{\mu_k'}}+\f{1}{ P} V( q_j^{k}(\mathbf u)-q_{j-1}^k(\mathbf u)) +\f {\mu_kPT^2}2(u_j^k)^2   + \sum_{\gamma=1}^{M}\Le(\f{(p_{\eta_{jk}}^\gamma)^2}{2Q_{jk}^\gamma}+T\eta_{jk}^\gamma\Ri)\Ri] \label{def_H'cl}
\end{align}
This can be verified through a direct proof:
\begin{align}
\f{\mr d \mc H'_{cl}}{\mr d t}
&=\sum_{k=1}^P\sum_{j=1}^N\Le[\f{ p_j^k}{{\mu_k'}}{\dot p}_j^k+\f{1}{ P} \sum_{l=1}^P\f{\p V(\mb q^{k}(\mb u))}{\p u_j^l}{\dot u}_j^l +\mu_kP T^2 u_j^k {\dot u}_j^k  +\sum_{\gamma=1}^{M}\Le(\f{p_{\eta_{jk}}^\gamma}{Q_{jk}^\gamma}\dot p_{\eta_{jk}}^\gamma+T\dot\eta_{jk}^\gamma\Ri)\Ri]= \nn \\
&\numeq{\ref{ref_eom_NHC}-\ref{NHC_latt} }\sum_{k=1}^P\sum_{j=1}^N\Bigg\{ \f{ p_j^k}{{\mu_k'}}\Le(-\mu_kPT^2u^k_j-\f 1 P \sum_{l=1}^P\f{\p V(\mb q^l(\mb u))}{\p u_j^k}-\f{p_{\eta_{jk}^1}}{Q_{jk}^1}p_j^k\Ri)+\f 1 P\sum_{l=1}^P\f{\p V(\mb q^{k}(\mb u))}{\p u_j^l}\f{p^l_j}{\mu_k'}+\mu_kPT^2 u_j^k \f{p^k_j}{\mu_k'} + \label{inv_kl}\\
&+\Le[\f{p_{\eta_{jk}^1}}{Q_{jk}^1}\Le(\f{({p}^k_j)^2}{\mu'_k}-T-\f{p_{\eta_{jk}^2}}{Q_{jk}^2}p_{\eta_{jk}^1}\Ri)+\sum_{\gamma=2}^{M-1}\f{p_{\eta_{jk}}^\gamma}{Q_{jk}^\gamma}\Le(\f{(p_{\eta_{jk}^{\gamma-1}})^2}{Q_{jk}^{\gamma-1}}-T -\f{p_{\eta_{jk}^{\gamma+1}}}{Q_{jk}^{\gamma+1}}p_{\eta_{jk}^\gamma}\Ri)+  \f{p_{\eta_{jk}}^M}{Q_{jk}^M}\Le(\f{(p_{\eta_{jk}^{M-1}})^2}{Q_{jk}^{M-1}}-T\Ri)  + T\sum_{\gamma=1}^{M}\f{p_{\eta_{jk}^\gamma}}{Q_{jk}^\gamma} \Ri]\Bigg\}= \nn\\
&=-\f{(p_{\eta_{jk}^1})^2}{Q_{jk}^1}\f{p_{\eta_{jk}^2}}{Q_{jk}^2}+\sum_{\gamma=2}^{M-1}\f{(p_{\eta_{jk}^{\gamma-1}})^2}{Q_{jk}^{\gamma-1}}\f{p_{\eta_{jk}^\gamma}}{Q_{jk}^\gamma}-\sum_{\gamma=2}^{M-1}\f{(p_{\eta_{jk}^{\gamma}})^2}{Q_{jk}^{\gamma}}\f{p_{\eta_{jk}^{\gamma+1}}}{Q_{jk}^{\gamma+1}}+\f{(p_{\eta_{jk}^{M-1}})^2}{Q_{jk}^{M-1}}\f{p_{\eta_{jk}}^M}{Q_{jk}^M}=  \nn\\
&=\sum_{\gamma=2}^{M}\f{(p_{\eta_{jk}^{\gamma-1}})^2}{Q_{jk}^{\gamma-1}}\f{p_{\eta_{jk}^\gamma}}{Q_{jk}^\gamma}-\sum_{\gamma=1}^{M-1}\f{(p_{\eta_{jk}^{\gamma}})^2}{Q_{jk}^{\gamma}}\f{p_{\eta_{jk}^{\gamma+1}}}{Q_{jk}^{\gamma+1}}=0
\end{align}
Given that this is the only conservation law satisfied by the system, the propagation of the dynamics in \crefrange{ref_eom_NHC}{NHC_latt} is microcanonical on the extended phase space w.r.t. $\mathcal H'_{\mathrm{cl}}$. In particular, the constraint 
\begin{equation*}
f_E(\mb x_t)=\mc N\delta(\mc H'_{cl}(\mb x_t)-E)
\end{equation*}
for a suitable normalization constant $\mc N$ is satisfied for any phase $\mathbf x_t$, $t\in \mathbb R$. 
Given that the present dynamics is non-Hamiltonian, volumes in the phase space will not be in general preserved by the time evolution, i.e. $\mr d \mb x_t\neq\mr d \mb x_0$. In particular, the propagation can be seen as a parametric change of coordinates
\begin{equation*}
\mb x_t = \mb x_t(t,\mb x_0) 
\end{equation*}
with the related measure transformation
\begin{equation*}
\mr d \mb x_t=J(\mb x_t,\mb x_0)\mr d \mb x_0
\end{equation*}
defining the Jacobian $J(\mb x_t,\mb x_0)=\f{\p \mb x_t}{\p \mb x_0}$. Is shown in \cite{tuckerman_1999} that it satisfies the following dynamical equation
\begin{equation}\label{dyn_J}
\f{\mr d}{\mr d t}J(\mb x_t,\mb x_0)=\kappa(\mb x_t)J(\mb x_t,\mb x_0)
\end{equation}
where we introduced the phase space compressibility
\begin{align}
\kappa(\mb x)\equiv\nabla_{\mb x}\cdot {\mb{\dot x}} &= \nabla_{\mb x}\cdot\bs\xi(\mb x, t)=\sum_{j=1}^N\sum_{k=1}^P \Le[ \Le(\f{\p \dot p_{jk}}{\p p_{jk}}+ \f{\p \dot q_{jk}}{\p q_{jk}}\Ri)+\sum_{\gamma=1}^{M}\Le(\f{\p\dot \eta_{jk}^\gamma}{\p \eta_{jk}^\gamma}+\f{\p\dot p_{\eta_{jk}^\gamma}}{\p p_{\eta_{jk}^\gamma}}\Ri)\Ri] = \nn\\
&=\sum_{j=1}^N\sum_{k=1}^P \Le[-\f{p_{\eta_{jk}^1}}{Q_{jk}^1}-\f{p_{\eta_{jk}^2}}{Q_{jk}^2} -\sum_{\gamma=2}^{M-1}\Le(\f{p_{\eta_{jk}^{\gamma+1}}}{Q_{jk}^{\gamma+1}}\Ri)\Ri]=-\sum_{j=1}^N\sum_{k=1}^P\sum_{\gamma=1}^{M}\f{p_{\eta_{jk}^\gamma}}{Q_{jk}^\gamma}=-\sum_{j=1}^N\sum_{k=1}^P\sum_{\gamma=1}^{M}\dot \eta_{jk}^\gamma \label{kap}
\end{align}
As Hamiltonian part of the dynamics is compressible, its contribution vanishes identically. The characteristics methods allows to solve \cref{dyn_J}, yielding 
\begin{equation}\label{J1}
J(\mb x_t,\mb x_0)= \exp\Le[\int_0^t\mr d s\kappa(\mb x_s)\Ri]
\end{equation}
By defining a function $w=w(\mb x_t,t)$ such that 
\begin{equation}\label{def_w}
\kappa(\mb x_t)\equiv \Le.\f{\mr d w(\mb x_{t'},t')}{\mr d t'}\Ri|_{t'=t}
\end{equation}
we can rewrite \cref{J1} as
\begin{equation*}
J(\mb x_t,\mb x_0)= \exp\left[\int_0^t \mathrm d s\; \left.\frac{\mathrm d w(\mathbf x_{t'},t')}{\mathrm dt'}\right|_{t'=s}\right] = \f{e^{-w(\mb x_0,0)}}{e^{-w(\mb x_t,t)}}\equiv\f{\sqrt{g(\mb x_0,0)}}{\sqrt{g(\mb x_t,t)}}
\end{equation*}
The conservation law of the non-Lebesgue measure involves now an additional weight from the metric $g$:
\begin{equation*}
\sqrt{g(\mb x_t,t)}\mr d \mb x_t=\sqrt{g(\mb x_0,0)}\mr d \mb x_0
\end{equation*}
 The microcanonical partition function yields to the canonical ensemble the physical d.o.f.:
\begin{align*}
&\Omega_E\equiv\mc N\int \mr d \mb x\sqrt{g(\mb x, 0)} \;f_E(\mb x)\numeq{\cref{kap}}\mc N\int \mr d \mb x \;\exp\Le[ \sum_{j'=1}^N\sum_{k'=1}^P\sum_{\gamma'=1}^M\eta_{j'k'}^{\gamma'}\Ri]
\delta(\mc H'_{cl}(\mb x)-E)=\\
&=\mc N\int \mr d \mb x \exp\Le[ \sum_{j'=1}^N\sum_{k'=1}^P\sum_{\gamma'=1}^M\eta_{j'k'}^{\gamma'}\Ri]\delta\Le(\sum_{j=1}^N\sum_{k=1}^P\Le[\f{(p_j^k)^2}{2{\mu_k'}}+\f{1}{ P} V( q_j^{k}(\mathbf u)-q_{j-1}^k(\mathbf u)) +\f {\mu_k PT^2}2(u_j^k)^2   + \sum_{\gamma=1}^{M}\Le(\f{(p_{\eta_{jk}}^\gamma)^2}{2Q_{jk}^\gamma}+T\eta_{jk}^\gamma\Ri)\Ri]-E\Ri)=\\
&=\mc Ne^{E/T}\int \mr d \mb u \mr d \mb p\; e^{-\mc H_{cl}(\mb u, \mb p)/T}\int \mr d \mb p_{\bs\eta} \prod_{j=1}^N\prod_{k=1}^P\prod_{\gamma=1}^{M} \exp\Le[-\frac 1 T\f{(p_{\eta_{jk}}^\gamma)^2}{2Q_{jk}^\gamma}\Ri]
\end{align*}
A parallelization scheme for the NHC propagator is implemented by partitioning on different cores the time evolution over a time step $\delta_\alpha$ for different thermostats, each acting on a $(j,k)$-th d.o.f. Conversely, a parallelization according to the Suzuki-Yoshida factorization in \cref{SY_fact} would impose additional synchronization procedures, as the order of the application or $S_{jk}(\delta_\alpha)$ for different values of $\alpha$ is relevant. By labeling the cores indexes $n_c=1\cdots N_c$, we can factorize: 
\begin{align}\label{parall_nhc}
e^{i\mc L_{NHC}\Delta t} \equiv\prod_{\alpha=1}^{n_{sy}}\Le[\prod_{j=1}^N\prod_{k=1}^PS_{jk}(\delta_\alpha/2)\Ri]^{n_R} =\prod_{\alpha=1}^{n_{sy}}\Le[\prod_{i=1}^{NP}S_i(\delta_\alpha/2)\Ri]^{n_R}=\prod_{n_c=1}^{N_c}\prod_{i=i_{n_c}}^{i_{n_c+1}-1}\prod_{\alpha=1}^{n_{sy}} \Le[S_i(\delta_\alpha/2)\Ri]^{n_R}
\end{align}
where we fixed a row-major ordering of the indexes $i\equiv Nk+j$ and we labelled the subgroup of d.o.f. on the $n_c$-th core as $\{i_{n_c},\cdots,i_{n_c+1}-1 \}$.
After the propagation in \cref{parall_nhc}, the cores are then synchronized in order to update the phases $\{(q_i, p_i)\}$ and hence to propagate the Hamiltonian part of the dynamics $e^{i\mc L_2\Delta t/2}e^{i\mc L_1\Delta t}e^{i\mc L_2\Delta t/2}$. A second cores' splitting according to \cref{parall_nhc} is then applied. This completes the propagation of the timestep $\Delta t$, as defined in \cref{propag_Dt}.
\section{Harmonic limit}\label{app:harm_lim}
The distributions introduced in \cref{sect:statistics_theo} can be conveniently rewritten in terms coordinates of the normal modes in the limit of a harmonic potential. This is approximately satisfied in the low temperature limit, while it holds exactly in a case of harmonic potential, when $\alpha=\beta=0$. We discuss the two cases separately in the next subsections. A pedagogical discussion on the quantum harmonic chain can be found in \cite{johnson2002}.
\subsection{Low temperature limit}
In the $T\to0+$ limit and for $\alpha\ge 4$, the density matrix of the canonical mixture in the energy eigenstates
\begin{equation}\label{def_hat_rho}
\hat \rho_T = \frac 1{Z_T}\sum_n e^{-E_n/T}\ket{n}\bra{n}
\end{equation}
can be approximated by the projector on the ground state
\begin{equation}\label{def_hat_psi}
\hat \rho_{0} \equiv \frac 1{Z_{0}}\ket{0}\bra{0}, \hspace{10mm} Z_{0}=\tr\{\ket {0} \bra{0}\}
\end{equation}
as the contributions for $n>1$ are exponentially smaller than the first one. In this limit, the quadratic part of the potential suffices in approximating the total interaction. To show this, let us introduce the displacements from the bottom of the potential well
\begin{equation}\label{def_xj}
\hat x_j = \hat q_j -jr_{\min}\hat I 
\end{equation}
to expand the two body interaction at second order:
\begin{equation}\label{Vquad} 
\hat V(\hat q_1,\cdots,\hat q_N) = \sum_{j=0}^N V(\hat q_{j+1}-\hat q_j) = (N+1)\hat V(r_{min}\hat I )+\sum_{j=1}^N\left[\frac{\hat V''(r_{min}\hat I )}{2}(\hat  x_{j+1}-\hat  x_j)^2 +\mathcal O(\hat x_{j+1}-\hat x_j)^2\right]
\end{equation}
where $r_\min$ denotes the absolute minimum 
\begin{equation}
r_\min =(-\alpha-\sqrt{\alpha^2-4\alpha})/(2\alpha)
\end{equation}
\cref{Vquad} is diagonalized by the configurations of the normal modes
\begin{equation}\label{def_eta_x}
\hat \eta_j \equiv \sqrt{\frac 2{N+1}}\sum_{l=1}^N \hat x_l\sin\left(\frac{\pi jl}{N+1} \right) , \hspace{10mm} j=1\cdots N
\end{equation}
The frequency of the $j$-th mode is
\begin{equation}
 \omega_j = 2\sin\left(\frac{\pi j}{2(N+1)}\right)
\end{equation}
The correspondent of \cref{def_multiv_Q0} in the basis of the normal modes becomes
\begin{align}
Q_{\mathcal J}^0(\eta_{j_1},\cdots, \eta_{j_n}) &\equiv \frac 1 {Z_{0}}\tr\left\{ \ket {0} \bra{0} \prod_{m=1}^n\hat\delta(\hat \eta_{j_m}-\hat I \eta_{j_m}) \right\} =  \frac 1 {Z_{0}} \int_{\mathbb R^{N}}\prod_{i=1}^N\mathrm d \eta'_i \; \braket{\boldsymbol \eta'| 0} \braket{ 0| \boldsymbol \eta'} \prod_{m=1}^n\delta( \eta'_{j_m}- \eta_{j_m}) = \label{Qpsi1} \\
&=\frac 1 {Z_{0}} \int_{\mathbb R^{N}}\prod_{i=1}^N\mathrm d \eta'_i \;\left\lvert {\psi_0}(\boldsymbol \eta') \right\rvert^2\prod_{m=1}^n\hat\delta( \eta'_{j_m}- \eta_{j_m})= \frac 1 {Z_0} \int_{\mathbb R^{N-n}}\prod_{\substack{i=1\\i\notin\mathcal J}}^N\left.\mathrm d \eta'_i  \;\left\lvert \psi_0(\eta'_1\cdots \eta'_n) \right\rvert^2\right|_{\substack{\eta'_{j_m}=\eta_{{j_m}}\\m=1\cdots n}} \label{Qpsi2}
\end{align}
The Schr\"odinger equation for the groundstate of the harmonic system is solved by uncoupled phononic constributions
\begin{align}
\psi_0(\boldsymbol\eta) &=\prod_{j=1}^N\left(\frac{\omega_j}{\pi}\right)^{1/4}e^{-\frac{\omega_j}{2}\eta_j^2}\label{psi_0_sol} \\
Z_0&=\int_{\mathbb R^N}\mathrm d\boldsymbol{\eta}\;\llv \psi_0(\boldsymbol{\eta})\rrv^2 = \prod_{j=1}^N\left(\frac{\pi}{\omega_j}\right)^{1/2} \label{Z_0_sol} 
\end{align}
Inserting \cref{psi_0_sol} and \cref{Z_0_sol} in \cref{Qpsi2} yields 
\begin{equation*}
Q_{\mathcal J}^0(\eta_{j_1},\cdots, \eta_{j_n}) =\prod_{{j_m}\in\mathcal J} \left(\frac{\omega_{j_m}}{\pi}\right)^{1/2} e^{-\omega_{j_m}\eta_{j_m}^2}
\end{equation*}
\subsection{Harmonic potential}\label{app:harm_pot}
\begin{figure}[H]
		\centering
		\includegraphics[scale=0.55]{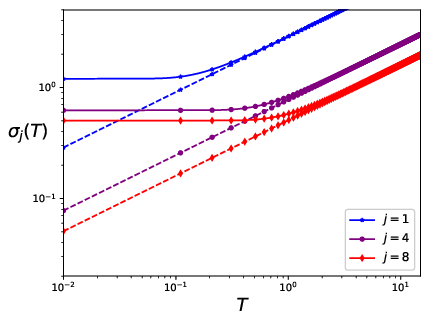}
		\caption{Variance of the distributions of the $j$-th normal mode in the classical (dotted line) and quantum (continuous line) statistics, at different temperatures}\label{fig:sigma_mod}
\end{figure}

In the limit $\alpha=\beta=0$ the FPU model is reduced to a quantum harmonic chain. The two-body potential can now be exactly diagonalized via the discrete Fourier transform of coordinates
\begin{equation}\label{def_eta_q}
\hat \eta_j \equiv \sqrt{\frac 2{N+1}}\sum_{l=1}^N \hat q_l\sin\left(\frac{\pi jl}{N+1} \right) , \hspace{10mm} j=1\cdots N
\end{equation}
This limit system happens to be still challenging from an analytical perspective \cite{lievens2008}. In the present appendix we derive a general expression for the distribution of a subset of normal modes of a Canonical mixture, and we use the result for the calculation of the average energy per normal mode. \\ 
The harmonic Hamiltonian is rewritten in terms of creation and annihilation operators as
\begin{equation}\label{def_H_harm}
\hat H = \sum_{j=1}^N \omega_j\left(\hat a^+_j\hat a_j+\frac 12 \right)\equiv\sum_{j=1}^N \omega_j\hat a^+_j\hat a_j+\mathcal E_0
\end{equation} 
where $\mathcal E_0$ denotes the zero-point energy. The contribution of the zero-point energy vanishes identically in the thermal traces due to the normalization by the partition function of the density matrix. This can be expanded in the basis of the Fock states as
\begin{align*}
\hat \rho_T &= \frac 1 {Z_T} \sum_{\mathbf l\in\mathbb N_0^N}e^{-\frac 1 T \sum_{j=1}^N\omega_j\hat a^+_j\hat a_j}\ket{\mathbf l}\bra{\mathbf l}=\frac 1{Z_T} \sum_{\mathbf l\in\mathbb N_0^N}\prod_{j=1}^N\sum_{s=0}^{+\infty}\left(-\frac {\omega_j}T\right)^s\frac 1 {s!}(\hat a^+_j\hat a_j)^s\ket{\mathbf l}\bra{\mathbf l} = \frac 1 {Z_T} \sum_{\mathbf l\in\mathbb N_0^N}e^{-\frac 1 T \sum_{j=1}^N\omega_jl_j}\ket{\mathbf l}\bra{\mathbf l}
\end{align*}
where $\mathbb N_0=\mathbb N\cup \{0\}$. We can then construct the analogue of the multivariate distribution in \cref{def_multiv_Q0} for the configurations of the normal modes, by defining:
\begin{align}\label{QJ_eta0}
Q_{\mathcal J}(\eta_{j_1},\cdots, \eta_{j_n}) &\equiv \tr\left\{ \hat \rho_T \prod_{m=1}^n\hat\delta(\hat \eta_{j_m}-\hat I \eta_{j_m}) \right\} =  \frac 1 {Z_{T}}\sum_{\mathbf l \in \mathbb N_0^N} \int_{\mathbb R^{N}}\mathrm d \boldsymbol\eta' \;e^{-\frac 1 T \sum_{j=1}^N\omega_jl_j}\braket{\boldsymbol \eta'| \mathbf l} \braket{\mathbf  l| \boldsymbol \eta'} \prod_{m=1}^n\hat\delta( \eta'_{j_m}- \eta_{j_m})
\end{align}
Given that the Hamiltonian in \cref{def_H_harm} is simply additive, the total wavefunction $\psi_{\mathbf l}({\boldsymbol \eta'})=\braket{\boldsymbol \eta | \mathbf l}$ is separable into single modes' solutions. It then follows \cite{picasso2016}: 
\begin{align}\label{psi_l2}
\braket{\boldsymbol \eta'| \mathbf l} \braket{\mathbf  l| \boldsymbol \eta'} = \llv \psi_{\mathbf l}(\boldsymbol \eta')\rrv^2 =  \prod_{j=1}^N \llv \psi_{l_j}({\eta'}_j)\rrv^2= \prod_{j=1}^N \llv\frac{{\omega_j}^{1/4}} {(2^{l_j}l_j!\sqrt{\pi})^{1/2}} H_{l_j}(\sqrt{\omega_j}{\eta'}_j)e^{-\omega_j{\eta'}_j^2/2}\rrv^2
\end{align}
which, inserted in \cref{QJ_eta0}, yields
\begin{align}
Q_{\mathcal J}(\eta_{j_1},\cdots, \eta_{j_n}) &= \frac 1 {Z_{T}}\sum_{\mathbf l \in \mathbb N_0^N}\prod_{j=1}^N \int_{\mathbb R^{N}}\mathrm d \boldsymbol\eta' \;e^{-\frac 1 T \omega_jl_j}\llv\frac {{\omega_j}^{1/4}} {(2^{l_j}l_j!\sqrt{\pi})^{1/2}} H_{l_j}(\sqrt{\omega_j}{\eta'}_j)e^{-\omega_j{\eta'}_j^2/2}\rrv^2 \prod_{m=1}^n\hat\delta( {\eta'}_{j_m}- \eta_{j_m})=\nn \\
&= \frac 1 {Z_{T}}\sum_{\mathbf l \in \mathbb N_0^N}\left[\prod_{j\in\mathcal J}\frac{e^{-\frac 1 T \omega_jl_j}\sqrt{\omega_j}}{2^{l_j}l_j!\sqrt{\pi}}H^2_{l_j}(\sqrt{\omega_j}{\eta}_j)e^{-\omega_j\eta_j^2}\right] \left[\prod_{\substack{m=1\\m\notin\mathcal J}}^N\frac{e^{-\frac 1 T \omega_ml_m}\sqrt{\omega_m}}{2^{l_m}l_m!\sqrt{\pi}}\int_{\mathbb R^{N-n}}\mathrm d \eta'_m\;H^2_{l_m}(\sqrt{\omega_m}{\eta'}_m)e^{-\omega_m{\eta'}_m^2}\right]=\nn\\
&= \frac 1 {Z_{T}}\left[\prod_{j\in\mathcal J}\sum_{l_j=0}^{+\infty}\frac{e^{-\frac 1 T \omega_jl_j}\sqrt{\omega_j}}{2^{l_j}l_j!\sqrt{\pi}}H^2_{l_j}(\sqrt{\omega_j}{\eta}_j)e^{-\omega_j\eta_j^2}\right] \left[\prod_{\substack{m=1\\m\notin\mathcal J}}^N\sum_{l_m=0}^{+\infty}e^{-\frac 1 T \omega_ml_m}\right]\label{QJ_eta1}
\end{align}
Each of the sums over the index $l_j$ in \cref{QJ_eta1} can be computed analytically, as
\begin{equation}\label{sum_H2}
\frac 1 {\sqrt{1-t^2}}\exp\left(\frac{2x^2 t}{1+t}\right)=\sum_{n=0}^{+\infty}\frac{H_n^2(x)}{2^n}\frac {t^n}{n!}, \hspace{10mm} \llv t \rrv < 1
\end{equation}
with
\begin{equation*}
t= e^{-\omega_j/T}\hspace{10mm} x = \sqrt{\omega_j}\eta_j
\end{equation*}
\cref{sum_H2} stems as a direct consequence of the definition of the Poisson kernel for Hermite polynomials \cite{he2018}. We can then simplify \cref{QJ_eta1} to
\begin{align}
Q_{\mathcal J}(\eta_{j_1},\cdots, \eta_{j_n}) &= \frac 1 {Z_{T}\pi^{n/2}}\left[\prod_{j\in\mathcal J}\frac{\sqrt{\omega_j}}{\sqrt{1-e^{-2\omega_j/T}}}\exp\left(-\omega_j\eta_j^2+\frac{2\omega_j\eta_j^2e^{-\omega_j/T}}{1+e^{-\omega_j/T}}\right)\right] \left[\prod_{\substack{m=1\\m\notin\mathcal J}}^N\frac 1 {1-e^{-\omega_m/ T}}\right]=\nn\\
&= \frac 1 {\pi^{n/2}}\prod_{j\in\mathcal J}\sqrt{\omega_j}\left(\frac{1-e^{-\omega_j/T}}{1+e^{-\omega_j/T}}\right)^{1/2}\exp\left[-\omega_j\left(1-\frac{2e^{-\omega_j/T}}{1+e^{-\omega_j/T}}\right)\eta_j^2\right]=\label{insert_ZT}\\
&= \frac 1 {\pi^{n/2}}\prod_{j\in\mathcal J}\sqrt{\omega_j}\left(\frac{1-e^{-\omega_j/T}}{1+e^{-\omega_j/T}}\right)^{1/2}\exp\left[-\omega_j\left(1-\frac{2e^{-\omega_j/T}}{1+e^{-\omega_j/T}}\right)\eta_j^2\right]=\nn \\
&= \prod_{j\in\mathcal J} \sqrt{\frac{\omega_j}{\pi}}\tanh^{1/2}\left(\frac{\omega_j}{2T}\right)\exp\left[-\omega_j\tanh\left(\frac{\omega_j}{2T}\right)\eta_j^2\right] \label{sol_Qeta}
\end{align}
In \cref{insert_ZT} we inserted the partition function 
\begin{equation*}
Z_T = \tr\left\{\hat \rho_T\right\} = \mathcal Q_{\emptyset} \numeq{\ref{QJ_eta1}} \prod_{m=1}^N\frac 1 {1-e^{-\omega_m/T}}
\end{equation*}
In the $T\gg1$ limit \cref{sol_Qeta} converges to the classical correspondent:
\begin{align*}
Q_{\mathcal J}(\eta_{j_1},\cdots, \eta_{j_n}) &\simeq_{T\gg 1} \prod_{j\in\mathcal J} \frac{\omega_j}{\sqrt{2\pi T}}e^{-\frac{\omega_j^2}{2T}\eta_j^2}\equiv \frac 1 {Z_{T}^{\eta, \mathrm {cl}}} \int_{\mathbb R^N}\mathrm d \boldsymbol \eta' \; \exp\left(-\frac 1 T \sum_{m=1}^N\frac{\omega_m^2}{2}{\eta'_m}^2\right)\prod_{j\in\mathcal J}\delta(\eta_j'-\eta_j)\equiv Q^{\mathrm{cl}}_{\mathcal J}(\eta_{j_1},\cdots, \eta_{j_n})
\end{align*}
where 
\begin{equation*}
Z_{T}^{\eta, \mathrm {cl}} = \prod_{j=1}^N\frac{\sqrt{2\pi T}}{\omega_j} 
\end{equation*}
We can identify the classical and quantum variances: 
\begin{align}
\sigma^c_j(T)& = \sqrt{T}/\omega_j \label{def_sigma_c}\\
\sigma^q_j(T)& = \left(2\omega_j\tanh\left(\frac{\omega_j}{2T}\right)\right)^{-1/2} \label{def_sigma_q}
\end{align}
The temperature dependence of the distribution for the modes $j=1,4,8$ is shown in \cref{fig:sigma_mod}.
While classically the variance vanishes at $T\to 0$, in the quantum regime it remains finite and is equal to $(2\omega_j)^{-1/2}$. This implies that the spatial part of the Hamiltonian shares half of the zero-point energy $\mathcal E_0^j=\omega_j/2$ related to the $j$-th normal mode:
\begin{align*}
\lim_{T\to 0^+} \frac{\omega_j^2}{2}\left\langle(\hat \eta_j)^2 \right\rangle_T = \frac{\omega_j}{4}=\frac{\mathcal E_0^j}2
\end{align*}
From the distribution of the normal modes in the harmonic limit \cref{sol_Qeta} we can determine the analogous distributions for the positions $\hat q_j$:
\begin{align*}
&\mathcal Q_{\mathcal J}^0(q_{j_1}, \cdots q_{j_n}) =\int_{\mathbb R^{N}}\prod_{l=1}^N\mathrm d\eta_{l}\;\sqrt{\frac{\omega_l}{\pi}}\tanh^{1/2}\left(\frac{\omega_l}{2T}\right) \;\exp\left[-\omega_l\tanh\left(\frac{\omega_l}{2T}\right)\eta_l^2\right]\prod_{j\in\mathcal J}\delta\left(q_j-\sqrt{\frac{2}{N+1}}\sum_{m=1}^N\eta_m\sin\left(\frac{\pi jm}{N+1}\right)\right)=\\
&=\frac 1{(2\pi)^n}\int_{\mathbb R^{N+n}}\prod_{j\in \mathcal J}\mathrm d k_j\; e^{ik_jq_j}\;\prod_{l=1}^N\mathrm d\eta_{l}\;\sqrt{\frac{\omega_l}{\pi}}\tanh^{1/2}\left(\frac{\omega_l}{2T}\right) \;\exp\left[-\omega_l\tanh\left(\frac{\omega_l}{2T}\right)\eta_l^2-ik_j\sqrt{\frac{2}{N+1}}\eta_l\sin\left(\frac{\pi jl}{N+1}\right)\right]
\end{align*}
The quadratic binomials in the exponential of the expression above, can be closed to 
\begin{align*}
&-\omega_l\tanh\left(\frac{\omega_l}{2T}\right)\eta_l^2-ik_j\sqrt{\frac{2}{N+1}}\eta_l\sin\left(\frac{\pi l}{N+1}\right)= \\
=&-\omega_l\tanh\left(\frac{\omega_l}{2T}\right) \left[\left(\eta_l+\frac {ik_j}2\sqrt{\frac{2}{N+1}}\frac{\sin\left(\frac{\pi l}{N+1}\right)}{\omega_l\tanh(\frac{\omega_l}{2T})}\right)^2 +\frac {k_j^2}{2(N+1)}\frac{\sin^2\left(\frac{\pi jl}{N+1}\right)}{\omega_l^2\tanh^2(\frac{\omega_l}{2T})}\right]
\end{align*}
yielding
\begin{align*}
&\mathcal Q_{\mathcal J}^0(q_{j_1}, \cdots q_{j_n})  =\frac 1{(2\pi)^n}\int_{\mathbb R^n}\prod_{j\in \mathcal J}\mathrm d k_j\; \exp\left(-\sum_{l=1}^N\frac{k_j^2\sin^2\left(\frac{\pi jl}{N+1}\right)}{2(N+1)\omega_l\tanh\left(\frac{\omega_l}{2T}\right)}+ik_j q\right)
\end{align*}
By defining 
\begin{equation}\label{sigma_q}
\sigma_{q_j}^{\text{qu}}(T)\equiv \left(\sum_{l=1}^N \frac{\sin^2\left(\frac{\pi jl}{N+1}\right)}{(N+1)\omega_l\tanh\left(\frac{\omega_l}{2T}\right)}\right)^{1/2}
\end{equation}
we can determine:
\begin{align}
&\mathcal Q_{\mathcal J}^0(q_{j_1}, \cdots q_{j_n}) = \frac 1 {(2\pi)^n}\int_{\mathbb R^n}\prod_{j\in \mathcal J}\mathrm d k_j\;\exp\left\{-\frac {\sigma_{q_j}^2(T)}2 k_j^2+ik_jq_j\right\}= \nn \\
&=\frac 1 {(2\pi)^n}\int_{\mathbb R^n}\prod_{j\in \mathcal J} \mathrm d k_j\;\exp\left\{-\frac {\sigma_{q_j}^2(T)}2\left[\left(k_j-\frac{iq_j}{\sigma_{q_j}^2(T)}\right)^2+\frac{q_j^2}{\sigma_{q_j}^4(T)}\right]\right\}=\prod_{j\in \mathcal J}\sqrt{\frac {1}{2\pi\sigma_{q_j}^2(T)}}\exp\left(-\frac {q_j^2}{2\sigma_{q_j}^2(T)}\right)\label{final_Qj}
\end{align}
\section{The force field}\label{app:force_field}
\begin{figure}[H]
	\centering
	\begin{minipage}{.45\textwidth}
		\centering
		\includegraphics[scale=0.55]{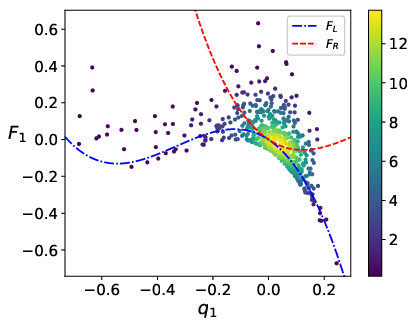}
		\caption{$P=1$, $T=0.01$, $\alpha=5$, $j=1$}\label{fig:F_j_1}
	\end{minipage}
	\hfill
	\begin{minipage}{.45\textwidth}
		\centering
		\includegraphics[scale=0.55]{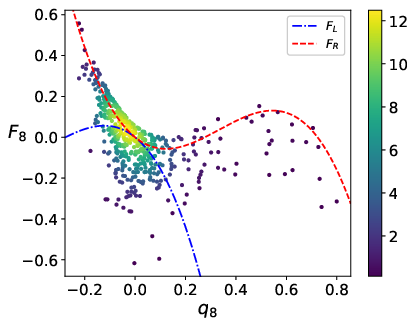}
		\caption{$P=1$, $T=0.01$, $\alpha=5$, $j=8$}\label{fig:F_j_2}
	\end{minipage}
\end{figure}

\begin{figure}[H]
	\centering
	\begin{minipage}{.45\textwidth}
		\centering
		\includegraphics[scale=0.55]{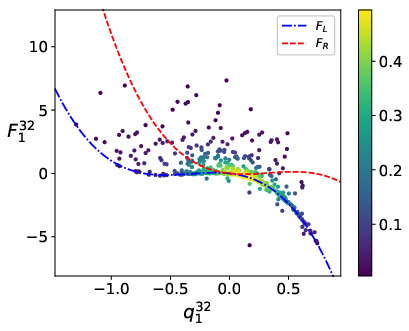}
		\caption{$P=64$, $T=0.01$, $\alpha=5$, $j=1$, $k=32$}\label{fig:F_j_3}
	\end{minipage}
	\hfill
	\begin{minipage}{.45\textwidth}
		\centering
		\includegraphics[scale=0.55]{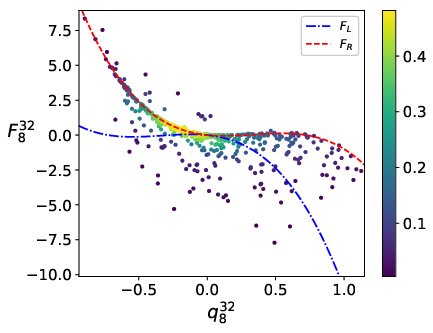}
		\caption{$P=64$, $T=0.01$, $\alpha=5$, $j=8$, $k=32$}\label{fig:F_j_4}
	\end{minipage}
\end{figure}

\begin{figure}[H]
	\centering
	\begin{minipage}{.45\textwidth}
		\centering
		\includegraphics[scale=0.55]{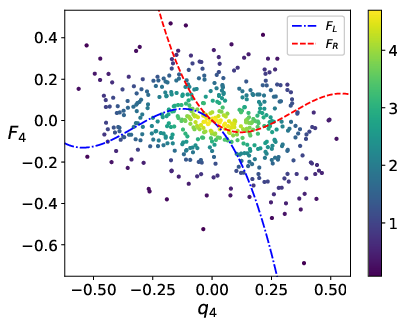}
		\caption{$P=1$, $T=0.01$, $\alpha=5$, $j=4$}\label{fig:F_k_id1}
	\end{minipage}
	\hfill
	\begin{minipage}{.45\textwidth}
		\centering
		\includegraphics[scale=0.55]{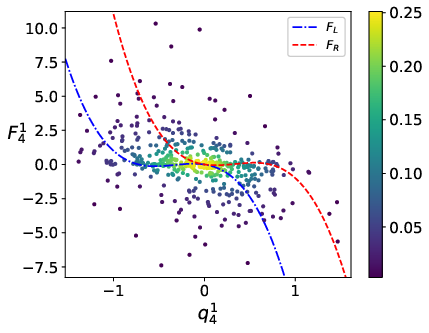}
		\caption{$P=64$, $T=0.01$, $\alpha=5$, $j=4$, $k=1$}\label{fig:F_k_id2}
	\end{minipage}
\end{figure}

\begin{figure}[H]
	\centering
	\begin{minipage}{.45\textwidth}
		\centering
		\includegraphics[scale=0.55]{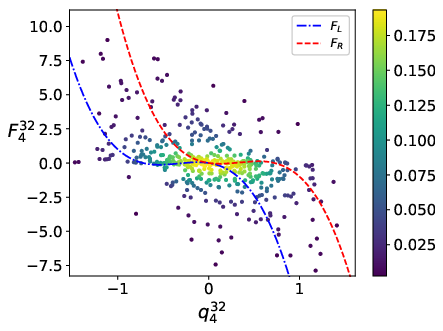}
		\caption{$P=64$, $T=0.01$, $\alpha=5$, $j=4$, $k=32$}\label{fig:F_k_id3}
	\end{minipage}
	\hfill
	\centering
	\begin{minipage}{.45\textwidth}
		\centering
		\includegraphics[scale=0.55]{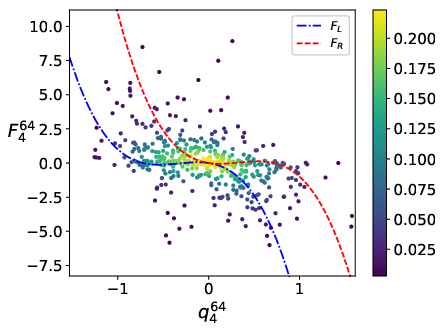}
		\caption{$P=64$, $T=0.01$, $\alpha=5$, $j=4$, $k=64$}\label{fig:F_k_id4}
	\end{minipage}
\end{figure}

In this appendix we present a complementary study on one-dimensional sections of the force field from the numerical samplings computed in the work. A comparison between classical and quantum results allows us to identify an additional criterion for to highlight quantum dispersion.  
We note that the expectation value of the force is identically zero, both in the classical and in the quantum regime:
\begin{align}
&\left\langle - \frac{\partial \hat V( \mathbf {\hat q})}{\partial \hat q_j} \right\rangle_T = \nn \\
&= - \lim_{P\to+\infty} \frac 1 {Z_{T,P}}\int_{\mathbb R^{2NP}} \mathrm d \mathbf q^1\cdots \mathrm  d \mathbf q^P \mathrm d \mathbf p^1\cdots \mathrm d \mathbf p^P \; \left.\exp\Le\{-\frac 1 T\mathcal H_{\mathrm{cl}}(\mathbf q^1\cdots \mathbf q^P, \mathbf p^1\cdots \mathbf p^P)\Ri\}
\frac 1 P \sum_{k=1}^P \frac{\partial V( \mathbf { q}^k)}{\partial  q_j^k}\right|_{\substack{\mathbf q^{P+1}= \mathbf q^{1}\\ q_0^l=q_{N+1}^ l=0,\;\;l=1\cdots P}}= \nn\\
&\equiv\lim_{P\to+\infty} \frac T {Z_{T,P}}\int_{\mathbb R^{2NP}}
\mathrm d q_1\cdots \mathrm  d q_{NP} \mathrm d p_1\cdots \mathrm d p_{NP} \;  \sum_{k=1}^P \frac{\partial}{\partial  q_{Nk+j}} \left.\exp\Le\{-\frac 1 T\mathcal H_{\mathrm{cl}}( q_1\cdots q_{NP}, p_1\cdots  p_{NP})\Ri\}
\right|_{\substack{ q_m^{P+1}= q_m^{1},\;\;m=1\cdots N\\ q_0^l=q_{N+1}^ l=0,\;\;l=1\cdots P}} = \label{null_F_rp}\\
&= \lim_{P\to+\infty} \frac T {Z_{T,P}} \sum_{k=1}^P \int_{\mathrm R^{2NP-1}}\left[\prod_{\substack{i=1\\i\neq Nk+j}}^{NP}\mathrm d q_i\right]\left[\prod_{l=1}^{NP}\mathrm d p_l\right]\left.\left[\exp\Le\{-\frac 1 T\mathcal H^{\mathrm{cl}}( q_1\cdots q_{NP}, p_1\cdots  p_{NP})\Ri\}\right]_{q_{Nk+j}={-\infty}}^{q_{Nk+j}={+\infty}}
\right|_{\substack{ q_m^{P+1}= q_m^{1},\;\;m=1\cdots N\\ q_0^l=q_{N+1}^ l=0,\;\;l=1\cdots P}}=0\label{short_bc}
\end{align}
The identity in \cref{null_F_rp} follows as the linear force from the polymer ring is identically zero:
\begin{align*}
&\left.\sum_{k,l=1}^P \frac{\partial}{\partial  q_j^k}(q_j^{l+1}-q_j^l)^2\right|_{q_j^{P+1}=q_j^{1}} = \left. 2\sum_{k,l=1}^P  (\delta_{l+1,k}-\delta_{l,k}) (q_j^{l+1}-q_j^l)\right|_{q_j^{P+1}=q_j^{1}} = \\ 
&\left.2\sum_{k=1}^P\left[ \sum_{l=1}^P\delta_{l+1,k}(q_j^{l+1}-q_j^l) -\sum_{m=2}^{P+1}\delta_{m-1,k}(q_j^{m}-q_j^{m-1}) \right] \right|_{q_j^{P+1}=q_j^{1}} = 0 
\end{align*} 
By imposing $P\equiv 1$ in the identities in \cref{short_bc}, we can see that the result is analogously satisfied for the classical limit of the system. In \cref{null_F_rp} we conveniently introduced a \textit{row-major} ordering for the Cartesian components of the configurations:
\begin{align}\label{row_maj}
\left\{ q_j^k \right\}_{\substack{j=1\cdots N \\k=1\cdots P}}\equiv \left\{q_{N(k-1)+j\equiv i}\right\}_{i=1\cdots NP}
\end{align}
\cref{short_bc} vanishes as the distribution of a globally confining potential is null at the boundaries of the phase space. The sections of the force field are sampled by collecting from a PIMD run uncorrelated values of the estimator 
\begin{equation*}
F_j(\mathbf q) =-\frac 1 P\sum_{k=1}^P \frac{\partial V( \mathbf{ q}^k)}{\partial q_j^k}
\end{equation*}
and plotting them versus the configuration $q_j$ corresponding to that phase space point. The result for the first and last moving particles, are shown in  \crefrange{fig:F_j_1}{fig:F_j_4}. We compare the classical and quantum regime by fixing the number of replicas to $P=1$ and $P=64$. In order reduce the density of points on the figures, only a specific value of $k\in \{1,\cdots,P\}$ has been plotted for $P=64$. We expect this choice not to imply any bias in the statistics: as mentioned in the discussion after \cref{def_QJk}, the thermal expectations of identical observables at different instances are equivalent. A numerical confirmation of such symmetry is given in the later figures \crefrange{fig:F_k_id1}{fig:F_k_id3}. 
The advantage of studying the statistics of the two extremes of the chain allows us to get analytical estimates for the boundaries of the classical force field. In particular, the forces exerted by the left (L)/right (R) walls on the first/last moving particles are respectively
\begin{align}
F_L(q_1^k) &= -\frac{\partial}{\partial q_1^k} V(q_1^k-q_0^k) = -V'(q_1^k)= -q_1^k-\alpha(q_1^k)^2-\alpha(q_1^k)^3 \label{def_F_L}\\
F_R(q_{N}^k) &= -\frac{\partial}{\partial q_N^k}V(q_{N+1}^k-q_N^k) = V'(-q_N^k)=- q_N^k +\alpha(q_N^k)^2-\alpha(q_N^k)^3 \hspace{10mm} \forall \;\;k= 1\cdots P \label{def_F_R}
\end{align}
We can notice that $F_L$ in \cref{fig:F_j_1} and \cref{fig:F_j_3} follows the lower boundaries of the force field. The negative shift of the numerical samples w.r.t. this limit case is ascribed to a collective behavior of all the other particles of the chain, which reduces the push exerted by the left soft wall. The present observation is coherent to the shift in the configurational distribution in the main text, in the discussion about \cref{fig:Q_j_1} and \cref{fig:Q_j_2}. An analogue and opposite argument applies to the right boundary in \cref{fig:F_j_2} and \cref{fig:F_j_4}. The action of the remaining $N-1$ d.o.f. on the force at the extremes persists in the quantum phase space. The main difference w.r.t. the classical curves is the fact that a higher ratio of points is sampled below the limiting force in \cref{fig:F_j_3}, and viceversa in \cref{fig:F_j_3}. This signals again the higher freedom exhibited by the quantum statistics. Finally, in \crefrange{fig:F_k_id1}{fig:F_k_id4} are shown the samples of the force of the central particle $j=4$, where the boundary effects are minimized. The same distributions obtained by collecting different values of $k=1,P/2$ and $P$ is shown for $P=64$; this is done in order to test numerically the assumption of equivalence of the replicas' statistics. The anisotropy of the samplings at the center of the lattice is smaller w.r.t. the boundaries, due to the screening of the boundary effects exerted by the other d.o.f..
\section{Harmonic limit of the positional auto-correlation function}\label{app:harmonic_lim_RP}
\subsection{Calculation from the RPMD time correlation}
In this appendix we derive the exact harmonic limit of the positional auto-correlation function, obtained from the propagation of the ring-polymer equations of motions \cref{RPMD1} and \cref{RPMD2}. We show that in this case the ring polymer is able to reproduce the exact result (compare \cref{K0_q} and \cref{K0_eta}). Hamilton's equations for the ring polymer are rewritten for the modes as
\begin{align}
\sqrt{\frac 2{N+1}} \sum_{j'=1}^N \sin\left(\frac{\pi j j'}{N+1}\right)\ddot \eta^k_{j'} &= - \sum_{j'=1}^N \left\{P^2T^2 \sqrt{\frac 2{N+1}}\sin\left(\frac{\pi j j'}{N+1}\right)\left[\left(2+\frac {\omega_{j'}^2}{P^2T^2}\right)\eta^k_{j'}-\eta^{k-1}_{j'}-\eta^{k+1}_{j'}\right]-\frac{\partial V^{\text{an}}(\boldsymbol \eta)}{\partial \eta_{j'}^k}\frac{\partial \eta_{j'}^k}{\partial q_{j}^k}\right\} = \\
&=  -\sqrt{\frac 2{N+1}} \sum_{j'=1}^N \sin\left(\frac{\pi j j'}{N+1}\right)\left\{P^2T^2 \left[\left(2+\frac {\omega_{j'}^2}{P^2T^2}\right)\eta^k_{j'}-\eta^{k-1}_{j'}-\eta^{k+1}_{j'}\right]-\frac{\partial V^{\text{an}}(\boldsymbol \eta)}{\partial \eta_{j'}^k}\right\} \label{eom_eta2} 
\end{align} 
where $V^{\text{an}}(\boldsymbol \eta)$ denotes the anharmonic part of the potential. We can now equate term by term the first and third identities, as the transform to the modes is invertible, hence its kernel is the null function on the discrete support $\{1,\cdots, N\}$. 
This yields
\begin{equation}\label{eom_eta3}
\ddot \eta^k_{j} = -P^2T^2 \left[\left(2+\frac {\omega_j^2}{P^2T^2}\right)\eta^k_{j}-\eta^{k-1}_{j}-\eta^{k+1}_{j}\right]-\frac{\partial V^{\text{an}}(\boldsymbol \eta)}{\partial \eta_{j}^k}
\end{equation}
Given that \cref{eom_eta3} preserves a Hamiltonian structure, we have shown that the normal modes of the physical coordinates identify a canonical transformation in the classical sense, even in the extended phase space of the classical isomorphism. Additionally, the periodic boundary conditions of the ring polymer are preserved by the change of coordinates:
\begin{align}
\eta_j^{P+1}= \sqrt{\frac 2{N+1}} \sum_{j'=1}^N\sin \left(\frac{\pi jj'}{N+1}\right) q_{j'}^{P+1}=\eta_{j}^1 \;\;\;\forall\; \;k=1\cdots N
\end{align}
In the harmonic limit, where $V^{\text{an}} \equiv 0$, the system in \cref{eom_eta3} can be rewritten in vectorial form as
\begin{align}\label{solve_lin}
\boldsymbol {\ddot \eta} \equiv- \mathbf K\boldsymbol \eta
\end{align}
where the components of $\mathbf K\in M_{NP}(\mathbb R)$ are defined by
\begin{align}\label{K_iip}
(\mathbf K)_{ii'} = P^2 T^2 
\left[\left(2+\frac {\omega_j^2}{P^2T^2}\right)\delta_{ii'}-\delta_{i,(i'-N)\%NP}-\delta_{i,(i'+N)\%NP}\right]
\end{align}
Via the matrix form of \cref{K_iip}
\begin{equation}\label{def_KC}
\mathbf K = P^2T^2\begin{pmatrix}
\mathbf C_0 & \mathbf C_1 & \cdots & \cdots & \mathbf C_{P-1} \\ 
\mathbf C_{P-1} & \mathbf C_0 & \mathbf C_1 & \cdots & \mathbf C_{P-2} \\ 
\vdots & \ddots & \mathbf C_0 & \ddots & \vdots \\ 
\vdots & \ddots & \ddots & \ddots & \vdots & \\ 
\mathbf C_1  & \cdots & \cdots & \mathbf C_{P-1} & \mathbf C_0 \\ 
\end{pmatrix}=\mathbf K^T
\end{equation}
we can identify its circulant structure in $M_P(M_N(\mathbb R))$, where 
\begin{equation}\label{vals_Ci}
\mathbf C_i =
\begin{cases}
\mathbf A, & i=0 \\
-\mathbf 1_N, & i=1, P-1 \\
\mathbf 0, & \text{elsewhere}
\end{cases}\in M_{N}(\mathbb R)
\end{equation}
$\mathbf 1_N$ denotes the identity matrix and 
\begin{align*}
\mathbf A_{jj'}= \left(2+\frac {\omega_j^2}{P^2T^2}\right)\delta_{jj'}
\end{align*}
It is possible to recast the search of a solution of the system of the $NP$ linear equations in \cref{solve_lin} in terms of the eigenvalue problem 
\begin{equation}\label{eigenprobl}
\mathbf K \cdot \mathbf x^{(k)} = \boldsymbol\lambda_k \cdot \mathbf x^{(k)} \hspace{10mm}k=0\cdots P-1 \hspace{10mm}\boldsymbol\lambda_k \in M_{N}(\mathbb C) 
\end{equation}
Let us assume that we can determine a unitary matrix $\mathbf Q\in M_{NP}(\mathbb C)$, whose columns are the normalized eigenvectors $\mathbf x^{(k)}$. $\mathbf Q$ diagonalizes by construction the matrix $\mathbf K$:
\begin{align}
(\mathbf Q^\dagger \mathbf K \mathbf Q)_{kk'}&= \sum_{l,l'=0}^{P-1}\left(\mathbf x_k^{(l)}\right)^*\mathbf K_{ll'}\mathbf x_{l'}^{(k')}
= \sum_{l,l'=0}^{P-1}\left(\mathbf x_k^{(l)}\right)^*\left(\mathbf K\cdot\mathbf x^{(k')}\right)_l
\numeq{\ref{eigenprobl}} \sum_{l=0}^{P-1}\left(\mathbf x_k^{(l)}\right)^*\left(\boldsymbol \lambda_{k'} \mathbf x^{(k')}\right)_l= \nn \\
&= \boldsymbol \lambda_{k'}\sum_{l=0}^{P-1}\left(\mathbf x_{k}^{(l)}\right)^* \mathbf x_l^{(k')} \numeq{\ref{orthog_xk}} \boldsymbol\lambda_{k'}\delta_{kk'} \equiv \boldsymbol \Lambda_{kk'}  \label{QKQ}
\end{align} 
where $\boldsymbol\Lambda\in M_{NP}(\mathbb C)$ is diagonal. By defining
\begin{equation}\label{def_alpha}
\boldsymbol \alpha \equiv \mathbf Q^\dagger\boldsymbol {\ddot \eta}
\end{equation}
we can rewrite \cref{solve_lin} as
\begin{align}\label{diag_eom}
\boldsymbol {\ddot\alpha} =- \left(\mathbf Q^\dagger\mathbf K \mathbf Q \right)\mathbf Q^\dagger\boldsymbol \eta = -\boldsymbol \Lambda \boldsymbol \alpha
\end{align}
Given that $\boldsymbol \Lambda$ is diagonal, the dynamical equations of the $NP$ components of the coordinates $\boldsymbol \alpha$ are now uncoupled, hence they can be solved independently. We can check by direct substitution that the components of the eigenvectors $\mathbb x^{(k)}$ are
\begin{equation}\label{def_xk}
\mathbf x^{(k)}_{k'} \equiv \mathbf 1_N\frac{\Omega_P^{kk'}}{\sqrt P} =\mathbf x^{(k')}_{k}\in   M_{N}(\mathbb C), \hspace{10mm} k,k'=0\cdots P-1
\end{equation}
where 
\begin{equation} \label{def_OmegaP}
\Omega_P \equiv \exp(2\pi i/P)
\end{equation}
Indeed, we have:
\begin{align}
(\mathbf K \cdot \mathbf x^{(k)})_{k'} 
&= \frac {P^2T^2}{\sqrt P}\sum_{l=0}^{P-1} \mathbf C_{(l-k')\%P} \Omega_P^{kl} =\frac {P^2T^2}{\sqrt P}\Omega_P^{kk'} \sum_{l=0}^{P-1} \mathbf C_{(l-k')\%P}\Omega_P^{k(l-k')} = \label{C_jl_neg} \\
&=\frac {P^2T^2}{\sqrt P} \Omega_P^{kk'} \sum_{l=0}^{P-1}\mathbf C_l\Omega_P^{kl} =   \left( P^2T^2\sum_{l=0}^{P-1}\mathbf C_l\Omega_P^{kl}\right)\mathbf 1_N\frac {\Omega_P^{kk'}}{\sqrt P} \equiv  \boldsymbol \lambda_k  \mathbf x^{(k)}_{k'}, \hspace{10mm} k,k'=0\cdots P-1 \label{period_l}
\end{align}
$\mathbb Q$ denotes therefore the discrete Fourier transform (DFT) matrix. The eigenvectors in \cref{def_xk} are orthonormal w.r.t. the scalar product in $M_{P}( M_N(\mathbb C))$:
\begin{equation}\label{orthog_xk}
\left \langle\mathbf x^{(k)}, \mathbf x^{(k')} \right \rangle \equiv \sum_{l=0}^{P-1} \mathbf x_{l}^{(k)} \left(\mathbf x_l^{(k')}\right)^* = \sum_{l=0}^{P-1} \mathbf x_{l}^{(k)} \left(\mathbf x_{k'}^{(l)}\right)^*= \frac {\mathbf 1_N} P\sum_{l=0}^{P-1} e^{i\frac{2\pi}Pl(k-k')}=\mathbf 1_N\delta_{k,k'}
\end{equation}
In the last identity we used the expansion of the delta function on the Discrete Fourier Transform (DFT) kernel:
\begin{equation} \label{delta_DFT}
\sum_{l=0}^{P-1}e^{\pm i\frac{2\pi }P lm}=\sum_{l=0}^{P-1}\Omega_P^{\pm lm}=P\delta_{m,0}+(1-\delta_{m,0})\frac{1-(e^{\pm i\frac{2\pi m}P})^P}{1-e^{\pm i\frac{2\pi m}P}}=P\delta_{m,0}  \hspace{10mm} m =0,\cdots P-1 
\end{equation}
or analogously
\begin{align}\label{delta_DFT1}
&\sum_{l=1}^{P}e^{\pm i\frac{2\pi }P lm}=\sum_{l=0}^{P-1}e^{\pm i\frac{2\pi }P (l+1)m}=P\delta_{m,0}+e^{\pm i\frac{2\pi m}P}(1-\delta_{m,0})\frac{1-(e^{\pm i\frac{2\pi m}P})^P}{1-e^{\pm i\frac{2\pi m}P}}=P\delta_{m,0} \hspace{10mm} m =0,\cdots P-1 
\end{align}
Let us notice that the indices of $l-k'$ in \cref{C_jl_neg} are allowed to take negative values. This is consistent with \cref{def_KC}, provided the labels of the components are defined modulo $P$, as specified. The same periodicity is satisfied by the powers of $\Omega_P$ in \cref{period_l}. This periodicity allows us to change the summation index $l-k'\rightarrow l$ in \cref{C_jl_neg} and \cref{period_l}. We can identify from \cref{period_l} the eigenvalues
\begin{align}
\boldsymbol \lambda_k&\equiv
P^2T^2 \sum_{l=0}^{P-1} \mathbf C_l \Omega_P^{kl} 
= P^2T^2 \left(\mathbf C_0 \Omega_P^0+\mathbf C_1\Omega_P^k+\mathbf C_{P-1}\Omega_P^{k(P-1)}\right) \label{def_lambda_k_0} \\
&= P^2T^2 \left[\mathbf A - \mathbf 1_N\left(e^{i\frac{2\pi k}{P}}+e^{i\frac{2\pi k (P-1)}{P}}\right)\right] =P^2T^2 \left[ \mathbf A -2 \mathbf 1_N \cos\left(\frac{2\pi k}P\right)\right] \hspace{10mm}k=0\cdots P-1 \label{def_lambda_k}
\end{align}
The columns of the orthogonalization matrix $\mathbb Q$ are defined by the eigenvectors $\mathbb x^{(k)}$:
\begin{equation}\label{def_Q}
(\mathbf Q)_{ll'} = \frac 1{\sqrt P} \exp\left(i\frac{\pi ll'}{P}\right)
\end{equation}
We can rewrite \cref{diag_eom} components-wise as
\begin{align}
{\ddot \alpha}_{j}^k &=- \left(\boldsymbol\Lambda\boldsymbol \alpha\right)_{N(k-1)+j} = - \sum_{i=1}^{NP}\boldsymbol\Lambda_{N(k-1)+j, i} \alpha_i= - \boldsymbol\Lambda_{N(k-1)+j, N(k-1)+j} \alpha_{j}^k = - P^2T^2 \left[ 2+\frac {\omega_j^2}{ P^2T^2} -2\cos\left(\frac{2\pi (k-1)}P\right)\right]\alpha_{j}^k \nn = \\
&=  -  P^2T^2 \left[ 4\sin^2\left(\frac{\pi (k-1)}P\right)+\frac {\omega_j^2}{ P^2T^2} \right]\alpha_{j}^k = -\Omega_{j,k}^2\alpha_{j}^k \hspace{10mm}k=1\cdots P, \;\;j=1\cdots N\label{solve_harm}
\end{align}
where we defined
\begin{equation}\label{def_Omega_jk}
\Omega_{j,k} \equiv  PT \left[ 4\sin^2\left(\frac{\pi (k-1)}P\right)+\frac {\omega_j^2}{ P^2T^2} \right]^{1/2}\hspace{10mm}k=1\cdots P, \;\;j=1\cdots N
\end{equation}
This bead-dependent normal frequency collapses to the classical limit
\begin{equation}
\left.\Omega_{j,k}\right|_{P=1}=\omega_j
\end{equation}
as expected. 
\cref{solve_harm} is solved by
\begin{align}\label{solved_harm}
\alpha_{j}^k(t) = \alpha_{j}^k\cos\left(\Omega_{j,k} t\right)+ \frac{\dot \alpha_j^k}{\Omega_{j,k}}\sin\left(\Omega_{j,k} t\right)
\end{align}
The harmonic correlation function (for $\alpha=0$) is approximated in the RPMD scheme by
\begin{equation}\label{tildeK_eta_sol}
\tilde K^{0}_{T,\eta_j}(t) \equiv\frac 1{Z_{TP}}\int_{\mathbb R^{2NP}}\mathrm d \boldsymbol \eta^1\cdots  \mathrm d \boldsymbol \eta^P \mathrm d \boldsymbol {\dot \eta}^1\cdots  \mathrm d \boldsymbol {\dot \eta}^P \; e^{-\frac 1 {TP}\mathcal H^0_P(\boldsymbol { \eta}^1, \cdots , \boldsymbol { \eta}^P, \boldsymbol {\dot \eta}^1, \cdots, \boldsymbol {\dot \eta}^P)} \eta_{j,P}\; \eta_{j,P}(t)
\end{equation}
where 
\begin{align*}
&\mathcal H^0_P(\boldsymbol { \eta}^1, \cdots  \boldsymbol { \eta}^P, \boldsymbol {\dot \eta}^1, \cdots \boldsymbol {\dot \eta}^P)= \sum_{j=1}^N\sum_{k=1}^P \left[ \frac{(\dot \eta_j^k)^2}{2}+ \frac {P^2T^2}{2}(\eta_j^{k+1}-\eta_j^k)^2 +\frac {\omega_j^2}2 ( \eta_j^k)^2 \right]= \\
&=\frac 12 \boldsymbol{\dot \eta}^T\boldsymbol{\dot \eta}+
\frac 12\boldsymbol \eta^T \mathbf K \boldsymbol \eta = \frac 12 \boldsymbol{\dot \alpha}^T\boldsymbol{\dot \alpha}+
\frac 12\boldsymbol \alpha^T \boldsymbol \Lambda \boldsymbol \alpha
\end{align*}
and 
\begin{equation*}
\eta_{j,P}=\frac 1 {P}\sum_{k=1}^P \eta_{j}^k
\end{equation*}
We can return to the original set of configurations $\boldsymbol \eta$ by inverting the transformation of the normal modes of the ring polymer in \cref{diag_eom}, by using the unitarity of $\mathbf Q$:
\begin{align}
\boldsymbol \eta &= \mathbf Q \boldsymbol \alpha \label{eta_alpha1} \\
\eta_j^k &\equiv \left(\boldsymbol \eta^k\right)_j =\frac 1 {\sqrt P} \sum_{l=0}^{P-1} \left( e^{\frac {2\pi i}P(k-1)l}\boldsymbol \alpha^{l+1} \right)_j \hspace{10mm} k=1\cdots P \label{eta_alpha2}\\
\eta_{j,P} &= \frac 1{P^{3/2}}\sum_{k=1}^P\sum_{l=0}^{P-1} e^{\frac {2\pi i}P(k-1)l} \alpha_j^{l+1} \numeq{\ref{delta_DFT1}}\frac 1 {\sqrt P} \alpha_j^1\label{eta_alpha3}
\end{align}
By inserting \cref{solved_harm} and \cref{eta_alpha3} in \cref{tildeK_eta_sol}, we get:
\begin{align}
\boxed{\tilde K^0_{T,\eta_j}(t)} &=\frac 1{PZ_{TP}} \int_{\mathbb R^{2NP}}\mathrm d \boldsymbol \alpha^1\cdots  \mathrm d \boldsymbol \alpha^P \mathrm d \boldsymbol {\dot \alpha}^1\cdots  \mathrm d \boldsymbol {\dot \alpha}^P \; e^{-\frac 1 {PT}\mathcal H_P(\boldsymbol { \alpha}^1, \cdots,  \boldsymbol { \alpha}^P, \boldsymbol {\dot \alpha}^1, \cdots, \boldsymbol {\dot \alpha}^P)} (\alpha_{j}^1 )^2\; \cos\left(\Omega_{j,1} t\right) = \nn \\
&= \frac 1 {P}  \cos\left(\omega_{j} t\right) \left\langle\left(\alpha_{j}^1\right)^2\right\rangle_T= \boxed{\frac{T}{ \omega_j^2} \cos\left(\omega_{j} t\right)}\label{sol_Keta_RP}
\end{align}
where we noticed that $\lvert\det \mathbf Q\rvert=1$ and that
\begin{align}\label{alphall'}
\lla \alpha_{l}^1\; \alpha_{l'}^1\rra_T = \frac 1 {Z_T^{\boldsymbol \alpha}}\int \mathrm d \boldsymbol \alpha \; e^{-\frac 1{2PT}\boldsymbol \alpha^T\boldsymbol \Lambda \boldsymbol \alpha} \alpha_{l}^1\; \alpha_{l'}^1 =\frac{PT}{\omega_l^2}\delta_{ll'}
\end{align} 
The solution of the position-position correlation function in the ring polymer approximation is easily constructed via \cref{sol_Keta_RP}:
\begin{align}\label{sol_Kq_RP}
&\boxed{\tilde K^0_{T,q_j}(t) }= \lla q_{j,P}\; q_{j,P}(t)\rra_T = \frac 1 {P^2}\sum_{k,k'=1}^P \lla q_j^k\; q_j^{k'}(t)\rra_T=\frac 2 {(N+1)P^2}\sum_{k,k'=1}^P\sum_{l,l'=1}^N\sin\left(\frac{\pi jl}{N+1}\right)\sin\left(\frac{\pi jl'}{N+1}\right) \lla \eta_l^k\; \eta_{l'}^{k'}(t)\rra_T= \nn\\
&=\frac 2 {(N+1)}\sum_{l,l'=1}^N\sin\left(\frac{\pi jl}{N+1}\right)\sin\left(\frac{\pi jl'}{N+1}\right) \lla \eta_{l,P}\; \eta_{l', P}(t)\rra_T \numeq{\ref{eta_alpha3}} \boxed{\frac 2 {(N+1)P}\sum_{l,l'=1}^N\sin\left(\frac{\pi jl}{N+1}\right)\sin\left(\frac{\pi jl'}{N+1}\right) \tilde K^0_{T,\eta_j}(t)}
\end{align}
The benefit of the present analysis is that the harmonic limit for the correlation functions is identical in the classical and in the quantum dynamics. (This does not hold for the harmonic limit of the statistical distributions, derived in Appendix $D.3$.) It is therefore simple to compare the action of the nonlinearities in the classical and quantum regimes. \\  
\subsection{Exact result}\label{exact_corr}
In the following we derive the exact results for the correlation functions considered in the work, by directly by expanding th quantum mechanical definitions of the Kubo-transform correlations.
\begin{align}
&\boxed{K_{\eta_j,\eta_k}^0(t)}\equiv\frac 1{Z_\beta\beta} \int_0^\beta \mathrm d \lambda \; \tr \left\{e^{-(\beta-\lambda)\hat H_0} \hat \eta_j e^{(-\lambda+it) \hat H_0} \hat \eta_k e^{-i\hat H_0 t}\right\}=\frac 1{Z_\beta\beta} \int_0^\beta \mathrm d \lambda \;\sum_{\mathbf n}\bra{\mathbf n} e^{-(\beta-\lambda+it)\hat H_0} \hat \eta_j e^{(-\lambda+it)\hat H_0}  \hat \eta_k \ket{\mathbf n}= \nn\\
&=\frac{1}{2Z_\beta\beta\sqrt{\omega_j\omega_k}} \int_0^\beta \mathrm d \lambda \;e^{-(\beta-\lambda+it)\mathcal E_0}\sum_{\mathbf n}e^{-(\beta-\lambda+it)\sum_{l=1}^N\omega_l n_l}\bra{\mathbf n} ( \hat a_j^\dagger+ \hat a_j) e^{(-\lambda+it)\sum_{m=1}^N\omega_m\hat a^\dagger_m \hat a_m}  ( \hat a_k^\dagger+ \hat a_k)\ket{\mathbf n}e^{(-\lambda+it)\mathcal E_0}= \nn\\
&=\frac {e^{-\beta\mathcal E_0}}{2Z_\beta\beta\sqrt{\omega_j\omega_k}} \int_0^\beta \mathrm d \lambda \;\sum_{\mathbf n\in \mathbb N_0^N}e^{-(\beta-\lambda+it)\sum_{l=1}^N\omega_l n_l}\left(\bra{\mathbf n-\mathbf e_j}\sqrt{n_j}+ \bra{\mathbf n+\mathbf e_j}\sqrt{n_j+1}\right) e^{(-\lambda+it)\sum_{m=1}^N\omega_m \hat a^\dagger_m \hat a_m} \times\nn \\
&\times \left(\sqrt{n_k+1}\ket{\mathbf n+\mathbf e_k}+\sqrt{n_k}\ket{\mathbf n-\mathbf e_k} \right)= \nn \\
&=\frac { \delta_{jk}  e^{-\beta\mathcal E_0}}{2Z_\beta\beta\omega_j
} \int_0^\beta \mathrm d \lambda \;\sum_{\mathbf n\in \mathbb N_0^N}e^{-(\beta-\lambda+it)\sum_{l=1}^N\omega_l n_l}\left[n_j e^{(-\lambda+it)\left(\sum_{m=1}^N\omega_m n_m-\omega_k\right)}+(n_j+1)e^{(-\lambda+it)\left(\sum_{m=1}^N\omega_m n_m+\omega_k\right)}\right]= \nn\\
&=\frac {\delta_{jk} e^{-\beta\mathcal E_0}}{2Z_\beta\beta\omega_j} \int_0^\beta \mathrm d \lambda \;\sum_{\mathbf n\mathbf \in \mathbb N_0^N}e^{-\beta\sum_{l=1}^N\omega_l n_l}\left(n_j e^{(\lambda-it)\omega_j}+(n_j+1)e^{-(\lambda-it)\omega_j}\right)=\nn\\
&=\frac {\delta_{jk} e^{-\beta\mathcal E_0}}{2Z_\beta\beta\omega_j} \sum_{\mathbf n\mathbf \in \mathbb N_0^N}e^{-\beta\sum_{l=1}^N\omega_l n_l}\left(n_j e^{-it\omega_j}\frac{e^{\beta\omega_j}-1}{\omega_j}+(n_j+1)e^{it\omega_j}\frac{1-e^{-\beta\omega_j}}{\omega_j}\right)=\nn \\
&=\frac {\delta_{jk}  e^{-\beta\mathcal E_0}}{2Z_\beta\beta\omega_j^2} \sum_{\mathbf n\mathbf \in \mathbb N_0^N}e^{-\beta\sum_{l=1}^N\omega_l n_l}\left[n_j e^{-it\omega_j}\left(e^{\beta\omega_j}-1\right)+(n_j+1)e^{it\omega_j}\left(1-e^{-\beta\omega_j}\right)\right]= \nn\\
&=\frac {\delta_{jk} e^{-\beta\mathcal E_0}}{2Z_\beta\beta\omega_j^2} \left[\prod_{\substack{l=1\\l\neq j}}^N \sum_{n_l=0}^{+\infty} e^{-\beta\omega_l n _l}\right]\left[\left(\sum_{n_j=0}^{+\infty} n_j e^{-\beta \omega_j n_j}\right)e^{-it\omega_j}\left(e^{\beta \omega_j}-1\right)+\left(\sum_{n_j=0}^{+\infty} (n_j+1) e^{-\beta \omega_j n_j}\right)e^{it\omega_j}\left(1-e^{-\beta\omega_j}\right)\right]=\nn\\
&=\frac {\delta_{jk} e^{-\beta\mathcal E_0}}{2Z_\beta\beta\omega_j^2}\left[\prod_{\substack{l=1\\l\neq j}}^N\frac 1 {1-e^{-\beta \omega_l}}\right]\left[\left(\sum_{n_j=0}^{+\infty} n_j e^{-\beta \omega_j n_j}\right)e^{-it\omega_j}\left(e^{\beta \omega_j}-1\right)+\left(\sum_{n_j=0}^{+\infty} (n_j+1) e^{-\beta \omega_j n_j}\right)e^{it\omega_j}\left(1-e^{-\beta\omega_j}\right)\right]=\nn\\
&=\frac {\delta_{jk} e^{-\beta\mathcal E_0}}{2Z_\beta\beta\omega_j^2}\left[\prod_{\substack{l=1\\l\neq j}}^N\frac 1 {1-e^{-\beta \omega_l}}\right] \left[\frac{e^{-\beta\omega_j}}{(1-e^{-\beta\omega_j})^2}e^{-it\omega_j}\left(e^{\beta \omega_j}-1\right)+\frac 1{(1-e^{-\beta\omega_j})^2}e^{it\omega_j}\left(1-e^{-\beta\omega_j}\right)\right]= \label{geom_sum}\\
&=\frac {\delta_{jk} e^{-\beta\mathcal E_0}}{Z_\beta\beta\omega_j^2}\left[\prod_{\substack{l=1\\l\neq j}}^N\frac 1 {1-e^{-\beta \omega_l}}\right] \frac{1-e^{-\beta\omega_j}}{(1-e^{-\beta\omega_j})^2}\cos(\omega_j t)= \delta_{jk} e^{-\beta\mathcal E_0}\left[\prod_{\substack{l=1}}^N\frac 1 {1-e^{-\beta \omega_l}}\right]\frac{\cos(\omega_j t)}{Z_\beta\beta\omega_j^2}= \boxed{\delta_{jk} \frac{\cos(\omega_j t)}{\beta\omega_j^2}}\label{K0_eta}
\end{align}
In \cref{geom_sum} we noticed that 
\begin{align*}
\sum_{n_j=0}^{+\infty} n_j e^{-\beta\omega_j n_j} &= -\frac{\partial}{\partial (\beta\omega_j)} \frac 1{1-e^{-\beta\omega_j}}= \frac {e^{-\beta\omega_j}}{(1-e^{-\beta\omega_j})^2} \\
\sum_{n_j=0}^{+\infty} (n_j+1) e^{-\beta\omega_j n_j}&=\frac {e^{-\beta\omega_j}}{(1-e^{-\beta\omega_j})^2}+ \frac 1{1-e^{-\beta\omega_j}}= \frac {1}{(1-e^{-\beta\omega_j})^2}
\end{align*}
while in \cref{K0_eta} we substituted the partition function
\begin{equation}\label{Z_beta_harm}
Z_\beta = \tr \left\{e^{-\beta \hat H}\right\}= e^{-\beta \mathcal E_0}\sum_{\mathbf n \in \mathbb N_0^N} \bra{\mathbf n} e^{-\beta \sum_{l=1}^N \hat a_l^\dagger\hat a_l} \ket{\mathbf n}=e^{-\beta \mathcal E_0} \prod_{l=1}^N \sum_{n_l=0}^{+\infty} e^{-\beta n_l}= e^{-\beta \mathcal E_0} \prod_{l=1}^N \frac{1}{1-e^{-\beta\omega_l}}
\end{equation}
It turns out that, in the harmonic limit, the quantum Kubo-transformed auto-correlation for the configurations of the modes coincides with the classical counterpart. In particular, the latter is computed as
\begin{align}
&\left \langle \eta_j(t)\eta_k(0) \right \rangle_\beta = \left \langle \left[ \eta_j\cos(\omega_jt)+\frac 1{\omega_j}\sin(\omega_j t)\dot\eta_j\right] \eta_k \right \rangle_\beta  = \\
&=\delta_{jk} \frac {\cos(\omega_j t)} {Z_\beta^{\eta_j}} \left(-\frac{2}{\omega_j^2}\frac{\partial}{\partial \beta}\right)Z_\beta^{\eta_j} = \delta_{jk}\beta^{1/2} \cos(\omega_j t) \frac 1{\omega_j^2\beta^{3/2}}= \delta_{jk} \frac{\cos(\omega_j t)}{\beta \omega_j^2}
\end{align}
where
\begin{equation*}
Z_\beta^{\eta_j} \equiv\int_{\mathbb R} \mr d \eta_j \; e^{-\frac {\beta\omega_j^2}2\eta_j^2} = \sqrt{\frac{2\pi}{\beta\omega_j^2}}
\end{equation*}
\begin{align}
&\boxed{K_{q_j, q_k}^0(t) }=\frac 1{Z_\beta \beta }  \int_0^\beta \mathrm d \lambda \; \tr \left\{ e^{ -(\beta -\lambda) \hat H_0} \hat q_j e^{-\lambda \hat H_0} e^{i\hat H_0 t} \hat q_k e^{-i\hat H_0 t}\right\}= \\
&=\frac 2{(N+1)Z_\beta \beta} \sum_{l,l'=1}^N \sin \left(\frac{\pi j l}{N+1}\right)\sin \left(\frac{\pi k l'}{N+1}\right)\int_0^\beta \mathrm d \lambda \; \tr \left\{ e^{-(\beta -\lambda) \hat H_0} \hat \eta_l e^{-\lambda \hat H_0} e^{i\hat H_0 t} \hat \eta_{l'} e^{-i\hat H_0 t}\right\}= \nn \\
&=\frac {2\delta_{jk}}{(N+1)} \sum_{l=1}^N \sin \left(\frac{\pi j l}{N+1}\right)\sin \left(\frac{\pi k l}{N+1}\right)\frac{\cos(\omega_l t)}{\beta\omega_l^2}=\boxed{\frac {2\delta_{jk}}{(N+1)} \sum_{l=1}^N \sin \left(\frac{\pi j l}{N+1}\right)\sin \left(\frac{\pi k l}{N+1}\right)K_{\eta_l, \eta_l }^0(t)\label{K0_q}}
\end{align}
\end{appendices}


\begin{thebibliography}{10}

\bibitem{fermi_pasta_ulam1955}
E.~Fermi, J.~Pasta, and S.~Ulam, ``Studies of nonlinear problems {I}, {L}os
  {A}lamos {R}eport {LA} 1940, 1955,'' 1974.

\bibitem{gavallotti}
G.~Gallavotti, {\em The Fermi-Pasta-Ulam Problem: A Status Report}.
\newblock Berlin, Heidelberg: Springer, 2008.

\bibitem{berman2005}
G.~P. Berman and F.~M. Izrailev, ``The fermi–pasta–ulam problem: Fifty
  years of progress,'' {\em Chaos: An Interdisciplinary Journal of Nonlinear
  Science}, vol.~15, no.~1, p.~015104, 2005.

\bibitem{bachelard2008}
R.~Bachelard, C.~Chandre, D.~Fanelli, X.~Leoncini, and S.~Ruffo, ``Abundance of
  regular orbits and nonequilibrium phase transitions in the thermodynamic
  limit for long-range systems,'' {\em Phys. Rev. Lett.}, vol.~101, p.~260603,
  Dec 2008.

\bibitem{gaison2014}
J.~Gaison, S.~Moskow, J.~D. Wright, and Q.~Zhang, ``Approximation of polyatomic
  fpu lattices by kdv equations,'' {\em Multiscale Modeling and Simulation},
  vol.~12, pp.~953--995, 2014.

\bibitem{carati2007}
A.~Carati, L.~Galgani, A.~Giorgilli, and S.~Paleari, ``Fermi-pasta-ulam
  phenomenon for generic initial data,'' {\em Phys. Rev. E}, vol.~76,
  p.~022104, Aug 2007.

\bibitem{amati2019}
T.~S. G.~Amati, H.~Meyer, ``Memory effects in the fermi–pasta–ulam model,''
  {\em Journal of Statistical Physics}, vol.~174, pp.~219--257, 2019.

\bibitem{burin2019}
A.~L. Burin, A.~O. Maksymov, M.~Schmidt, and I.~Y. Polishchuk, ``{Chaotic
  Dynamics in a Quantum Fermi–Pasta–Ulam Problem},'' {\em Entropy},
  vol.~21, no.~1, 2019.

\bibitem{edwards1971}
J.~T. Edwards and D.~J. Thoules, ``{Regularity of the density of states in
  Anderson's localized electron model},'' {\em Journal of Physics C: Solid
  State Physics}, vol.~4, pp.~453--457, mar 1971.

\bibitem{leitner1997}
D.~M. Leitner and P.~G. Wolynes, ``{Quantization of the Stochastic Pump Model
  of Arnold Diffusion},'' {\em Phys. Rev. Lett.}, vol.~79, pp.~55--58, Jul
  1997.

\bibitem{ivic2006}
Z.~Ivić and G.~Tsironis, ``{Biphonons in the $\beta$-Fermi–Pasta–Ulam
  model},'' {\em Physica D: Nonlinear Phenomena}, vol.~216, no.~1, pp.~200 --
  206, 2006.
\newblock Nonlinear Physics: Condensed Matter, Dynamical Systems and
  Biophysics.

\bibitem{kibey2015}
A.~Kibey, R.~Sonone, B.~Dey, and J.~C. Eilbeck, ``{Quantization of
  $\beta$-Fermi–Pasta–Ulam lattice with nearest and next-nearest neighbor
  interactions},'' {\em Physica D: Nonlinear Phenomena}, vol.~294, pp.~43 --
  53, 2015.

\bibitem{danshita2014}
I.~Danshita, R.~Hipolito, V.~Oganesyan, and A.~Polkovnikov, ``{Quantum damping
  of Fermi–Pasta–Ulam revivals in ultracold Bose gases},'' {\em Progress of
  Theoretical and Experimental Physics}, vol.~2014, no.~4, p.~043I03, 2014.

\bibitem{feynman1948}
R.~P. Feynman, ``Space-time approach to non-relativistic quantum mechanics,''
  {\em Rev. Mod. Phys.}, vol.~20, pp.~367--387, Apr 1948.

\bibitem{tuckerman1993}
M.~E. Tuckerman, B.~J. Berne, G.~J. Martyna, and M.~L. Klein, ``{Efficient
  molecular dynamics and hybrid Monte Carlo algorithms for path integrals},''
  {\em The Journal of Chemical Physics}, vol.~99, no.~4, pp.~2796--2808, 1993.

\bibitem{chandler1981}
D.~Chandler and P.~G. Wolynes, ``Exploiting the isomorphism between quantum
  theory and classical statistical mechanics of polyatomic fluids,'' {\em The
  Journal of Chemical Physics}, vol.~74, no.~7, pp.~4078--4095, 1981.

\bibitem{carati2018}
A.~Carati and A.~Ponno, ``Chopping time of the fpu $\alpha$-model,'' {\em
  Journal of Statistical Physics}, vol.~170, pp.~883--894, Mar 2018.

\bibitem{carati2015}
A.~Carati, A.~Maiocchi, L.~Galgani, and G.~Amati, ``{The Fermi--Pasta--Ulam
  System as a Model for Glasses},'' {\em Mathematical Physics, Analysis and
  Geometry}, vol.~18, p.~31, Nov 2015.

\bibitem{martyna1996}
G.~J. Martyna, M.~E. Tuckerman, D.~J. Tobias, and M.~L. Klein, ``{Explicit
  reversible integrators for extended systems dynamics},'' {\em Molecular
  Physics}, vol.~87, no.~5, pp.~1117--1157, 1996.

\bibitem{hansen_mdonald1986}
J.-P. Hansen and I.~Ranald~McDonald, ``{Theory of simple liquids, Second
  Edition},'' 1986.

\bibitem{gosson2017}
M.~{de Gosson}, ``{Quantum Hamonic Analysis of the Density Matrix: Basics},''
  in {\em arXiv:1703.00889}, Mar. 2017.

\bibitem{hans_mc_ch4}
J.-P. Hansen and I.~R. McDonald, ``Chapter 4 - distribution function
  theories,'' in {\em Theory of Simple Liquids (Fourth Edition)} (J.-P. Hansen
  and I.~R. McDonald, eds.), pp.~105 -- 147, Oxford: Academic Press, fourth
  edition~ed., 2013.

\bibitem{czako2010}
G.~Czakó, A.~L. Kaledin, and J.~M. Bowman, ``{A practical method to avoid
  zero-point leak in molecular dynamics calculations: Application to the water
  dimer},'' {\em The Journal of Chemical Physics}, vol.~132, no.~16, p.~164103,
  2010.

\bibitem{czako2010_bis}
G.~Czakó, A.~L. Kaledin, and J.~M. Bowman, ``Zero-point energy constrained
  quasiclassical, classical, and exact quantum simulations of isomerizations
  and radial distribution functions of the water trimer using an ab initio
  potential energy surface,'' {\em Chemical Physics Letters}, vol.~500, no.~4,
  pp.~217 -- 222, 2010.

\bibitem{craig2004}
I.~R. Craig and D.~E. Manolopoulos, ``{Quantum statistics and classical
  mechanics: Real time correlation functions from ring polymer molecular
  dynamics},'' {\em The Journal of Chemical Physics}, vol.~121, no.~8,
  pp.~3368--3373, 2004.

\bibitem{habershon2013}
S.~Habershon, D.~E. Manolopoulos, T.~E. Markland, and T.~F. Miller,
  ``Ring-polymer molecular dynamics: Quantum effects in chemical dynamics from
  classical trajectories in an extended phase space,'' {\em Annual Review of
  Physical Chemistry}, vol.~64, no.~1, pp.~387--413, 2013.
\newblock PMID: 23298242.

\bibitem{tuckerman2010}
M.~Tuckerman, {\em Statistical Mechanics: Theory and Molecular Simulation}.
\newblock 2010.

\bibitem{welsch2016}
R.~Welsch, K.~Song, Q.~Shi, S.~C. Althorpe, and T.~F. Miller,
  ``{Non-equilibrium dynamics from RPMD and CMD},'' {\em The Journal of
  Chemical Physics}, vol.~145, no.~20, p.~204118, 2016.

\bibitem{mellet2015}
A.~Mellet and S.~Merino-Aceituno, ``Anomalous energy transport in fpu-$\beta$
  chain.,'' {\em Journal of Statistical Physics}, vol.~160, pp.~583--621, 2015.

\bibitem{danieli2017}
C.~Danieli, D.~K. Campbell, and S.~Flach, ``Intermittent many-body dynamics at
  equilibrium,'' {\em Phys. Rev. E}, vol.~95, p.~060202, Jun 2017.

\bibitem{wang2019}
J.~Wang, T.~xing Liu, X.~zhi Luo, X.-L. Xu, and N.~Li, ``Anomalous energy
  diffusion in two-dimensional nonlinear lattices,'' 2019.

\bibitem{benettin2008}
G.~Benettin and G.~Gradenigo, ``A study of the fermi–pasta–ulam problem in
  dimension two,'' {\em Chaos: An Interdisciplinary Journal of Nonlinear
  Science}, vol.~18, no.~1, p.~013112, 2008.

\bibitem{stoppato2016}
M.~Stoppato, ``{The quantum Fermi Pasta Ulam problem},'' 2016.

\bibitem{zanardi2004}
P.~Zanardi, D.~A. Lidar, and S.~Lloyd, ``{Quantum Tensor Product Structures are
  Observable Induced},'' {\em Phys. Rev. Lett.}, vol.~92, p.~060402, Feb 2004.

\bibitem{yoshida1988}
H.~Yoshida, ``Non-integrability of the truncated toda lattice hamiltonian at
  any order,'' {\em Communications in Mathematical Physics}, vol.~116,
  pp.~529--538, Dec 1988.

\bibitem{anderson1994}
A.~Anderson, ``Canonical transformations in quantum mechanics,'' {\em Annals of
  Physics}, vol.~232, no.~2, pp.~292 -- 331, 1994.

\bibitem{martyna1992}
G.~J. Martyna, M.~L. Klein, and M.~Tuckerman, ``Nosé–hoover chains: The
  canonical ensemble via continuous dynamics,'' {\em The Journal of Chemical
  Physics}, vol.~97, no.~4, pp.~2635--2643, 1992.

\bibitem{tuckerman2006}
M.~E. Tuckerman, J.~Alejandre, R.~L{\'{o}}pez-Rend{\'{o}}n, A.~L. Jochim, and
  G.~J. Martyna, ``{A Liouville-operator derived measure-preserving integrator
  for molecular dynamics simulations in the isothermal{\textendash}isobaric
  ensemble},'' {\em Journal of Physics A: Mathematical and General}, vol.~39,
  pp.~5629--5651, apr 2006.

\bibitem{hall1984}
R.~W. Hall and B.~J. Berne, ``Nonergodicity in path integral molecular
  dynamics,'' {\em The Journal of Chemical Physics}, vol.~81, no.~8,
  pp.~3641--3643, 1984.

\bibitem{tuckerman1999_comm}
M.~E. Tuckerman and G.~J. Martyna, ``{Comment on “Simple reversible molecular
  dynamics algorithms for Nosé–Hoover chain dynamics” },'' {\em The
  Journal of Chemical Physics}, vol.~110, no.~7, pp.~3623--3625, 1999.

\bibitem{yoshida1990}
H.~Yoshida, ``Construction of higher order symplectic integrators,'' {\em
  Physics Letters A}, vol.~150, no.~5, pp.~262 -- 268, 1990.

\bibitem{tuckerman_1999}
M.~E. Tuckerman, C.~J. Mundy, and G.~J. Martyna, ``{On the classical
  statistical mechanics of non-Hamiltonian systems},'' {\em Europhysics Letters
  ({EPL})}, vol.~45, pp.~149--155, jan 1999.

\bibitem{johnson2002}
S.~C. Johnson and T.~D. Gutierrez, ``Visualizing the phonon wave function,''
  {\em American Journal of Physics}, vol.~70, no.~3, pp.~227--237, 2002.

\bibitem{lievens2008}
S.~Lievens, N.~I. Stoilova, and J.~V. der Jeugt, ``A linear chain of
  interacting harmonic oscillators: solutions as a wigner quantum system,''
  {\em Journal of Physics: Conference Series}, vol.~128, p.~012028, aug 2008.

\bibitem{picasso2016}
L.~Picasso, {\em {Lectures in Quantum Mechanics}}.
\newblock 2016.

\bibitem{he2018}
Y.~He and F.~Yange, ``{Some recurrence formulas for the Hermite polynomials and
  their squares},'' {\em Open Mathematics}, vol.~16, pp.~553--560, 2018.

\end{thebibliography}
\end{document}